\documentclass[prd,twocolumn,nofootinbib,showpacs,superscriptaddress]{revtex4-1}

\usepackage{amsfonts}
\usepackage{amsmath}
\usepackage{amssymb}
\usepackage{bm}
\usepackage{dcolumn}
\usepackage{graphicx}
\usepackage{graphics}
\usepackage[latin1]{inputenc}
\usepackage{latexsym}
\usepackage{rotating}
\usepackage{hyperref}
\usepackage[all]{hypcap} 
\usepackage{xspace} 
\usepackage[usenames]{color}
\usepackage{mathrsfs}

\usepackage{ulem}
\definecolor {darkgreen}{rgb}{0.2,0.7,0.2}

\newcommand\be{\begin{equation}}
\newcommand\ba{\begin{eqnarray}}
\newcommand\ee{\end{equation}}
\newcommand\ea{\end{eqnarray}}
\newcommand\bw{\begin{widetext}}
\newcommand\ew{\end{widetext}}

\newcommand{\nn}{\nonumber}

\newcommand{\MIN}{{\mbox{\tiny min}}}
\newcommand{\MAX}{{\mbox{\tiny max}}}

\newcommand{\N}{{\mbox{\tiny N}}}
\newcommand{\mrm}{\mathrm}

\newcommand{\LE}{{\mbox{\tiny LE}}}
\newcommand{\txt}[1]{{\textrm{\tiny{#1}}}}
\newcommand{\ItoQ}{$\bar{I}$--$\bar{Q}$\xspace}
\newcommand{\OmegaEq}{\Omega_{\txt{Eq}}^{\txt{Surf}}}
\newcommand{\OmegaBk}{\Omega_{\txt{bk}}}

\newcommand{\Eq}{{\mbox{\tiny Eq}}}

\begin{document}
\title{Why I-Love-Q}    

\author{Kent Yagi}
\affiliation{Department of Physics, Montana State University, Bozeman, MT 59717, USA.}

\author{Leo C. Stein}
\affiliation{Center for Radiophysics and Space Research, Cornell
    University, Ithaca, NY 14853 USA}

\author{George Pappas}
\affiliation{School of Mathematical Sciences, The University of Nottingham,
University Park, Nottingham NG7 2GD, United Kingdom}

\author{Nicol\'as Yunes}
\affiliation{Department of Physics, Montana State University, Bozeman, MT 59717, USA.}

\author{Theocharis A. Apostolatos}
\affiliation{Section of Astrophysics, Astronomy and Mechanics, Department of Physics, University of Athens, Panepistimiopolis Zografos GR15783, Athens, Greece}

\begin{abstract} 

Black holes are said to have no hair because all of their multipole moments can be expressed in terms of just their mass, charge and spin angular momentum. 
The recent discovery of approximately equation-of-state-independent relations among certain multipole moments in neutron stars suggests that they are also approximately bald. 
We here explore the yet unknown origin for this universality. 
First, we investigate which region of the neutron star's interior and of the equation of state is most responsible for the universality.  
We find that the universal relation between the moment of inertia and the quadrupole moment is dominated by the star's \emph{outer-core}, a shell of width  (50--95)\% of the total radius, which corresponds to the density range $(10^{14}$--$10^{15})$g/cm$^3$. 
In this range, realistic neutron star equations of state are not sufficiently similar to each other to explain the universality observed.
Second, we study the impact on the universality of approximating stellar isodensity contours as self-similar ellipsoids.   
An analytical calculation in the non-relativistic limit reveals that the shape of the ellipsoids \emph{per se} does not affect the universal relations much, but relaxing the self-similarity assumption can completely destroy it. 
Third, we investigate the eccentricity profiles of rotating relativistic stars and find that the stellar eccentricity is roughly constant, with variations of roughly (20--30)\% in the region that matters to the universal relations.
Fourth, we repeat the above analysis for differentially-rotating, non-compact, regular stars and find that this time the eccentricity is not constant, with variations that easily exceed 100\%, and moreover, universality is lost.
These findings suggest that universality arises as an \emph{emergent approximate symmetry}: as one flows in the stellar-structure phase space from non-compact star region to the relativistic star region, the eccentricity variation inside stars decreases, leading to approximate self-similarity in their isodensity contours, which then leads to the universal behavior observed in their exterior multipole moments.

\end{abstract}

\pacs{04.30.Db,97.60.Jd}
\date{\today}
\maketitle

\section{Introduction}

Astrophysical observations of neutron stars (NSs) may reveal one of the most important ``known unknowns'' of nuclear physics: the relation between density and pressure, the so-called \emph{equation of state} (EoS), of supra-nuclear matter~\cite{lattimer_prakash2001,lattimer-prakash-review,Lattimer:2012nd}. One can constrain the EoS by measuring at least two NS observables that depend strongly on the star's internal structure. The most well-studied observables are the NS mass and radius (see e.g.~\cite{ozel-review} and references therein). Although the latter has not been measured to better than 20\% accuracy, a Bayesian analysis has allowed some constraints on the EoS~\cite{steiner-lattimer-brown,Lattimer:2013hma} and on certain quantities in nuclear physics, such as the nuclear symmetry energy~\cite{lattimer-lim,Lattimer:2014sga}.   

Some relations among certain NS observables depend very weakly on the EoS. Such approximate universality has been found, for example, between the NS binding energy and compactness~\cite{1989ApJ...340..426L,1997PhR...280....1P,lattimer_prakash2001}, between the mass-shedding (Keplerian) frequency for rotating configurations and compactness for non-rotating configurations~\cite{2004Sci...304..536L,2009A&A...502..605H}, among NS oscillation modes~\cite{andersson-kokkotas-1998,benhar2004,tsui-leung,lau}, among certain tidal parameters~\cite{Yagi:2013sva}, among gravitational-wave (GW) observables from NS binaries~\cite{kiuchi,kyutoku,Kyutoku:2011vz,2012PhRvL.108a1101B,Bernuzzi:2014kca,Takami:2014zpa} and among the compactness, a dimensionless spin parameter and the effective gravitational acceleration on the surface of a rapidly-rotating NSs~\cite{AlGendy:2014eua}. 

A stronger, yet still approximate universality has been recently found among the moment of inertia ($I$), the tidal and rotational Love numbers (or tidal deformability for the former) and the quadrupole moment ($Q$) of NSs and quark stars (QSs)~\cite{I-Love-Q-Science,I-Love-Q-PRD} (see also~\cite{1994ApJ...424..846R,lattimer_prakash2001,bejger,lattimer-schutz,urbanec} for universal relations among $I$, $Q$ and the NS compactness). The relations were confirmed by~\cite{lattimer-lim} through a detailed study of different EoSs, by~\cite{maselli} for NS binary system that are strongly and dynamically tidally deformed, and by~\cite{I-Love-Q-B} for magnetized NSs, provided they are not slowly-spinning magnetars. These relations also hold for rapidly-rotating stars, as demonstrated numerically by~\cite{Pappas:2013naa,Chakrabarti:2013tca,Yagi:2014bxa} and analytically in~\cite{Stein:2013ofa,piecewise-extension}, with the latter in the Newtonian limit.

The I-Love-Q relations have several important applications. First, a measurement of any one of these quantities automatically allows for the determination of the other two, without having to know the EoS. Second, these relations allow for model-independent and EoS-independent strong-field test of General Relativity (GR). For example, Refs.~\cite{I-Love-Q-Science,I-Love-Q-PRD} showed that measuring the NS Love number with GW observations and the moment of inertia from binary pulsars can place very stringent constraints on parity-violations in gravity~\cite{jackiw,CSreview}. The relations were recently studied in~\cite{Sham:2013cya,Pani:2014jra} for other classes of modified theories of gravity. Third, the I-Love-Q relations break degeneracies between certain parameters in astrophysical and GW observations. For example, the Love-Q relation can be used to break the degeneracy between $Q$ and spins in GW observations of spin-aligned binary NSs~\cite{I-Love-Q-Science,I-Love-Q-PRD}. Moreover, the I-Q relation can be used to break degeneracies in X-ray pulse profile observations with NICER~\cite{2012SPIE.8443E..13G} and LOFT~\cite{2012AAS...21924906R,2012SPIE.8443E..2DF}, as shown in~\cite{Psaltis:2013fha,Baubock:2013gna}. 

The approximately universal I-Love-Q relations resemble the celebrated \emph{no-hair} relations for black holes (BHs)~\cite{MTW,robinson,israel,israel2,hawking-uniqueness0,hawking-uniqueness,carter-uniqueness}. Astrophysical (uncharged) BHs are said to have no hair because all of their \emph{multipole moments}, i.e.~the coefficients in a multipolar expansion of the gravitational field far from the source, can be written completely in terms of just their mass and spin angular momentum~\cite{MTW,robinson,israel,israel2,hawking-uniqueness0,hawking-uniqueness,carter-uniqueness}. Ordinary stars or compact stars need not be bald, and thus, their multipole moments could depend strongly on the star's EoS. 

Nonetheless, the approximate universal relations between $I$ and $Q$ and a new relation between $Q$ and the current octupole~\cite{Pappas:2013naa} for compact stars suggest the existence of  an approximate no-hair relation. Recently, in fact, approximately universal no-hair relations were found for lower multipole moments $(\ell \leq 10)$ in terms of the first three (the mass monopole, the current dipole and the mass quadrupole), albeit in the non-relativistic, Newtonian limit~\cite{Stein:2013ofa} and with a certain elliptical isodensity approximation~\cite{Lai:1993ve} that models stellar isodensity contours as self-similar ellipsoids. Some of these results have been recently confirmed numerically and in full GR for the mass hexadecapole and the mass quadrupole moments~\cite{Yagi:2014bxa} of compact stars.

But why does such approximate universality hold in the first place? When discovering the I-Love-Q relations, Refs.~\cite{I-Love-Q-Science,I-Love-Q-PRD} suggested two non-excluding hypothesis. One of them was that universality holds because all EoSs are somewhat similar in the NS region that dominates the I-Love-Q calculation, i.e.~in the outer core. Indeed, our ignorance of the EoS is smaller toward the crust than toward the inner core of NSs. The other hypothesis was that universality holds because the I-Love-Q relations must approach the BH limit as the stellar compactness increases, and of course, for BHs these relations must be universal. These explanations are not quite satisfactory, however, because the EoS does vary in the outer core and NSs are not nearly as compact as BHs; in fact, there is no continuous limit from a NS sequence to a BH. 

\subsection{Methodology and Executive Summary}
\label{sec:intro-summary}

The purpose of this paper is to gain a better understanding of why approximate universal relations among multipole moments, in particular between $I$ and $Q$, hold for NSs. Since rotational Love numbers are directly related to $I$ and $Q$~\cite{I-Love-Q-PRD}, such reasoning will also explain the origin of the universal relation between the rotational Love number and $I$ or $Q$.
 We achieve this goal by investigating the universal relations in detail and addressing the following questions: 
\begin{itemize}
\item[(I)] Which part of the EoS and NS interior region is most responsible for the universality? Are the
realistic EoSs similar to each other in this region? 
\item[(II)] Which assumption in the elliptical isodensity approximation is most important in the universality? 
\item[(III)] Does the elliptical isodensity approximation hold for realistic relativistic stars? What do the eccentricity radial profiles of such stars look like?
\item[(IV)] Does the elliptical isodensity approximation hold for realistic non-compact stars? Are there universal relations for such stars?
\end{itemize}

The first question is tackled as follows. We consider slowly-rotating NSs and QSs in the Hartle-Thorne approximation~\cite{hartle1967,hartlethorne} with an extended piecewise polytropic EoS~\cite{Read:2008iy}, characterized by five free parameters: $(p_1,\Gamma_1,\Gamma_2,\Gamma_3,\rho_{1})$: $p_1$ changes the overall normalization of the EoS, $\Gamma_1$ affects both the inner and outer core regions, $\Gamma_2$ and $\Gamma_3$ only modify the inner core region, and $\rho_{1}$ determines the transition density between the $\Gamma_1$ and $\Gamma_2$ regions. A version of such a piecewise polytropic EoS has been shown to recover realistic EoSs for suitable choices of the free parameters~\cite{Read:2008iy}. We vary $(p_1,\Gamma_1,\Gamma_2,\Gamma_3,\rho_1)$ from a fiducial set that recovers the realistic SLy EoS~\cite{SLy} to study how each parameter affects the universal I-Q relation, the radial profiles of the energy density and (the integrand of) the moment of inertia and quadrupole moment.

We find that, although all parameters have a large effect on the mass-radius relation, $p_1$, $\Gamma_{3}$ and $\rho_1$ hardly affect the relation at all, while $\Gamma_1$ and $\Gamma_2$ affect the I-Q relation the most for low- and high-compactness stars respectively and in a linear fashion. These results are confirmed by computing the radial profile of the integrand of the moment of inertia and quadrupole moment, which are most affected when we vary $\Gamma_{1}$ and $\Gamma_{2}$. We find that the moment of inertia and quadrupole moment are dominated by contributions within a shell inside the star of size 50--95\% of the total radius. 

\begin{figure}[tb]
\centering
\includegraphics[width=\columnwidth,clip=true]{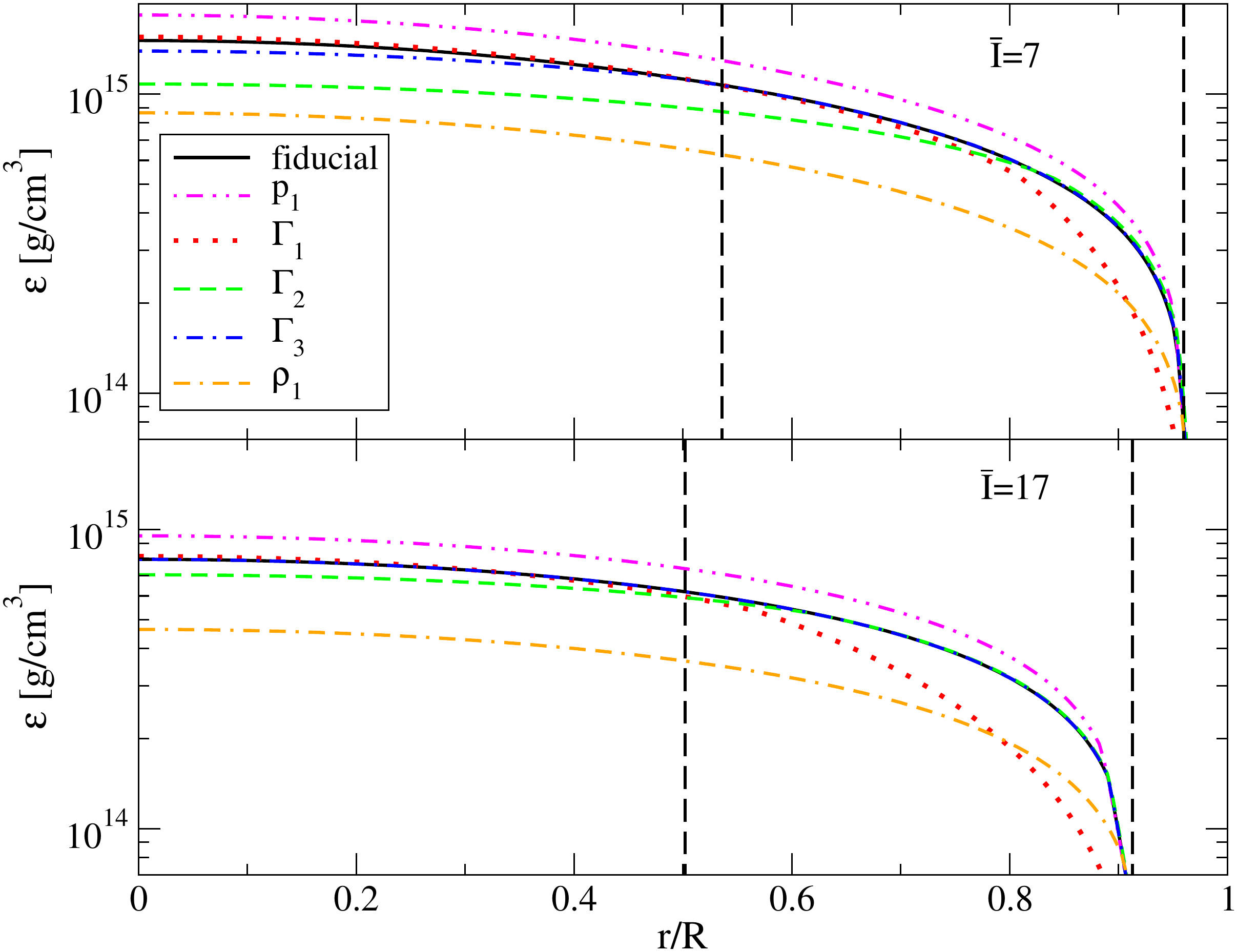}
\caption{
(Color online) 
 Energy-density profile $\epsilon$ of non-rotating NSs with the dimensionless moment of inertia of $\bar{I}=7$ ($M \approx 1.8M_\odot$) (top) and $\bar{I}=17$ ($M \approx 1.1M_\odot$) (bottom) for the fiducial SLy EoS, and with 30\% variations of piecewise polytrope parameters $p_1$, $\Gamma_1$, $\Gamma_2$, $\Gamma_3$ and $\rho_1$. $R$ in the horizontal axis refers to the NS radius for non-rotating configurations. The region in between the vertical dashed lines corresponds to the one that contributes the most to the calculation of $I$ and $Q$. Observe that $\bar{I}$ is mostly affected by the EoS in the range $\epsilon = (10^{14}-10^{15})$g/cm$^3$. 
\label{fig:density-profile}
}
\end{figure}

Figure~\ref{fig:density-profile} presents the radial profile of the energy density for non-rotating NS configurations with two different choices of dimensionless moment of inertia $\bar{I} = I/M^{3}$, where $M$ is the stellar mass. We vary each piecewise polytrope parameter by 30\% from the corresponding value of the fiducial SLy EoS. The 90\% contribution to the moment of inertia and the quadrupole moment comes from the radial region in vertical dashed lines. One sees that the I-Q universality is mostly affected by the energy densities in the range $10^{14}$--$10^{15}$g/cm$^3$. Although the NS EoSs all have a similar slope within such a region, they can differ by as much as $\sim 17\%$, and hence this fact alone cannot explain the $\mathcal{O}(1)$\% universality in the I-Q relation. 

Since the region that matters the most for the universal relations is quite far from the NS inner core, we suspect that a Newtonian analysis might suffice to understand the reasons behind the universality. To determine if this is the case, we construct rapidly-rotating NS solutions using the \texttt{RNS} code~\cite{stergioulas_friedman1995}. These solutions confirm that although relativistic and rotational effects make stars more centrally-condensed, the energy density profiles are not modified much in the region that matters. Thus, an analysis carried out in the non-relativistic limit should be sufficient to understand why universality holds, which then brings us to the second question.  

This question is tackled by studying the universal relations in the non-relativistic, {\emph{Newtonian}} limit, extending the work in~\cite{Stein:2013ofa} by relaxing the elliptical isodensity approximation. This approximation consists of three main conditions~\cite{Lai:1993ve}: (i) that constant density contours are self-similar surfaces; (ii) that such surfaces are ellipsoids; and (iii) that the isodensity profile in terms of the isodensity radius is identical to that of a non-rotating star of the same volume. We relax each of these, one by one, and find that condition (ii) does not affect the universal relations at all, but conditions (i) and (iii) can destroy them. In particular, if one allows the stellar eccentricity to depend on the radial coordinate rather than being a constant, the EoS universality can be lost. This suggests that the self-similarity of the elliptical isodensity approximation plays an important role in the universality. 

\begin{figure}[tb]
\centering
\includegraphics[width=\columnwidth,clip=true]{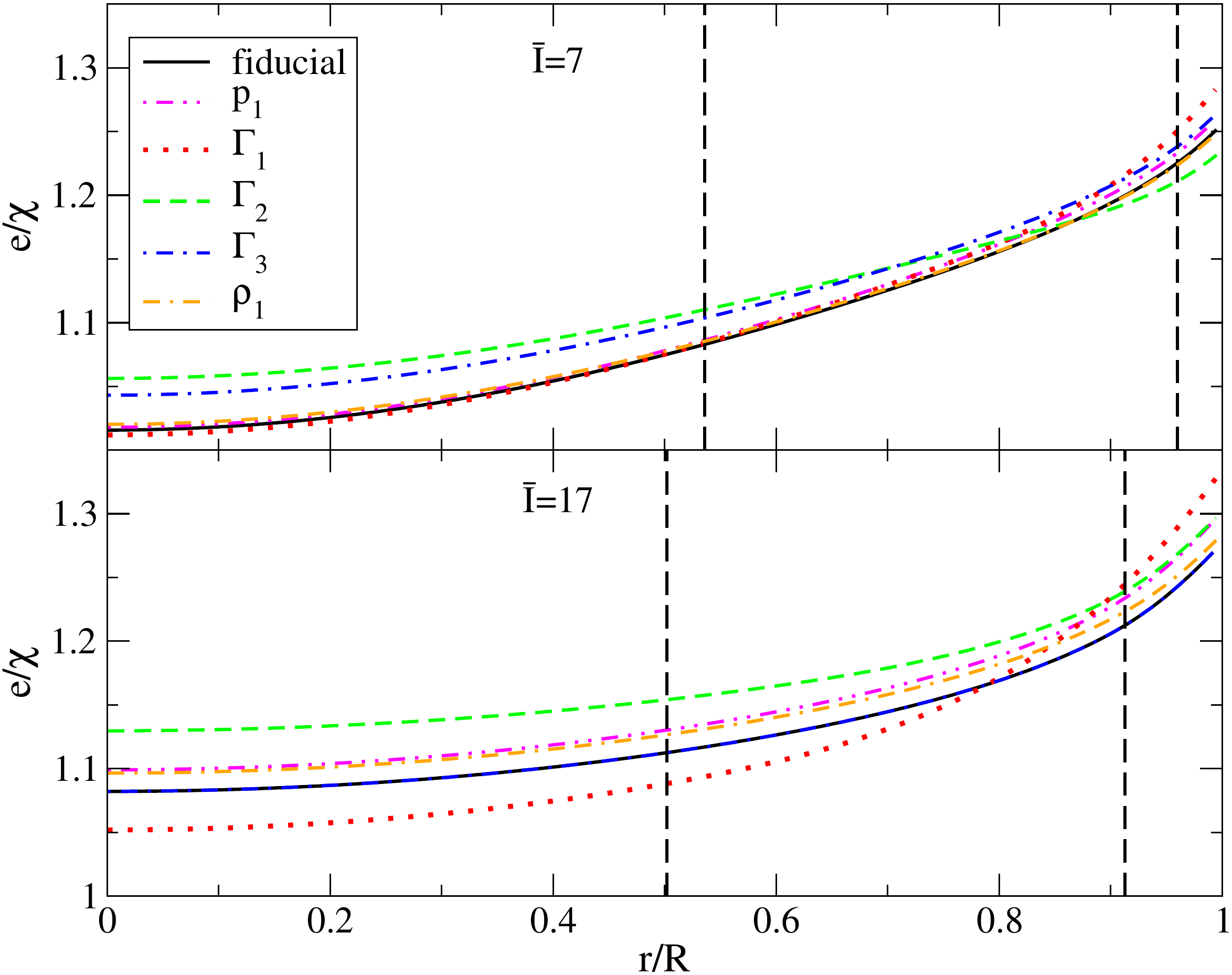}
\caption{
(Color online) 
 Radial profile of the stellar eccentricity (normalized by the dimensionless spin parameter $\chi$) for slowly-rotating NSs with $\bar{I}=7$ (top) and $\bar{I}=17$ (bottom) for the SLy EoS, and with 30\% variations of $p_1$, $\Gamma_1$, $\Gamma_2$, $\Gamma_3$ and $\rho_1$. The meaning of vertical lines are the same as in Fig.~\ref{fig:density-profile}. Notice that since we are working within the slow-rotation approximation, $e/\chi$ does not depend on the NS spin. Observe that the eccentricity varies by $\sim 10\%$ within the region that matters to the universality.
\label{fig:ecc-profile}
}
\end{figure}

But are the isodensity contours of realistic NSs approximately self-similar? This is the focus of the third question we tackle, which we answer by constructing both slowly-rotating and rapidly-rotating NS solutions and extracting the stellar eccentricity in terms of the radial coordinate. Figure~\ref{fig:ecc-profile} shows the eccentricity profile for slowly-rotating NSs with various EoSs. One sees that in the region that matters [i.e.~$r \in (50\%,95\%) R$, with $R$ the stellar radius], the eccentricity only changes by $\sim 10 \%$ for slowly-rotating stars. For rapidly-rotating NSs, we find that the eccentricity variation inside the star is always below $(20-30)\%$. We extend the Newtonian analysis described in the previous paragraph by using such realistic eccentricity profiles to correct the elliptical isodensity approximation. We find that this leads to changes of $\sim 10$\% at most relative to the universal relations obtained with constant eccentricity, e.g.~if the \ItoQ relations are universal to roughly 1\%, the variations in the eccentricity correct these relations to 0.1\%. This supports the validity of the approximation for realistic NSs.

Finally, we study whether non-compact stars are also approximately elliptically self-similar and whether the universal relations hold in this case. This is the basis of the fourth question described earlier, which we tackle by constructing realistic, differentially-rotating, non-compact stellar solutions with the publicly-available \textsc{ESTER} code~\cite{2013LNP...865...49R,2013A&A...552A..35E} and extracting the eccentricity profile. We find that the eccentricity variation easily exceeds 100\% inside the star, varying much more than in the relativistic, compact-star case. If the elliptical isodensity approximation is (at least in part) responsible for the universality, then we would expect universality to be lost in such stars. We find that this is indeed the case: the $I$-$Q$ relation is highly sensitive to variations in the opacity law, which we use as a proxy for EoS variation. 

\subsection{A Phenomenological Picture for Universality}

These findings suggest that the EoS universality in NSs and QSs arises due to an (approximate) \emph{emergent symmetry}. Emergent symmetries are common in quantum field theory and condensed matter physics, for example in the study of chiral spin liquids~\cite{PhysRevB.85.195126}. The main idea here is that as some set of parameters (usually the energy or temperature of the system) are tuned beyond a given threshold, the description of the system acquires an approximate symmetry that is not present in general. 

\begin{figure}[tb]
\centering
\includegraphics[width=\columnwidth,clip=false]{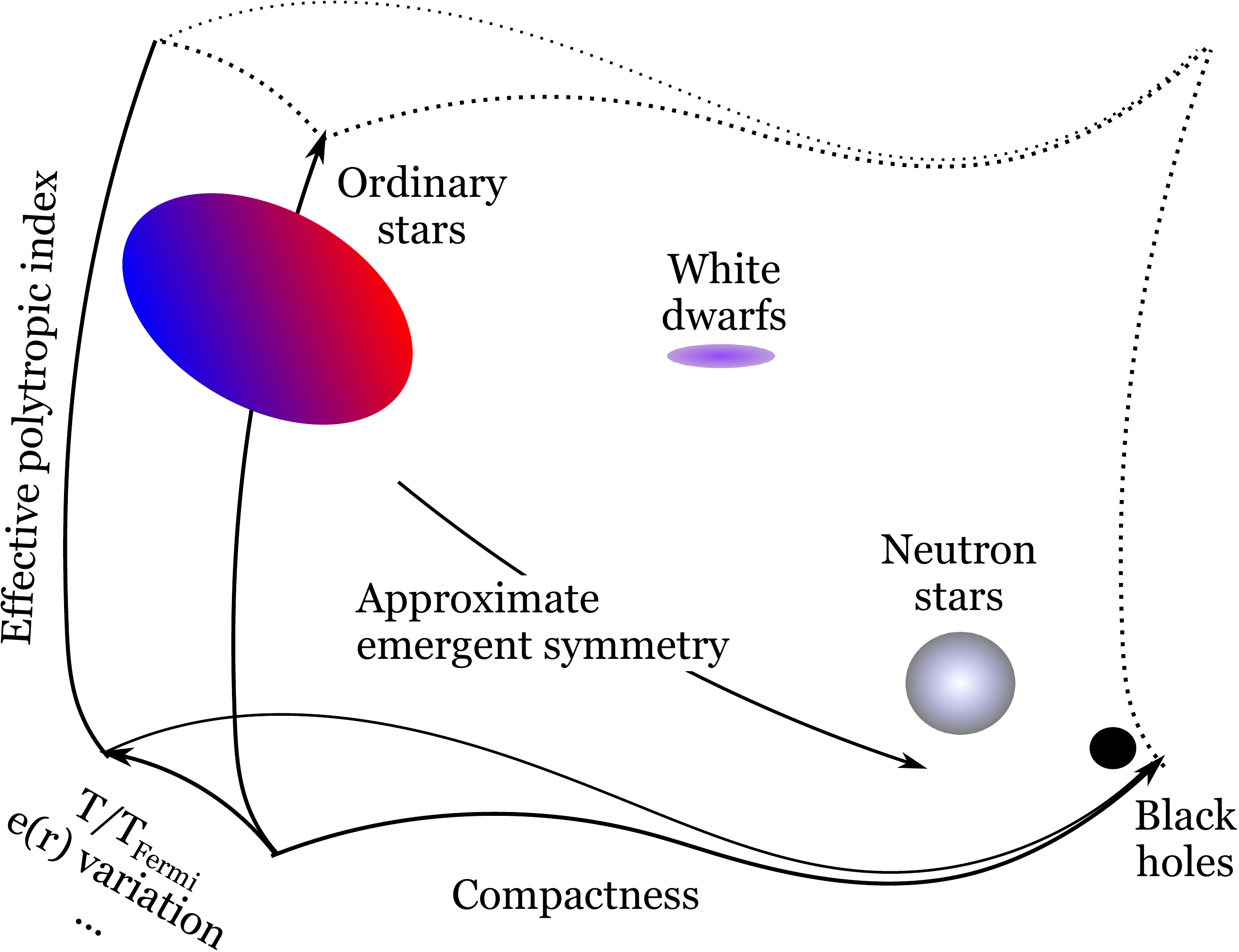}
\caption{
(Color online) 
Schematic diagram of the stellar phase space. Compact stars live in one corner of this space, while non-compact stars live in another corner. As one flows from the latter to the former along a particular two-dimensional subspace, spanned by compactness $C$ and polytropic index $n$, an approximate symmetry arises: isodensity contours become approximately self-similar. This approximate symmetry is then responsible for the universality in the exterior multipole moments of compact stars. 
\label{fig:schematic-diag-emergence}
}
\end{figure}

Consider a phase space with coordinates that represent a quantity that
characterizes a star, such as its compactness, temperature, magnetic field strength, etc.
Consider now a two-dimensional subspace, with one coordinate being the stellar compactness and 
the other the EoS polytropic index $n$. NSs live in the large compactness and small $n$ region, 
while regular, non-compact stars live in the smaller compactness and larger $n$ region, 
as depicted in Fig.~\ref{fig:schematic-diag-emergence}.
As one flows from the non-compact to the relativistic, compact star region in this projected two-dimensional subspace, 
eccentricity radial profiles become less variable and nearly constant throughout the star, i.e.~a radial remapping leaves the eccentricity profile approximately invariant. This suggests the emergence of an 
approximate symmetry associated with this isodensity self-similar invariance, 
which in turn, leads to the universality in the multipole relations for relativistic stars. 
As one further increases the compactness, one approaches the BH region in phase space,
where the universality becomes exact, as expressed by the no-hair theorems. 

\subsection{Organizations and Conventions}

The remainder of this paper describes the main results presented above in much more detail. 
Section~\ref{sec:piecewise} answers the first question. Focusing mainly on slowly-rotating NSs and QSs, we investigate how each piecewise EoS parameter affects the universal I-Q relation. We determine which radial region contributes to $I$ and $Q$ the most and compare the result to the radial profile of the energy density to determine which part of the EoS matters the most. We then compare how similar NS EoSs are within this region. We also construct rapidly-rotating NS solutions and see how the rotational and relativistic effects affect the density profile. 
Section~\ref{sec:isodensity} answers the second question. We relax each of the conditions in the elliptical isodensity approximation and see how they affect the universality for uniformly-rotating Newtonian polytropes.
Section~\ref{sec:ecc-profile} answers the third question. We extract the eccentricity profile of both slowly- and rapidly-rotating NSs and study the validity of the elliptical isodensity approximation.
Section~\ref{sec:regular-stars} answers the fourth question. We construct a realistic, differentially-rotating, non-compact star and investigate the universality on the multipole relations. We also look at the eccentricity profile and compare it with that of realistic relativistic stars.
We conclude in Sec.~\ref{sec:emergent} by combining these findings into one picture that proposes an emergent approximate symmetry as a possible explanation for the universality.
Section~\ref{sec:future} suggests a few possible directions for future work. 
All throughout, we use geometric units: $c=1=G$.

\section{The Region that Matters}
\label{sec:piecewise}

What part of the NS interior and what part of the NS EoS contributes the most to the multipole moments of the star, and thus, affects the three-hair relations the most~\cite{I-Love-Q-Science,I-Love-Q-PRD,Pappas:2013naa,Stein:2013ofa}? In this section, we address this question by studying the multipole moments of NSs characterized be a piecewise EoS. We vary each piece of the EoS and find that the multipole moments are most affected by what is going on in the so-called outer core, i.e.~in a shell with inner radius of roughly $0.5 R$ and outer radius $0.95 R$, with $R$ the stellar radius for non-rotating configurations. This region corresponds to densities in roughly the interval $(10^{14}$--$10^{15}) \; {\rm{g/cm^{3}}}$. In this section, we will provide evidence for these results, mainly focusing on slowly-rotating NSs and QSs. In the last subsection, we construct rapidly-rotating NS solutions and see how the rotational and relativistic effects change the stellar energy density profile.

\subsection{Piecewise Polytropic EoSs}

\begin{figure}[tb]
\centering
\includegraphics[width=\columnwidth,clip=true]{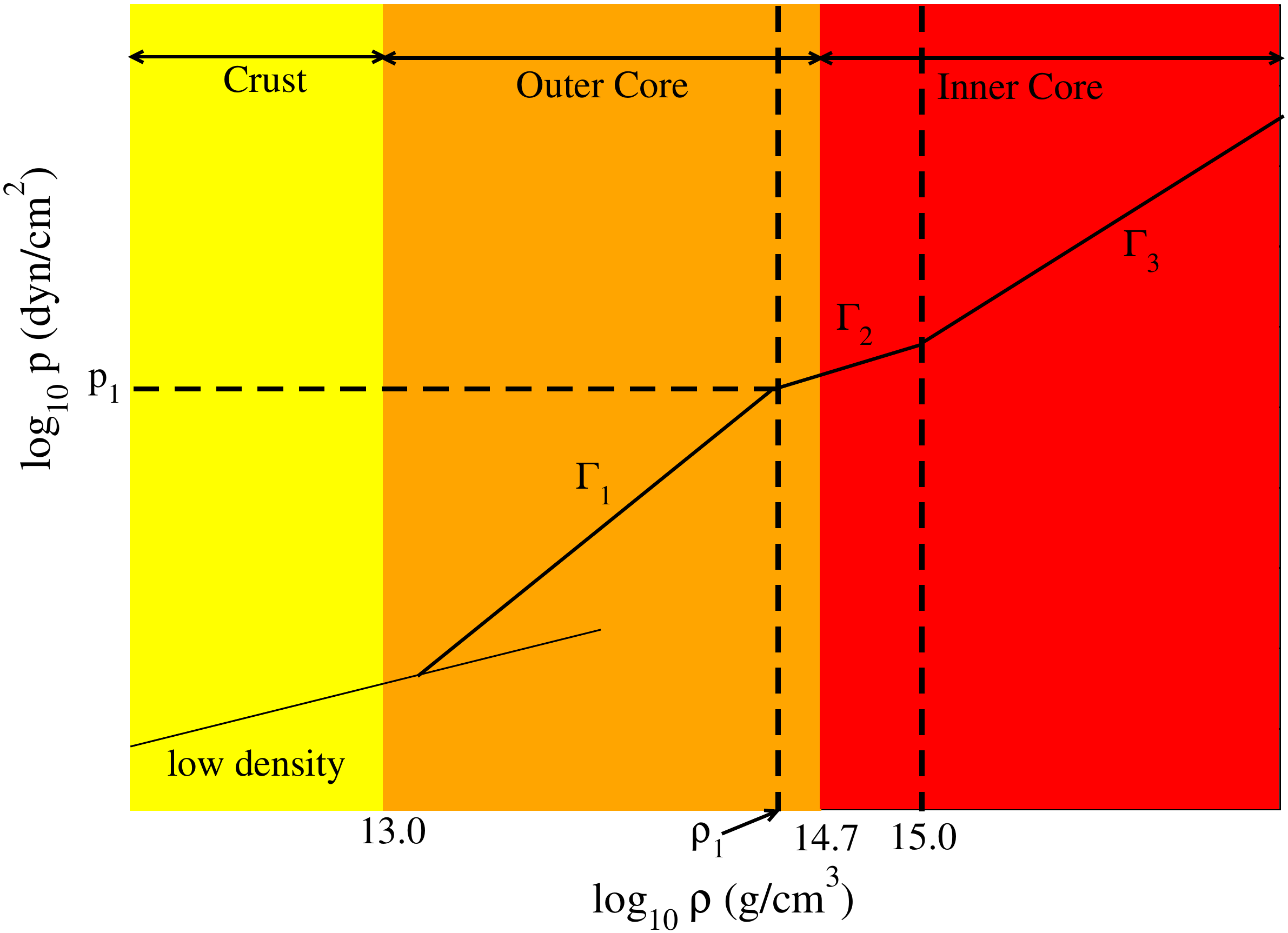}
\caption{
(Color online) Schematic diagram of the piecewise EoS and the different regions of a NS's interior shaded in yellow, orange and red.
\label{fig:schematic}
}
\end{figure}

Let us divide the pressure ($p$)--rest mass density ($\rho$) phase space into three regions: the crust, the outer core and the inner core. Figure~\ref{fig:schematic} shows a schematic diagram of this classification. The lowest density region corresponds to the crust, which transitions into the outer core at a density of roughly $10^{14} \mrm{g/cm^{3}}$. The inner and outer cores have higher densities and pressures than the crust, where the transition point is roughly at $10^{14.7} \mrm{g/cm^{3}}$. 

Although there is no precise definition of these regions, the choices made above make physical sense. The crust-outer core transition is where roughly the lattice of neutron-rich nuclei become superfluid neutrons that coexist with type II superconducting protons~\cite{shapiro-teukolsky,lattimer-prakash-review,read-love}. The outer and inner core transition, which may or may not exist in some stars depending on their mass, is where the superfluid neutrons condense into pions or transition to a neutron solid or quark matter, or some other phase distinct from a neutron superfluid~\cite{shapiro-teukolsky,lattimer-prakash-review,read-love}. The inner core is the region least well-understood from nuclear physics experiments, since it is so far-removed from the scales we can probe in the laboratory. 

In each of these regions, we model the EoS through polytropic equations. In the crust we will use four polytropes, in the outer and inner core three polytropes. The full EoS is then a piecewise function with 7 polytropes of the form
\be
\label{piecewisepoly}
p(\rho) = 
\begin{cases}
p_{\mrm{crust}}(\rho), & (\rho < \rho_0)\,, \\
(p_1/\rho_{1}^{\Gamma_1}) \; \rho^{\Gamma_1}, & (\rho_0 < \rho < \rho_{1})\,, \\
(p_1/\rho_{1}^{\Gamma_2}) \; \rho^{\Gamma_2}, & (\rho_{1} <  \rho < \rho_{2})\,,  \\
(\rho_2/\rho_{1})^{\Gamma_2} (p_{1}/\rho_{2}^{\Gamma_{3}})   \; \rho^{\Gamma_3}, & (\rho_{2} < \rho)\,, 
\end{cases}
\ee
where $(\Gamma_{i},p_{1},\rho_1)$ are free constants, $\rho_{2} = 10^{15} \; \mrm{g/cm^3}$, $\rho_0$ is the solution to $p_{\mrm{crust}}(\rho_0) = p_1 (\rho_0/\rho_{1})^{\Gamma_1}$, with $p_{\mrm{crust}}(\rho)$ given in~\cite{Read:2008iy}. Such a piecewise EoS depends on 5 free constants $(p_{1},\Gamma_{1},\Gamma_{2},\Gamma_{3},\rho_1)$, where $\Gamma_1$ mainly affects the EoS slope in the outer core, $\Gamma_{2}$ and $\Gamma_3$ mainly affect the EoS slope in the inner core, $p_1$ is an overall scaling factor that corresponds to the pressure at $\rho=\rho_1$ and $\rho_1$ changes the transition density of the first two regions. Equation~\eqref{piecewisepoly} with a fixed $\log_{10} \rho_1 = 14.7$ has been shown to reproduce a great number of realistic EoSs~\cite{Read:2008iy}, e.g.~the SLy EoS~\cite{SLy} is reproduced with $(\log_{10}p_1, \Gamma_1, \Gamma_2, \Gamma_3) = (34.384,3.005,2.988,2.851)$ to $1\%$ accuracy.\footnote{$\log_{10} \rho_1$ and $\log_{10} p_1$ mean $\log_{10} [\rho_1/1\mrm{(g/cm^3)}]$ and $\log_{10} [p_1/1\mrm{(dyn/cm^2)}]$ respectively.} The range of values of $(p_1, \Gamma_1, \Gamma_2, \Gamma_3)$ that reproduce all known realistic EoSs is, in fact, within approximately 30\% from those of the SLy EoS~\cite{Read:2008iy}, which is why we choose the latter as our fiducial EoS. We here enlarged the model of~\cite{Read:2008iy} by introducing an additional parameter ($\rho_1$) to increase the degrees of freedom in the variability of the EoS. 

For future convenience, let us relate the rest-mass density $\rho$ to the energy density $\epsilon$. From the first law of thermodynamics, the latter is given by~\cite{Read:2008iy}
\be
\epsilon_{i} (\rho) = (1 + a_i) \rho +  \frac{K_i}{\Gamma_i-1} \rho^{\Gamma_i}
\ee
in the $i^{\rm th}$ region, where the coefficients $a_i$ are given below Eq.~(4) of~\cite{Read:2008iy}. Notice that it is $\epsilon$ that sources the Einstein equations, while it is $\rho$ that enters the non-relativistic, Newtonian definition of the multipole moments.

\subsection{Mass-Radius and \ItoQ Relations}
\label{sec:IQrel}

We will study the universal relations between the NS moment of inertia and quadrupole moment in slowly- and
uniformly-rotating, non-magnetized NSs. We construct these stars by
solving the Einstein equations perturbatively in a slow-rotation
expansion to quadratic order in spin, following
Hartle and Thorne~\cite{hartle1967,hartlethorne}. At zeroth order in spin, we calculate
the stellar mass $M$ and radius $R$ of the non-rotating
configuration. At first order in spin, we extract the moment of
inertia
\begin{equation}
I = \frac{S_1}{\Omega}\,,  
\end{equation}
where $S_1$ and $\Omega$ are the spin angular momentum and the angular
velocity respectively. At second order in spin, we calculate the
quadrupole moment $Q$~\cite{hartle1967,hartlethorne}.

\begin{figure}[tb]
\centering
\includegraphics[width=\columnwidth,clip=true]{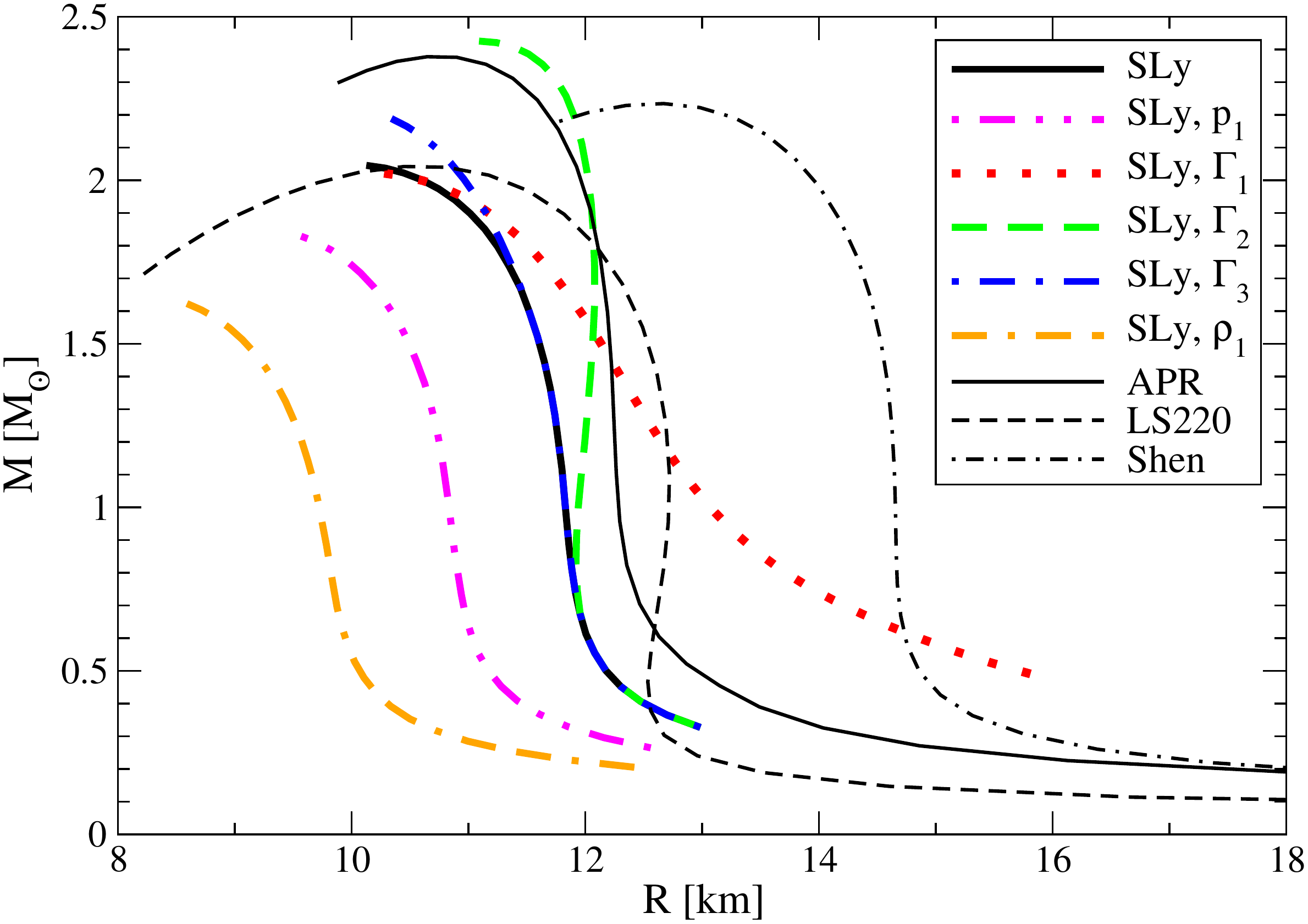}
\caption{
(Color online) Mass-radius relation for the fiducial SLy EoS and that obtained by varying $p_1$, $\Gamma_1$, $\Gamma_2$, $\Gamma_3$ and $\rho_1$ by 30\%. Notice that such a variation has a large impact on the mass-radius relation. We also present the mass-radius relation for the APR, LS220 and Shen EoSs for reference.
\label{fig:MR}
}
\end{figure}

\begin{figure*}[tb]
\centering
\includegraphics[width=\columnwidth,clip=true]{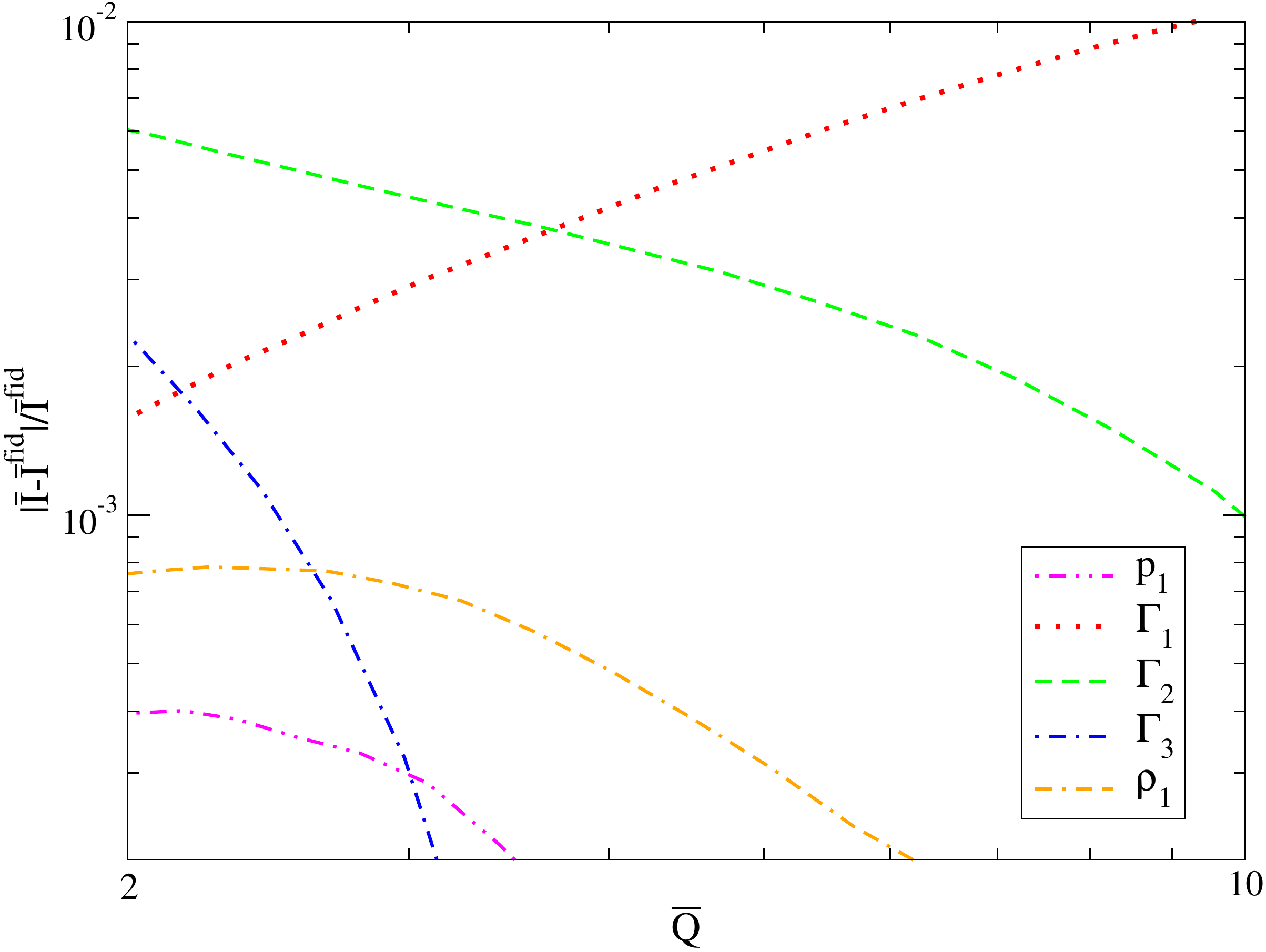}
\includegraphics[width=\columnwidth,clip=true]{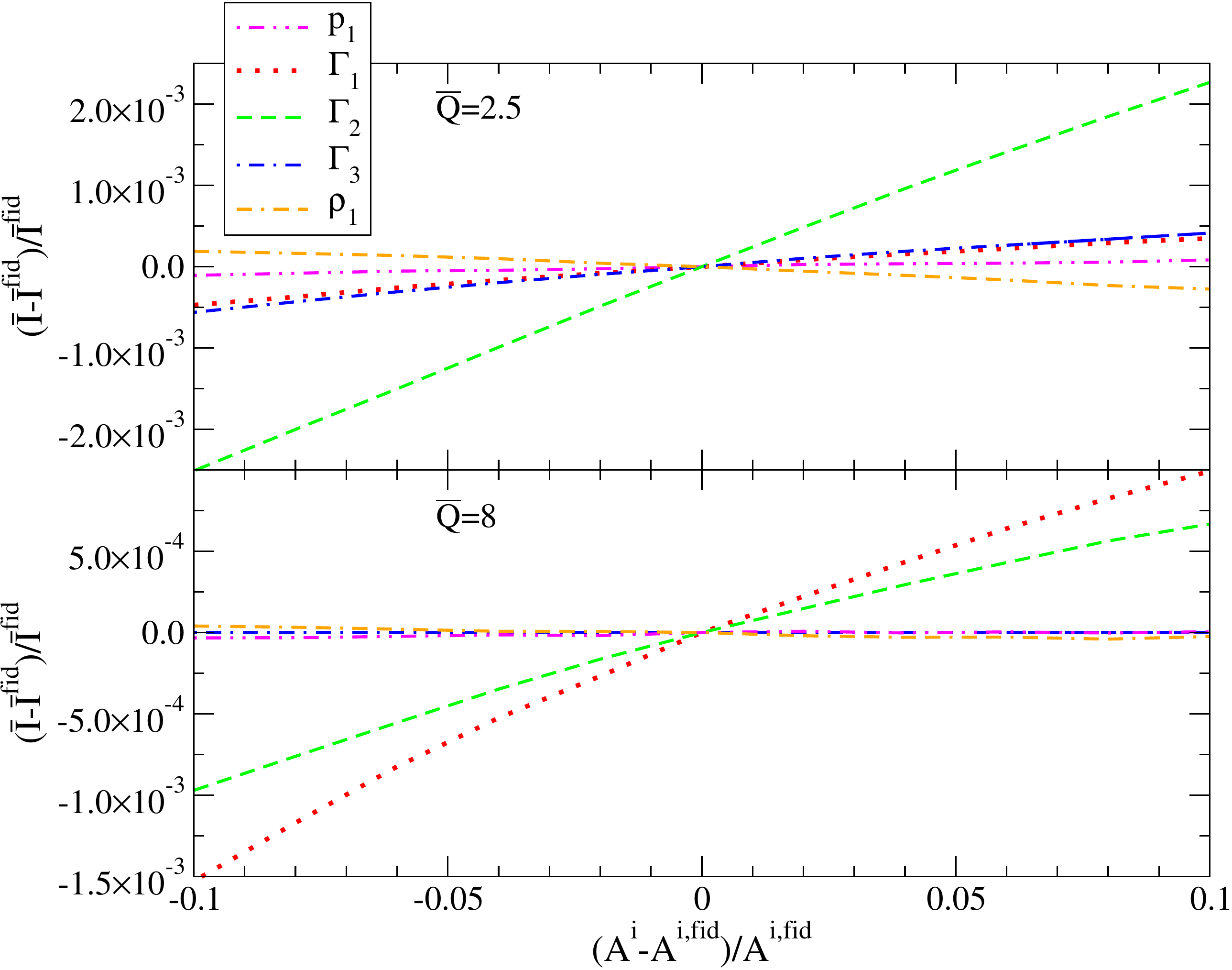}
\caption{
(Color online) Left: Fractional difference on the \ItoQ relation between the varied and fiducial SLy EoSs. For the former, we vary $ p_1$, $\Gamma_1$, $\Gamma_2$, $\Gamma_3$ and $\rho_1$ by 30\%. Right: (Top) Fractional difference of the \ItoQ relation at $\bar{Q}=2.5$ with varied piecewise EoS parameters from those of the fiducial SLy EoS. The horizontal axis represents the
fractional deviation of  $A^i=(p_1, \Gamma_1, \Gamma_2, \Gamma_3, \rho_1)$. Observe that the fractional difference scales almost linearly with the piecewise EoS parameters. (Bottom) Same as the top panel, but for $\bar{Q}=8$.
\label{fig:DeltaI-Q}
}
\end{figure*}

Before looking at $I$ and $Q$, let us investigate how the non-rotating configurations are affected by modifying the piecewise EoS coefficients. Let us then vary these coefficients individually by $30\%$ from the fiducial SLy values, i.e.~we separately set $\log_{10}{p_{1}} = 34.2291$, or $\Gamma_{1}=2.1035$, or $\Gamma_{2}=3.8844$, or $\Gamma_{3}=3.7063$, or $\log_{10} \rho_1 = 14.8139$, while keeping the other parameters equal to their SLy values. These variations are consistent with the range of variability associated with the different EoSs~\cite{Read:2008iy}. Figure~\ref{fig:MR} shows the mass-radius relation for various modified SLy EoSs, together with the mass-radius relation for the APR~\cite{APR}, LS220~\cite{LS} and Shen~\cite{Shen1,Shen2} EoSs. We impose a neutrino-less, beta-equilibrium condition for the latter two EoSs. Observe that $\Gamma_3$ and $\Gamma_2$ only affect the relation for $M>1.7M_\odot$ and $M>0.6M_\odot$, respectively, while $\Gamma_1$ has a larger effect on the lower mass region. Since $p_1$ and $\rho_1$ affect the overall magnitude of the EoS, they scale the mass-radius relation rather than change its shape. 

Let us now study how the \ItoQ universal relations are affected by changing the piecewise polytropic coefficients. We work with the dimensionless quantities
\be
\bar{I} \equiv \frac{I}{M^3}\,, \quad \bar{Q} \equiv -\frac{Q}{M^3 \chi^2}\,, 
\ee
where the dimensionless spin parameter is defined to be
\begin{equation}
\chi \equiv \frac{S_1}{M^2}\,.
\end{equation}
For reference, a value of $\bar{Q} = 1$ and $10$ roughly correspond to a
non-rotating star mass of $2 M_{\odot}$ and $1 M_{\odot}$
respectively, using the SLy EoS. 

The left panel of Fig.~\ref{fig:DeltaI-Q}
shows the fractional difference on the \ItoQ relation relative to the
fiducial EoS due to changing the piecewise polytrope coefficients. 
For stars with $M > 1.8 M_{\odot}$, $\Gamma_2$ affects the \ItoQ relation the most, whereas when $M < 1.8
M_{\odot}$, $\Gamma_1$ affects the relation the most, in both cases to
$\mathcal{O}(1\%)$. Such a small fractional difference is precisely why the
universality is \emph{approximate} and it is consistent with
that reported in~\cite{I-Love-Q-Science,I-Love-Q-PRD}. Modifications to $p_{1}$, $\Gamma_{3}$
and $\rho_{1}$ affect the \ItoQ relations the least, by only $0.1\%$. In
particular, $p_{1}$ and $\rho_1$ essentially scale the EoS, which does not affect
$\bar{I}$ and $\bar{Q}$ as a function of $C$~\cite{I-Love-Q-PRD};  
the \ItoQ relation is insensitive to variations of these parameters.

The right panel of Fig.~\ref{fig:DeltaI-Q} presents the fractional difference of the \ItoQ relation at $\bar{Q}=2.5$ (top) and $\bar{Q}=8$ (bottom) from the fiducial SLy EoS against various piecewise EoS parameters, $A^i=(p_1, \Gamma_1, \Gamma_2, \Gamma_3, \rho_1)$. Observe that the fractional difference scales almost linearly with the EoS parameters. This figure shows that changing the EoS parameters by 10\% only modifies the relation by $\sim 0.2\%$ at most. We have checked that the fractional difference in the relation when varying $\Gamma_1$ and $\Gamma_2$ simultaneously (the two most important parameters regarding universality) is essentially the same as varying $\Gamma_1$ and $\Gamma_2$ separately and then adding their fractional difference linearly.   

The right panel of Fig.~\ref{fig:DeltaI-Q} also shows that decreasing the EoS slopes $\Gamma_{i}$ from the fiducial values affects the universality more than if we increase the slopes. Increasing $\Gamma_{i}$ corresponds to decreasing the polytropic index $n$, since $\Gamma = 1 + 1/n$. For example, the SLy values of $(\Gamma_1, \Gamma_2, \Gamma_3) = (3.005,2.988,2.851)$ correspond to $(n_{1},n_{2},n_{3}) = (0.499,0.503,0.540)$, and thus, decreasing $\Gamma_{i}$ corresponds to increasing $n$ above $0.5$, and vice-versa. Strange quark stars are well-modeled by polytropes with index $n \sim 0$, while white dwarfs are well modeled with polytropes of index $n>2$. Thus, as we increase $\Gamma_{i}$ we make the star look more like a QS, while when we decrease $\Gamma_{i}$, it looks more like a white dwarf. Notice, interestingly, that the universality is affected more as we approach the white-dwarf branch than as we approach the QS branch. We think this is because as we approach this branch, the elliptical isodensity approximation becomes less valid, which significantly affects the universality, as we will see later in this paper. 

\begin{figure*}[tb]
\centering
\includegraphics[width=\columnwidth,clip=true]{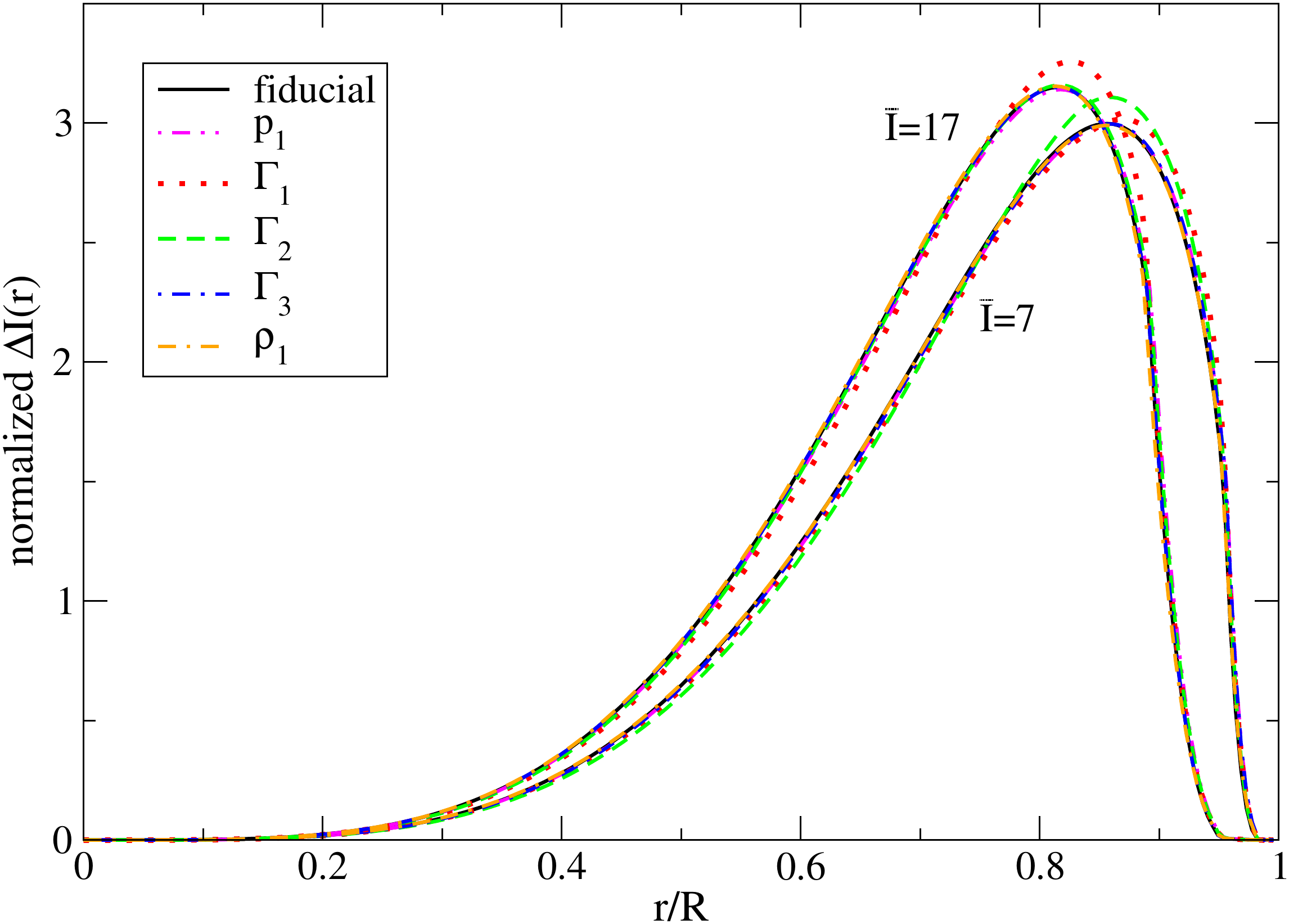}
\includegraphics[width=\columnwidth,clip=true]{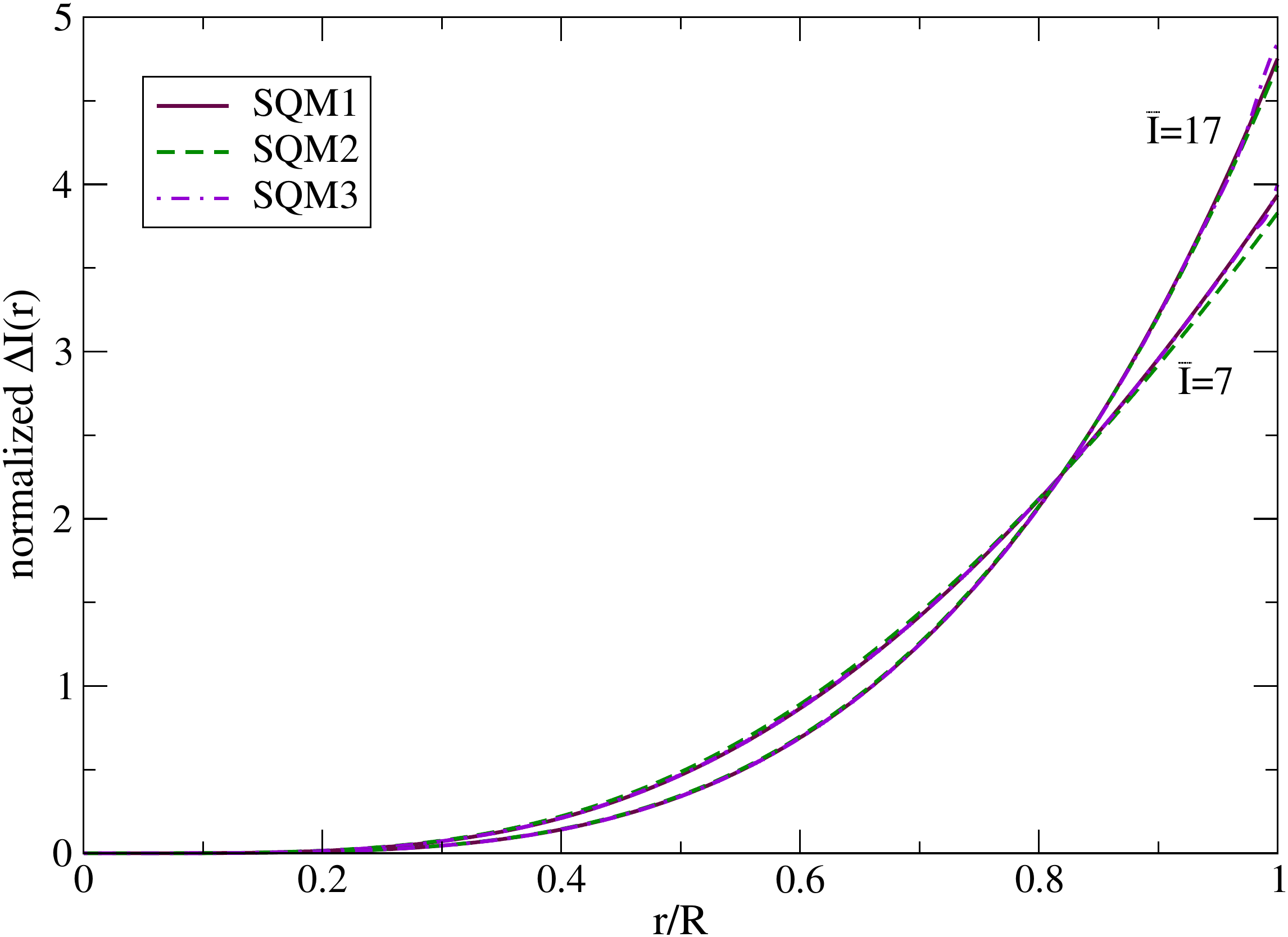}
\caption{
(Color online) The radial profile of the normalized integrand of $I$ at $\bar{I}=7$ and $\bar{I}=17$ for NSs with the fiducial SLy EoS and QSs with three different QS EoSs~\cite{SQM} (right). The other curves in the left panel correspond to varying the EoS  parameters by 10\% from the fiducial ones. Observe that the integrands are not very sensitive to the variation of the EoS. 
\label{fig:Iintegrand}
}
\end{figure*}

\subsection{Radial Integrand of the NS and QS Moment of Inertia and Quadrupole Moment}

In order to understand which part of the stellar interior region affects the universal relations the most, let us look at the radial profile of the integrand of the moment of inertia for a fixed $\bar{I}$. For slowly-rotating stars in full GR, $I$ is given in integral form as~\cite{hartle1967}
\be
\label{eq:I-integral}
I =  \int^R_0 \Delta I (r) dr \,, \quad \Delta I (r)  \equiv  - \frac{2}{3} r^3 \frac{d j}{dr} \frac{\omega (r)}{\Omega}\,,
\ee
where $R$ is the stellar radius for a non-rotating configuration,
\be
j(r) \equiv e^{-[\nu(r) + \Lambda(r)]/2}\,,
\ee
and $\nu$, $\Lambda$ and $\omega$ are metric functions related to the linear-in-spin line element via
\ba
\label{eq:metric}
ds^2 &=& -e^{\nu(r)} dt^2 + e^{\Lambda (r)} dr^2 \nn \\
& & + r^2 \left\{ d\theta^2 + \sin^2 \theta \left[d\phi - [\Omega - \omega (r)] dt \right]^2 \right\}\,.
\ea
Equation~\ref{eq:I-integral} reduces to~\cite{hartle1967}
\be
\label{eq:I-newtonian}
I^\N = \frac{8\pi}{3}  \int^R_0 r^4 \rho (r) dr
\ee
in the Newtonian limit.

Figure~\ref{fig:Iintegrand} shows $\Delta I$ as a function of fractional radius for NSs (left) with the fiducial SLy EoS and EoS parameters that deviate by 10\% from the fiducial EoS, and for QSs (right) with three different QS EoSs, with $\bar{I}=7$ and $\bar{I}=17$ held fixed. We normalized the integrands such that the area in Fig.~\ref{fig:Iintegrand} is unity. Observe that the integrand is dominated by the region within 50\%--95\% of the total radius. Observe also that the integrands are almost unaffected by the variation of the EoS parameters for a fixed $\bar{I}$. These results extend and confirm those found in~\cite{I-Love-Q-PRD}, where a related integrand was investigated in the Newtonian limit for a fixed NS compactness. 

\begin{figure}[tb]
\centering
\includegraphics[width=\columnwidth,clip=true]{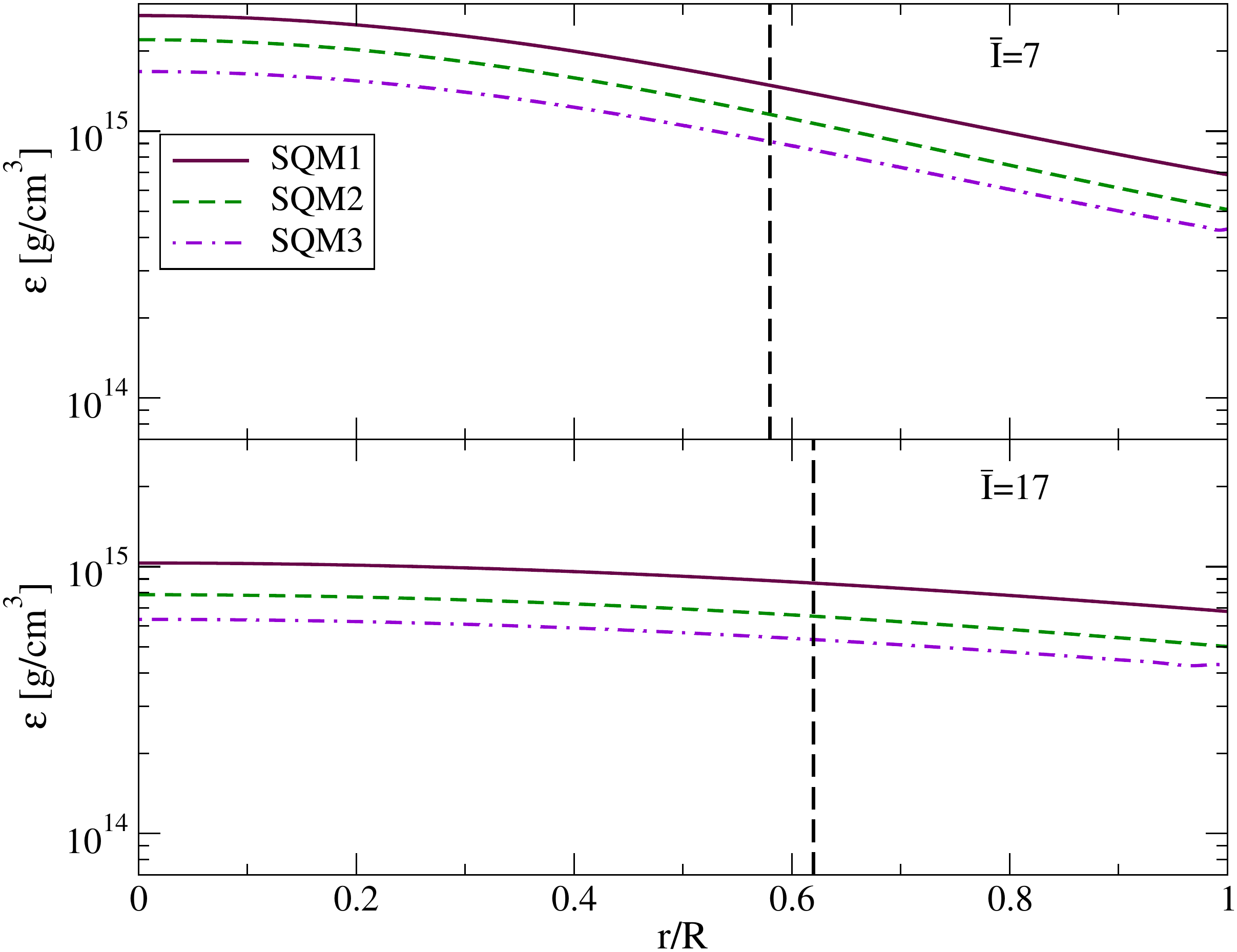}
\caption{
(Color online) 
 Energy density profile (same as Fig.~\ref{fig:density-profile}) but for QSs. Observe that the density is almost constant throughout the star and non-vanishing at the stellar surface. The region above the vertical dashed line gives a 90\% contribution to the QS moment of inertia and quadrupole moment.
\label{fig:density-profile-QS}
}
\end{figure}

In order to quantify which part of the stellar interior contributes the most to the integrand, we identify the region, delimited by $r_\MIN$ and $r_\MAX$, inside which the integrand integrates to 90\% of the total $I$:
\be
\int^{r_\MAX}_{r_\MIN} \Delta I(r) dr = 0.9 \int^R_0 \Delta I (r) dr\,, 
\ee
with the symmetry constraint that $\Delta I (r_\MIN) = \Delta I (r_\MAX)$. In the NS case with the fiducial SLy EoS, we find $(r_\MIN,r_\MAX) = (0.54,0.96) R$ and $(0.50,0.91) R$ for $\bar{I}=7$ and $\bar{I}=17$ respectively. In the QS case, $(r_{\MIN},r_{\MAX})=  (0.58,1)R$ and $(r_{\MIN},r_{\MAX})=  (0.62,1)R$ for $\bar{I} = 7$ and $\bar{I} = 17$ respectively. In the latter case, the integrand increases monotonically as a function of $r/R$ because the density does not drop to zero at the stellar surface, as shown in Fig.~\ref{fig:density-profile-QS}. QSs correspond to nearly constant density stars, especially when $\bar{I}$ is large, where the stellar compactness is relatively small. These results confirm that $\bar{I}$ is mostly affected by the region in the approximate range (50--95)\% and (60--100)\% of the stellar radius in the NS and QS cases respectively.

One can also consider the radial profile of the integrand of $Q$. When the stellar eccentricity is constant throughout the star, $Q$ is given by $Q \propto e^2 \int \rho \; r^4 dr$ (see Eqs.~(5) and~(7) in~\cite{Stein:2013ofa}) for Newtonian, slowly-rotating polytropes. In reality, eccentricity is a function of radius and one has to include $e(r)^2$ in the radial integral. Therefore, the integrand of $Q$ is $\Delta Q \sim \rho(r) r^4 e(r)^2$, modulo an overall constant. We checked that the radial profile of such integrand is almost identical to the one of $\Delta I (r)$ in Fig.~\ref{fig:Iintegrand}. This means that both $\bar{I}$ and $\bar{Q}$ are mostly affected by the region in the range (50--95)\% of the radius.

That the \ItoQ relations are dominated by the EoS in the region roughly between $50\%$ and $95\%$ of the NS's total radius was first pointed out in~\cite{I-Love-Q-Science,I-Love-Q-PRD}. Unfortunately, that paper referred to this region as the ``outer-layers'' of the NSs, which has recently been taken to mean the NS crust. Clearly the region $r/R \in (0.50,0.95)$ does not comprise the NS crust, but rather the outer core and the very outer part of the inner core.

With this at hand, we can now determine the $(p,\epsilon)$ region of the EoS that dominates the universal relations. Figures~\ref{fig:density-profile} and~\ref{fig:density-profile-QS} show the energy density profile of NSs and QSs respectively with $\bar{I}=7$ (top) and $\bar{I}=17$ (bottom) for various modified EoSs, together with $r_\MIN/R$ and $r_\MAX/R$ plotted as vertical dashed lines. Recall that larger values of $\bar{I}$ correspond to smaller stellar compactnesses $C = M/R$. The figure shows that $\bar{I}$ and $\bar{Q}$ are most affected by the EoS in the range $\epsilon = (10^{14}-10^{15})$g/cm$^3$ and $\epsilon = (4 \times 10^{14}$--$1.5 \times 10^{15})$g/cm$^3$ in the NS and QS cases respectively.

\begin{figure}[tb]
\centering
\includegraphics[width=\columnwidth,clip=true]{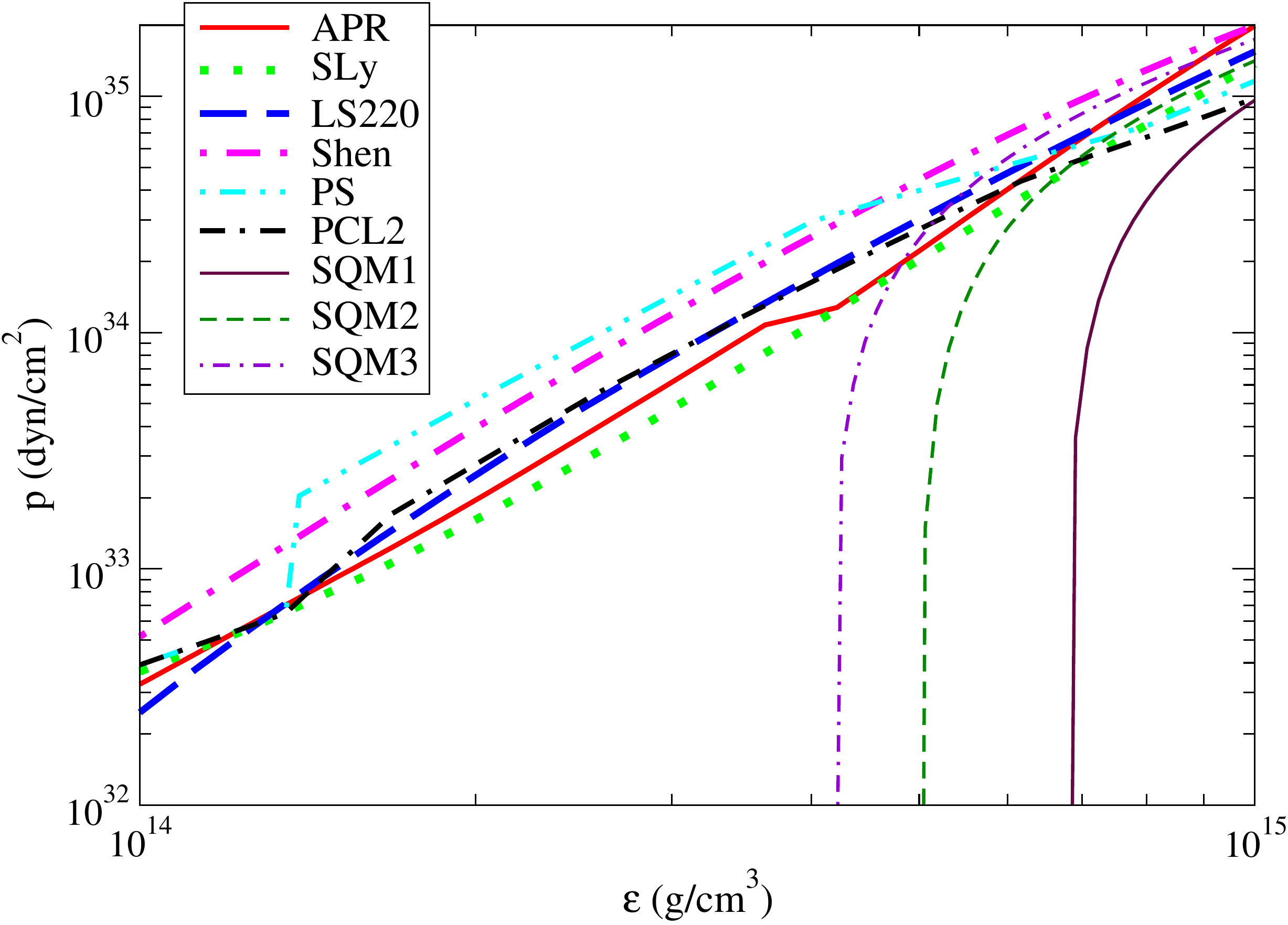}
\caption{
(Color online) 
 EoS in the $(p,\epsilon)$ region that dominates the calculation of the moment of inertia. 
\label{fig:p-rho-zoom}
}
\end{figure}

One possible explanation for the approximate EoS universality is that all EoSs are approximately the same in this density region. We found in Sec.~\ref{sec:IQrel} that the EoS slope is more important than the overall EoS magnitude. Therefore, one needs to look at the variation in the EoS slope in the density region that matters, as shown in Fig.~\ref{fig:p-rho-zoom}. In order to quantify this variation, we fitted a polytrope of the form $p = \bar K \epsilon^{\bar \Gamma}$ to several tabulated NS EoSs in this region (see Table~\ref{table:slope} for the best-fit values of $\bar \Gamma$ and the estimated errors). Observe that the variation in the slope can be as high as $\sim 17\%$, but it would increase significantly if one included QSs. Thus, the approximate similarity of the EoS (or its slope) in the density region that matters cannot be the direct explanation for the approximate EoS universality we observe. As we will see in Sec.~\ref{sec:ecc-profile}, such $\mathcal{O}(10\%)$ variation in the EoS slope produces a very small variation in the stellar eccentricity profile, which we propose as the main origin of the approximate universality.
{\renewcommand{\arraystretch}{1.2}
\begin{table}
\begin{centering}
\begin{tabular}{lr@{$\pm$}l}
\hline
\hline
\noalign{\smallskip}
\multicolumn{1}{c}{EoS} &
\multicolumn{2}{c}{$\langle\bar \Gamma\rangle$} \\
\hline
\noalign{\smallskip}
 APR~\cite{APR} & $2.79$&$0.016$ \\
 SLy~\cite{SLy} & $2.70 $&$0.037$ \\
 LS220~\cite{LS} & $2.77 $&$0.0086$ \\
 Shen~\cite{Shen1,Shen2}  & $2.62 $&$0.0064$ \\
 PS~\cite{PS} & $2.35 $&$0.076$ \\
  PCL2~\cite{SQM} & $2.57 $&$0.064$\\
\noalign{\smallskip}
\hline
\hline
\end{tabular}
\end{centering}
\caption{
(Color online)
Best-fit value of the EoS slope $\bar \Gamma$ for various realistic NS EoSs in the energy density range $10^{14}$--$10^{15}$g/cm$^3$, where the moment of inertia and quadrupole moment are affected the most.
\label{table:slope} }
\end{table}
}

\subsection{Relativistic and Rotational Effects on the Density Profiles}
\label{subsec:rel-rot-effects}

Do the conclusions derived in the previous subsection continue to hold for relativistic stars with a rapid rotation? In this subsection, we tackle this question, but to keep it simple, we parameterize the EoS through the following polytrope:
\be
\label{eq:poly2}
p = \bar{K} \epsilon^{1 + 1/\bar{n}}\,,
\ee
where we recall that $\epsilon$ is the stellar energy density. $\bar{K}$ and $\bar{n}$ are the overall constant and polytropic index respectively. Notice that $\bar{n}$ is different from $n$ introduced at the end of Sec.~\ref{sec:IQrel} as the former is the power index for $\epsilon$ while the latter is the one for $\rho$. With this EoS, we now calculate the rotational corrections to the energy density radial profile to see if such effects limit the validity of the arguments posed in the previous subsection. We also study the relativistic effects to see the validity of Newtonian calculations that we will present in the next section.

Let us first concentrate on relativistic modifications to non-rotating configurations. In GR, the equation of hydrostatic equilibrium in spherical symmetry is modified from its Newtonian form to the Tolman-Oppenheimer-Volkov (TOV) equation:
\be
\frac{dp}{dr} = -\frac{m \epsilon}{r^2} \left( 1 + \frac{p}{\epsilon} \right) \left( 1 + \frac{4\pi r^3 p}{m} \right) \left(1-\frac{2m}{r}\right)^{-1}\,.
\ee
Relativistic corrections arise because all energy densities gravitate (the first two terms) and the geometry is modified (the last term). $m(r)$ in the above equation is related to $\Lambda (r)$ in Eq.~\eqref{eq:metric} via
\be
e^{-\Lambda (r)} = 1 - \frac{2 m(r)}{r}\,,
\ee
where we use Schwarzschild-like coordinates.

The TOV equation can be written in a dimensionless form if one makes the following substitutions: $\epsilon = \epsilon_c \bar \vartheta^{\bar{n}}$, $r = \alpha\xi$, $p = \bar{K} \epsilon_c^{1+1/\bar{n}} \bar \vartheta^{\bar{n}+1}$ and $m = \epsilon_c \alpha^3 \bar{m}$, where $p_c$ and $\epsilon_c$ are the pressure and energy density at the stellar center and $\alpha^2 \equiv (\bar{n}+1)\bar{K} \epsilon_c^{1/\bar{n}-1}$. The new dimensionless equation has the form,\footnote{An equivalent equation is found in~\cite{1964ApJ...140..434T}.}
\ba
\frac{d\bar \vartheta}{d\xi} &=& -\frac{\bar{m}}{\xi^2} \left( 1 + \lambda \bar \vartheta \right) \left( 1 + \lambda \frac{4\pi \bar \vartheta^{\bar{n}+1}\xi^3}{\bar{m}} \right) \nn \\
& & \times \left(1-2(\bar{n}+1) \lambda \frac{\bar{m}}{\xi}\right)^{-1}\,, 
\ea
where $\lambda \equiv p_c/\epsilon_c$ is a measure of how relativistic a particular configuration is, and thus, it parametrizes the deviations from the Newtonian approximation. One recovers the (Newtonian) Lane-Emden (LE) equation when $\lambda=0$, with only a difference in scales by a factor of $\sqrt{4\pi}$ due to different definitions of $\alpha$. 

\begin{figure}
\centering
\includegraphics[width=.45\textwidth]{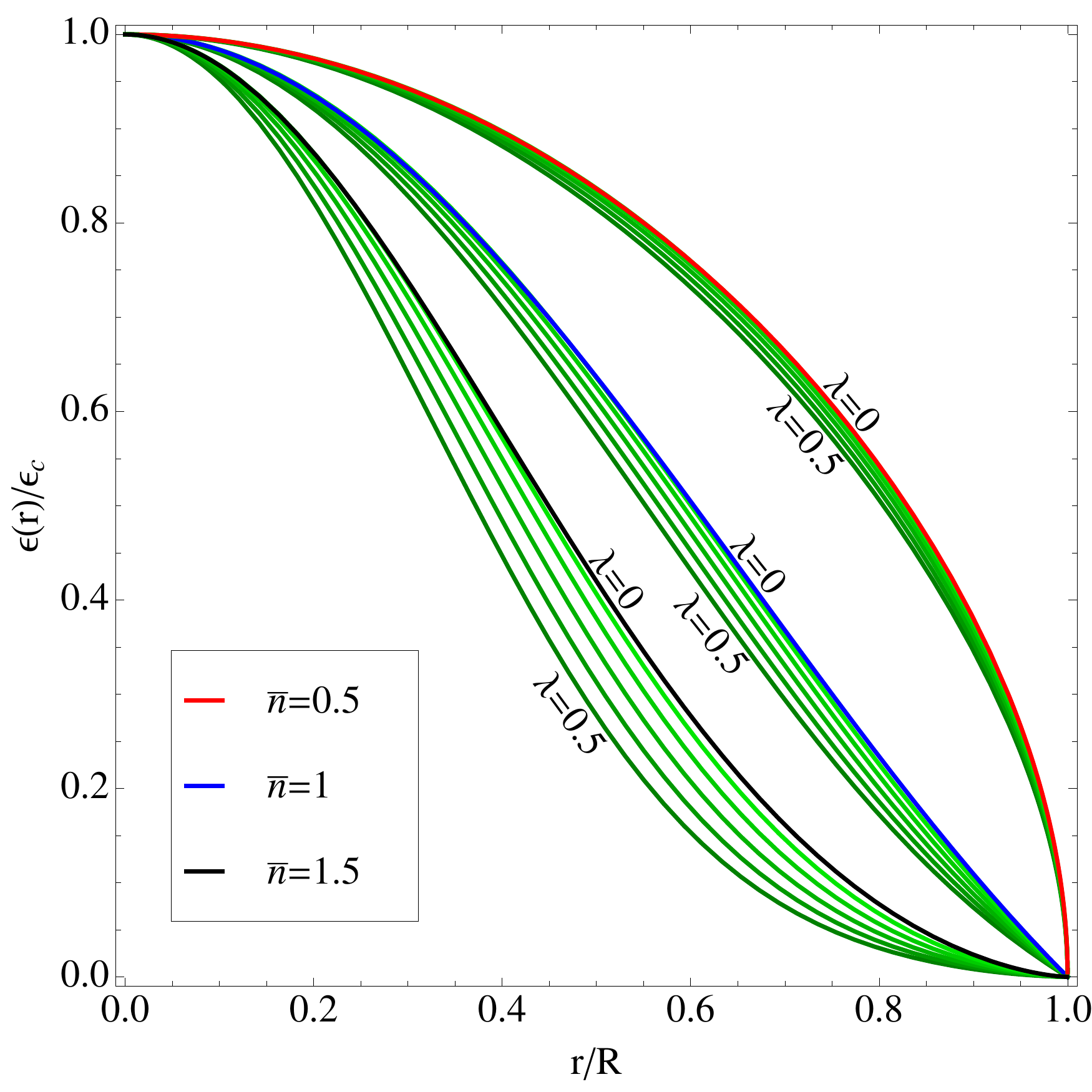}
\caption{
(Color Online) 
Energy density profile for non-rotating polytropes in the Newtonian limit ($\lambda=0$) and with relativistic corrections using $\lambda = [0,0.5]$ in increments of 0.1 from top to bottom and various polytropic indices $\bar{n}$. Observe how relativistic effects make the profiles more centrally-condensed.  
\label{densityFigGR}}
\end{figure}

Figure~\ref{densityFigGR} shows the energy density profiles for Newtonian models ($\lambda$ =0) and relativistic models with $\lambda = [0,0.5]$ in increments of 0.1, for EoSs with polytropic index $\bar{n}=0.5$, $1$, and $1.5$. The $\lambda=0.5$ case corresponds approximately to the maximum mass model for these polytropes. Observe that as relativistic effects become stronger, the density profiles become more centrally condensed. This effect is more pronounced for higher $\bar{n}$ values, which is consistent with~\cite{1964ApJ...140..434T}.

Let us now study rotational modifications to the energy density profile. To do so, we construct sequences of rotating models using the \texttt{RNS} code~\cite{stergioulas_friedman1995}. Calculations are performed in quasi-isotropic coordinates, but we will present the results in Schwarzschild-like coordinates so that one can compare them to the slow-rotation results presented earlier. Specifically, for different central densities,
starting from values which correspond to Newtonian stars up to values that correspond to the relativistic models of maximum mass, we construct sequences of rotating polytropes with rotation rates up to the Kepler (mass-shedding) limit. 

\begin{figure}
\centering
\includegraphics[width=.215\textwidth]{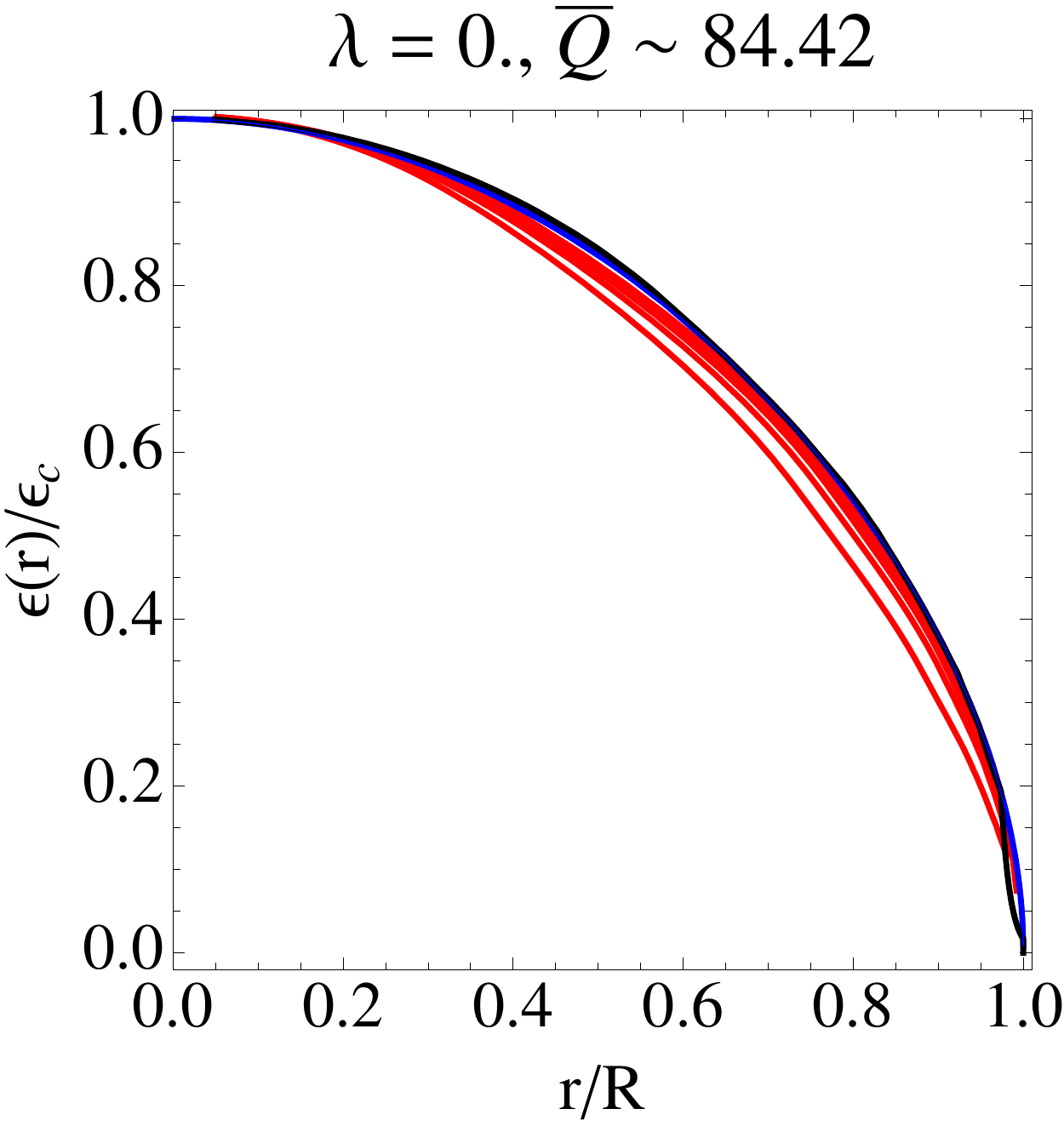}
\includegraphics[width=.215\textwidth]{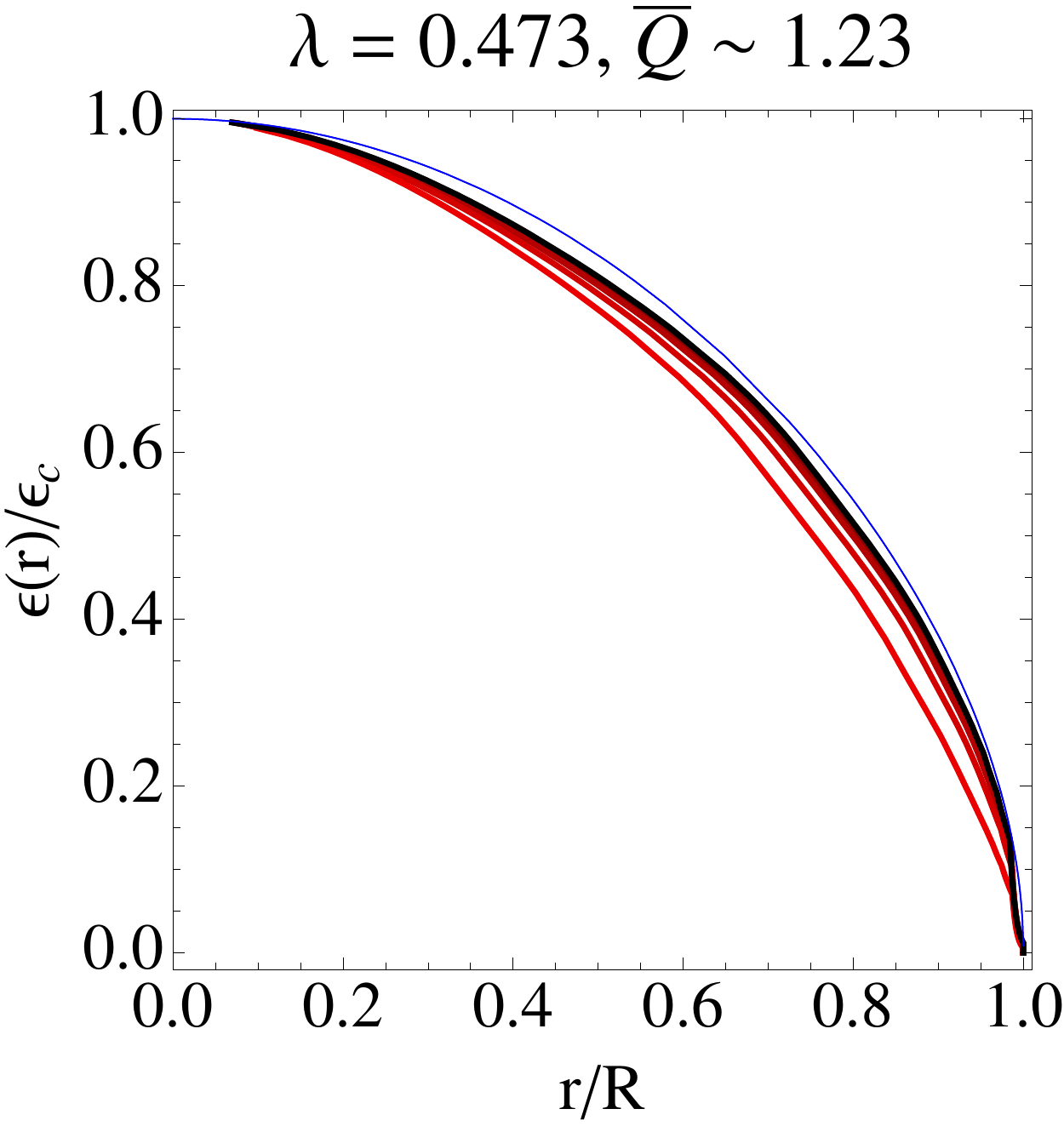} 
\\
\includegraphics[width=.215\textwidth]{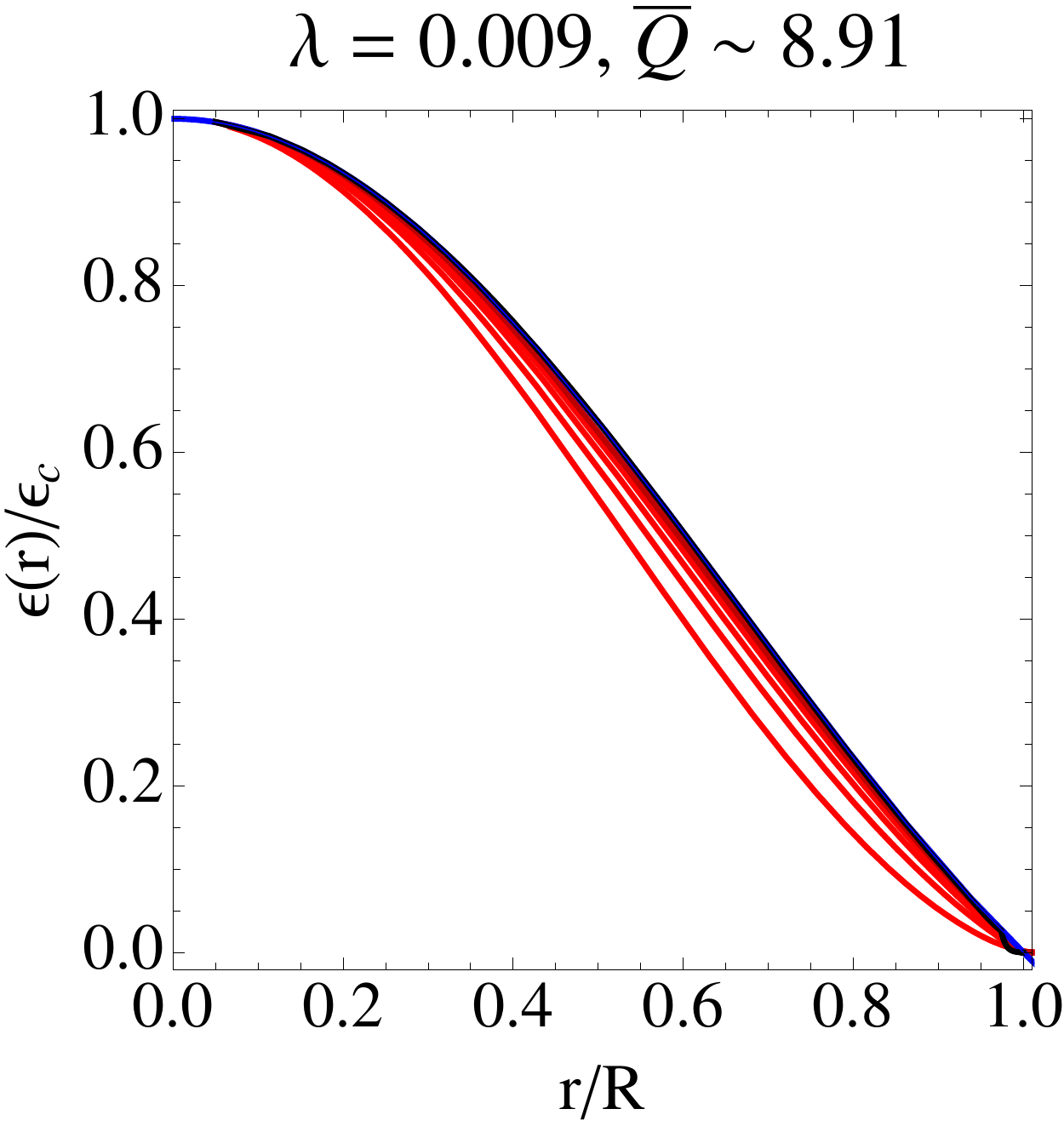}
\includegraphics[width=.215\textwidth]{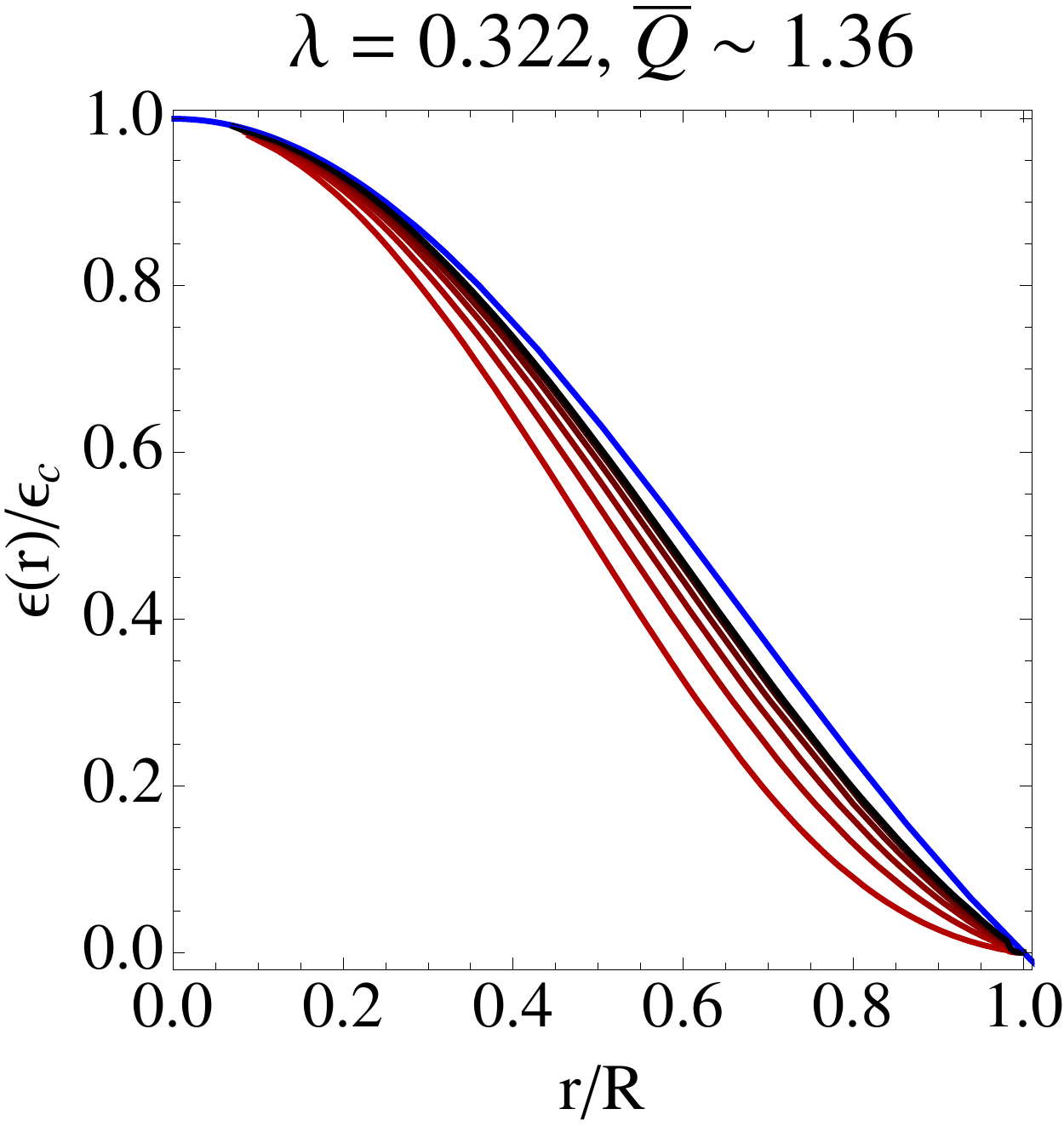}
\caption{
(Color Online) 
The energy density profile for polytropes with an index $\bar{n}=0.5$ (top panels) and $1$ (bottom panels), for two extreme values of $\lambda$ and various rotation rates. The blue and black curves are the Newtonian Lane-Emden and relativistic TOV models of non-rotating configurations respectively. Five red curves present the relativistic rotating models with rotation of $\chi = 4.19, 7.52, 10.05, 12.25, 14.21$\footnote{Newtonian stars have large radius, and hence have large (dimensionless) angular momentum that can easily exceed unity.} for $\lambda =0$ (top, left), $\chi = 0.25, 0.45, 0.58, 0.69, 0.76$ for $\lambda =0.473$ (top, right), $\chi = 0.49, 0.87, 1.15, 1.38, 1.56$ for $\lambda = 0.009$ (bottom, left) and $\chi = 0.2, 0.35, 0.46, 0.54, 0.59$ for $\lambda = 0.322$ (bottom, right) from top to bottom. 
\label{RNS-density-N05}}
\end{figure}

Figure~\ref{RNS-density-N05} shows the energy density profile for
$\bar{n}=0.5$ (top panels) and 1 (bottom panels) with various rotation
rates. Rotation can change density profiles of non-rotating
configurations at most by $\sim 30\%$. Observe that rotation creates
more centrally-condensed configurations. This effect is not very
strong for Newtonian models but becomes more prominent for
relativistic ones. One might think that the centrifugal force arising
from rotation would make objects less centrally condensed, rather than
more centrally condensed. However, the profiles we are studying here
are normalized density as a function of normalized equatorial radius,
$r_{\Eq}/R_{\Eq}$. Along the equatorial plane, all points are pushed
outward by rotation. The amount by which they are pushed out increases
with radius, in agreement with the eccentricity increasing
with radius. Thus, the denominator $R_{\Eq}$ increases faster than the
numerator $r_{\Eq}$, so in terms of the ratio $r_{\Eq}/R_{\Eq}$, a
value of given normalized density moves inwards, i.e.~the profile
becomes more centrally condensed.

We have then seen that relativistic corrections and rotational corrections do modify the density profiles of stellar configurations. These modifications, however, are of ${\cal{O}}(10 \%)$ relative to the results obtained in the Newtonian, non-relativistic limit. Thus, these corrections do not alter the reasoning presented in the previous subsection. For example, in Fig.~\ref{fig:density-profile} and~\ref{fig:density-profile-QS} we presented the density profile of slowly-rotating NSs and QSs. For rapidly rotating models, these profiles would change by a \emph{relative} factor of ${\cal{O}}(10\%)$, but the boundaries of the radial and density regions that matter the most for the existence of approximate universality would essentially not be modified.  

\section{Relaxing the Elliptical Isodensity Approximation}
\label{sec:isodensity}

In the previous section, we established that the region that matters the most regarding universality is the outer core of the star. With this knowledge at hand, let us try to understand what approximate symmetries are present that could be responsible for the universality observed. To do so, we will work in the non-relativistic, Newtonian limit, as then we can do all calculations analytically and understand the symmetries more clearly. Relativistic corrections were already studied in Sec.~\ref{subsec:rel-rot-effects} and in~\cite{Yagi:2014bxa}.

One of the key approximations used to derive the no-hair-like relations for NSs and QSs in the non-relativistic limit~\cite{Stein:2013ofa} was the elliptical isodensity approximation~\cite{Lai:1993ve}. This approximation has three main ingredients:
\begin{enumerate}
\item {\bf{Self-similar Isodensity Condition}}. That the density profile is a sequence of \emph{self-similar surfaces} with the same, constant stellar eccentricity 
\be
e = e_{0} = {\rm{const.}}
\ee
This implies that the radius- and polar angle-dependent density profile can be approximated as a function of a single radial variable, $\tilde{r}$, i.e.~$\rho(r,\theta) = \rho({\tilde{r}})$. Although we have introduced an eccentricity here, we have not yet restricted the isodensity contours to ellipsoids. 
\item {\bf{Elliptical Condition}}. That the shape of the isodensity surfaces is an ellipsoid. This implies that there exists a change of coordinates, from spherical to elliptical, such that the moment-integrals separate:
\be
R_{*}^{\mrm{ellip}}(\theta) = \left(\frac{\sin^{2}{\theta}}{a_{1}^{2}} + \frac{\cos^{2}{\theta}}{a_{3}^{2}}\right)^{-1/2}\,,
\ee
where $R_*^{\mrm{ellip}}$ denotes the elliptical stellar surface and $a_{1}$ and $a_{3}$ are the semi-major and semi-minor axis of the ellipsoid. This condition, in particular, excludes triaxial surfaces.
\item {\bf{Spherical Density Profile Condition}}. That once we have transformed to the radial coordinate $\tilde{r}$, the rescaled density $\vartheta = (\rho/\rho_c)^{1/n}$, where $\rho_c$ is the central density, is a LE function, i.e. the rescaled density for the non-rotating Newtonian polytropes: 
\be
\label{eq:theta-spherical}
\vartheta(\tilde{r}) = \vartheta_{\LE}(\tilde{r})\,.
\ee
\end{enumerate}

In order to understand which condition is responsible for the approximate EoS universality observed in~\cite{Stein:2013ofa}, we systematically relax each of them and assess their relative importance. We find that the particular shape of the self-similar surfaces [condition (2)] does not impact the universality at all. On the other hand, the self-similar assumption [condition (1)] and the spherical density approximation [condition (3)] do affect the universality dramatically. We will see below that breaking either of these conditions can destroy the approximate EoS universality.    

\subsection{Universal Relations with the Elliptical Isodensity Approximation}
\label{sec:3-hair}

Before we begin to break the conditions of the elliptical isodensity approximation, let us review how the approximate EoS universal no-hair relations are derived in the Newtonian limit. Consider a uniformly-rotating, unmagnetized and cold Newtonian star with a polytropic EoS, given by
\be
p = K \rho^{1+1/n}\,.
\ee
The mass and current multipole moments in the non-relativistic, Newtonian limit are~\cite{Ryan:1996nk}:
\ba
\label{eq:MM-def}
M_\ell &=& 2 \pi \int^{1}_{-1} \; d\mu \int^{R_*(\mu)}_0 \!\!\! dr \; \rho(r,\mu) \;  P_\ell (\mu)  \; r^{\ell +2} \,, \\
\label{eq:SS-def}
S_\ell &=& \frac{4 \pi \Omega}{\ell+1}  \int^{1}_{-1}  \; d\mu \int^{R_*(\mu)}_0 \!\!\! dr \; \rho(r,\mu) \; \frac{d P_\ell (\mu)}{d\mu}  \;  (1-\mu^2)  \; r^{\ell +3}\,,\nn \\
\ea
where $\mu = \cos \theta$, $R_*(\mu)$ is the stellar surface, $P_\ell(\mu)$ are Legendre polynomials and $\Omega$ is the star's angular frequency. For future convenience, let us introduce here the dimensionless multipole moments 
\begin{align}
\label{eq:Mbar-def}
\bar{M}_{2 \ell + 2} &= \left(-1\right)^{\ell+1} \frac{M_{2 \ell + 2}}{M^{2 \ell +3} \chi^{2 \ell + 2}}\,,
\\
\label{eq:Sbar-def}
\bar{S}_{2 \ell + 1} &= \left(-1\right)^{\ell} \frac{S_{2 \ell + 1}}{M^{2 \ell +2} \chi^{2 \ell + 1}}\,,
\end{align}
where $\chi \equiv S_{1}/M^{2}$. Notice that $M_{2} = Q$ is the (mass) quadrupole moment, while $S_{1}$ is the (current) dipole moment and $M_{0} = M$ is the (mass) monopole.

Let us begin by re-deriving the first universal relation of~\cite{Stein:2013ofa}, using only the self-similar isodensity condition [condition (1)]. This assumption allows us to separate the moment integrals into radial and angular parts. Introducing the coordinate system $x^{i} = \tilde{r}\Theta(\mu) n^{i}$, with $n^{i} = (\sin{\theta} \cos{\phi},\sin{\theta} \sin{\phi}, \cos{\theta})$, for some function $\Theta$ of $\mu$ only (due to the self-similar isodensity condition), we find
\begin{equation}
\label{eq:mass-def}
M_{\ell} = 2 \pi \; I_{\ell,3} \; R_{\ell}\,, \quad 
S_{\ell} = \frac{4 \pi \ell}{2 \ell + 1} \Omega \; \delta I_{\ell} \; R_{\ell+1}\,,
\end{equation}
where the radial and angular integrals are given by
\be
R_\ell = \int_0^{a_1} \rho(\tilde{r}) \tilde{r}^{\ell +2} \; d\tilde{r}\,, \quad I_{\ell,k} = \int_{-1}^{1} \Theta(\mu)^{\ell+k} \; P_{\ell}(\mu) \; d\mu\,,
\ee
and $\delta I_{\ell} = I_{\ell-1,5} - I_{\ell+1,3}$, with $a_1$ the stellar semi-major axis, which satisfies $a_1 \Theta (\mu) = R_*(\mu)$. One can then immediately derive the first universal relation by taking the ratio of the moments:
\be
\label{universal-1}
\frac{\bar{M}_{2 \ell + 2}}{\bar{S}_{2 \ell + 1}} = \left[\frac{4 \ell+3}{6 \ell + 3} \; \frac{\delta I_{1}}{\delta I_{2 \ell+1}} \; \frac{I_{2\ell+2,3}}{I_{2,3}}\right] \bar{M}_{2}\,.
\ee
This relation does not depend on the polytropic index $n$ since it does not explicitly depend on $R_\ell$, but it does depend on the spin (or equivalently on $e$) in general. 

Let us now use the elliptical condition [condition (2)] in Eq.~\eqref{universal-1}. Doing so, the $\Theta(\mu)$ function is uniquely given by~\cite{Stein:2013ofa}
\begin{equation}
\label{eq:Theta-original}
\Theta(\mu) \equiv \sqrt{\frac{1-e^2}{1-e^2 (1-\mu^2)}}\,,
\end{equation}
where $e$ is the eccentricity of the ellipsoids. Using this, the first universal relation becomes 
\be
\label{universal-1-original}
\frac{\bar{M}_{2 \ell + 2}}{\bar{S}_{2 \ell + 1}} =  \bar{M}_{2}\,.
\ee
This is the final form of the first universal relation, which has now become not only EoS-independent but also spin-independent. Notice that the third condition of the elliptical isodensity approximation was never needed.    

Let us now consider the second universal relation. To derive this, let us first use the self-similar isodensity condition [condition (1)] and define 
\be
\label{R-rescaling}
R_{\ell} = \rho_{c} \left(\frac{a_{1}}{\xi_{1}}\right)^{\ell+3} {\cal{R}}_{n,\ell}\,,
\ee 
where $\xi = (\xi_{1}/a_{1}) \tilde{r}$ is a dimensionless radius with $\xi = \xi_{1}$ the surface of the star and 
\be
\label{eq:Rcal-def}
{\cal{R}}_{n,\ell} = \int_{0}^{\xi_{1}} \left[\vartheta(\xi)\right]^{n} \xi^{\ell+2} d\xi\,,
\ee
with $\vartheta(\xi) = (\rho/\rho_{c})^{1/n}$. 
We can re-express $\rho_c$ in terms of the mass monopole [Eq.~\eqref{eq:mass-def}] to find
\be
\label{eq:rhoc}
\rho_{c} = \frac{M}{2 \pi I_{0,3}} \left( \frac{\xi_1}{a_1} \right)^3 \frac{1}{\mathcal{R}_{n,0}}\,.
\ee
Similarly, we can re-express $a_1/\xi_1$ in terms of $S_{2 \ell +1}/S_1$. After doing so, we find the second universal relation:
\be
\label{universal-2}
\bar{M}_{2 \ell + 2} = \bar{A}_{n,\ell} \; \left(\bar{S}_{2 \ell + 1}\right)^{1 + 1/\ell}\,,
\ee
where we have defined the coefficient
\be
\bar{A}_{n,\ell} =  \left(\frac{4 \ell + 3}{6 \ell + 3}\right)^{1 + 1/\ell} \frac{I_{2 \ell+2,3}}{I_{0,3}} \left(\frac{\delta I_{1}}{\delta I_{2 \ell+1}}\right)^{1+1/\ell} \tilde{\mathcal{R}}_{n, \ell}\,,
\label{barA-coeff}
\ee
with
\be
\label{eq:Rtilde}
 \tilde{\mathcal{R}}_{n, \ell} \equiv \frac{{\cal{R}}_{n,2}^{1+1/\ell}}{{\cal {R}}_{n,0} \; {\cal{R}}_{n,2 \ell+2}^{1/\ell}}\,.
\ee
Observe that $\bar{A}_{n,\ell}$ depends on the angular integral, and hence it depends on the spin in general. Notice that we have only applied the first condition of the elliptical isodensity approximation to derive Eqs.~\eqref{universal-2}--\eqref{eq:Rtilde}. 

Let us now use the second and third conditions. If one uses the second one through Eq.~\eqref{eq:Theta-original}, then Eq.~\eqref{barA-coeff} reduces to
\be
\bar{A}_{n,\ell} = \frac{(2 \ell +3)^{1/\ell}}{3^{(1+1/\ell)}} \tilde{\mathcal{R}}_{n, \ell}\,.
\label{barA-coeff-original}
\ee
On top of this, if one assumes the third condition through Eq.~\eqref{eq:theta-spherical}, Eq.~\eqref{eq:Rtilde} reduces to
\be
 \tilde{\mathcal{R}}_{n, \ell}^\LE \equiv \frac{{\cal{R}}_{n,2}^\LE{}^{1+1/\ell}}{|\vartheta'_\LE (\xi_1)| \xi_1^2 {\cal{R}}^\LE_{n,2 \ell+2}{}^{1/\ell}}\,,
\ee
where we used the LE equation:
\be
\frac{1}{\xi^2} \frac{d}{d \xi} \xi^2 \frac{d}{d \xi} \vartheta = - \vartheta^n\,, 
\ee
and the index ``LE'' is to remind the reader that the quantities are evaluated assuming that $\vartheta(\xi)$ is the LE function. $\mathcal{R}_{n,\ell}^\LE$ corresponds to Eq.~\eqref{eq:Rcal-def} with $\vartheta(\xi)$ replaced by $\vartheta_\LE(\xi)$.

Reference~\cite{Stein:2013ofa} showed that when one applies all of the three conditions, $\bar{A}_{n,\ell}$ depends on the EoS, here parametrized through $n$, by at most $\mathcal{O}(10\%)$ for $\ell \leq 4$ with $n=[0.3,1]$. This implies that the universal relations among lower $\ell$ multipoles depend only weakly on the EoS. Moreover, through this analysis one finds that Eq.~\eqref{barA-coeff-original} is spin-independent.

The \ItoQ relation discussed in the previous section can be reproduced as follows. First, using the self-similar isodensity condition [condition (1)] and Eq.~\eqref{eq:mass-def}, one finds
\be
\frac{\bar M_2}{\bar I} = -\frac{3 I_{2,3}}{2 \delta I_1 \chi^2}\,.
\ee
If one further imposes the elliptical condition [condition (2)], one finds~\cite{Stein:2013ofa}
\be
\frac{\bar M_2}{\bar I} = \frac{e^2}{2 \chi^2}\,.
\ee
In order to see the EoS-universality of the \ItoQ relation for a fixed $\chi$, one needs to express $e$ in terms of $\chi$, $n$ and $\bar{I}$. Such an expression is given in Ref.~\cite{Stein:2013ofa} in the elliptical isodensity approximation, using the relation between $\Omega$ and $e$ found in Ref.~\cite{Lai:1993ve}. 

In the remainder of this section, however, we will focus on how the universal relation among multipole moments is affected by relaxing the elliptical isodensity approximation. Reference~\cite{Stein:2013ofa} showed that the \ItoQ relation has a similar (but slightly smaller) EoS-variation to the $\bar S_3$--$\bar M_4$ relation, with the EoS-dependence of the latter encoded in $\bar A_{n,1}$. We thus expect that the the \ItoQ relations will be affected by the relaxation of the elliptical isodensity approximation in roughly the same way as the universal relation among multipole moments is affected. We will study the latter in detail in the next subsections, but we leave a detailed analysis of how the \ItoQ relations are affected by the elliptical isodensity approximation to future work. 

\subsection{Relaxing the Elliptical Condition}

\begin{figure}[tb]
\centering
\includegraphics[width=0.75\columnwidth,clip=true]{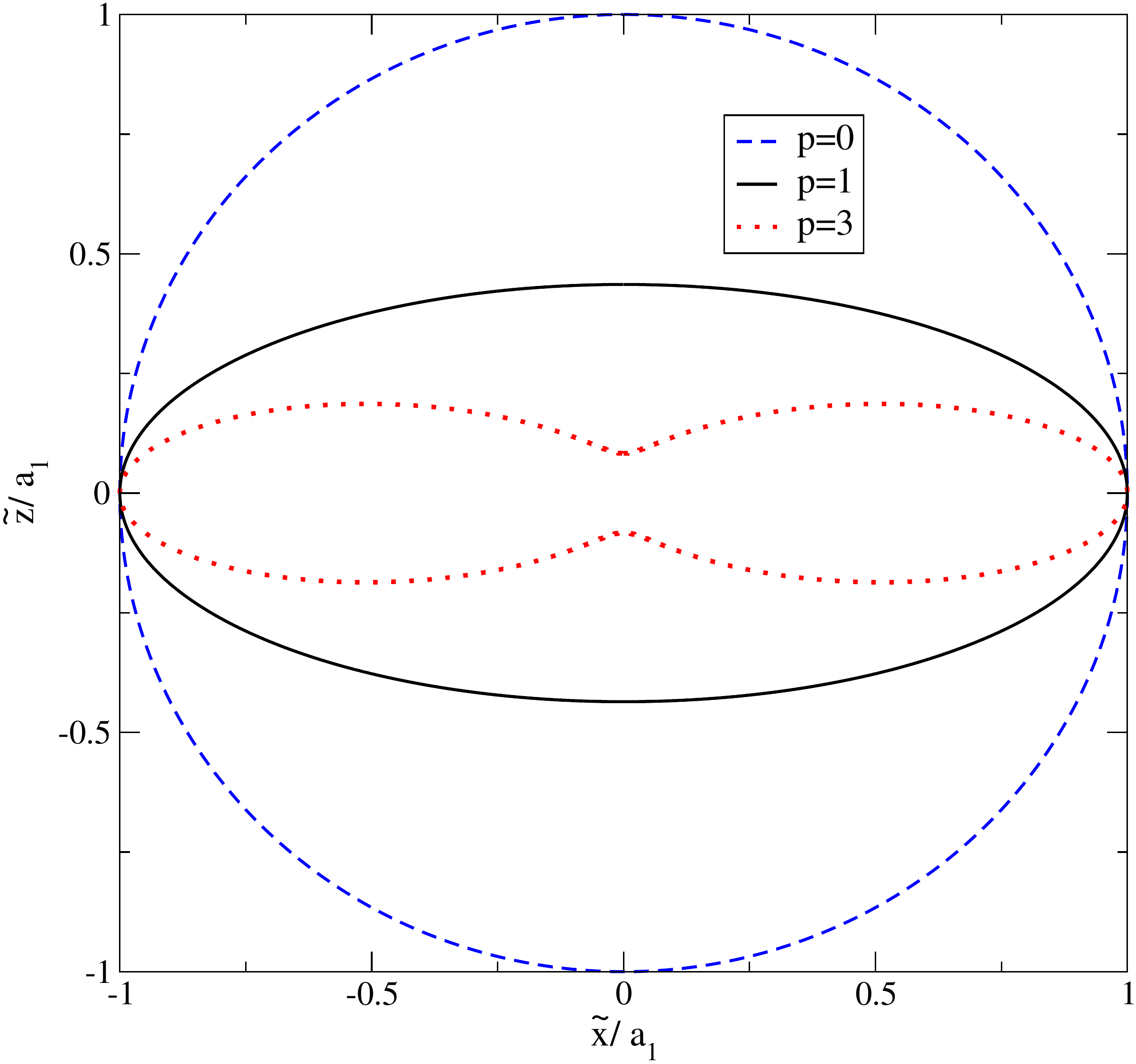}
\caption{
(Color online) Stellar shape in the $\tilde{x}$-$\tilde{z}$ plane given by Eq.~\eqref{eq:Theta-p} for different values of $p$ and $e = 0.9$. 
\label{fig:ellipses}
}
\end{figure}

Let us now relax each of the conditions made above in turn to see how it affects the universality. We start off by breaking the elliptical condition, since this is easiest. We can achieve this by choosing the function $\Theta(\mu)$ to be something other than Eq.~\eqref{eq:Theta-original}. This function specifies the shape of the self-similar, isodensity contours. One could, for example, choose
\be
\label{eq:Theta-p}
\Theta(\mu) = \left[\frac{1 - e^{2}}{1 - e^{2} \left(1 - \mu^2\right)}\right]^{p/2}\,,
\ee
for $p$ an odd, positive integer, i.e. $p \in {\mathbb{N}}_{\rm odd}$. When $p=1$, one recovers an elliptical coordinate system, while for $p>1$ the star becomes more peanut-shaped (in the $\tilde{z}$-$\tilde{x}$ plane), and for $p<1$ it becomes more spherical, as shown in Fig.~\ref{fig:ellipses}. Notice that when $e = 0$, the star is spherical irrespective of the choice of $p$, since then $\Theta = 1$. Interestingly, as $p$ increases, the peanut-shape becomes more and more pronounced. Clearly then, the cases when $p > 1$ should be taken as toy problems only. In order to keep the analysis generic, we do not specify the form of $\Theta$ in this subsection.

Modifying the functional form of $\Theta$ only changes the angular integral $I_{\ell,k}$ in the multipole moment expression in Eq.~\eqref{eq:mass-def}. Thus, the relations among multipoles are still valid if they are expressed in terms of $I_{\ell,k}$, without explicitly solving $I_{\ell,k}$. It follows that the universal relations in Eqs.~\eqref{universal-1} and~\eqref{universal-2} with Eq.~\eqref{barA-coeff} still hold even if one relaxes the elliptical condition. 

The first universal relation is automatically EoS-independent, so let us focus on the EoS-dependence of the second one. Let us consider the fractional difference of the barred coefficients from some mean $\langle \bar{A}_{n, \ell} \rangle$ on $n$. Since the EoS-dependence is encoded in $\tilde{\mathcal{R}}_{n, \ell}$ of Eq.~\eqref{barA-coeff}, one easily finds
\be
\label{eq:Abar-frac-diff}
\frac{\bar{A}_{n,\ell}-\langle \bar{A}_{n, \ell} \rangle}{\langle \bar{A}_{n, \ell} \rangle} = \frac{\tilde{\mathcal{R}}_{n, \ell} - \langle \tilde{\mathcal{R}}_{n, \ell} \rangle}{\langle \tilde{\mathcal{R}}_{n, \ell} \rangle}\,.
\ee
Since $\langle \tilde{\mathcal{R}}_{n, \ell} \rangle$ does not depend
on $\Theta$, the fractional difference of $\bar{A}_{n,\ell}$ is the
same even if one relaxes the elliptical condition. This shows
that the shape of the self-similar surfaces does
not affect the universality at all.

\subsection{Relaxing the Spherical Density Profile Condition}

Let us now consider the relaxation of the spherical density profile condition, namely the assumption that $\vartheta (\xi)$ is a LE function. One could, for example, let 
\be
\vartheta(\xi) = \vartheta_\LE (\xi) \left[1 + \epsilon_1 \vartheta_1 (\xi)\right]\,,
\ee
where $\epsilon_1 \ll 1$ is some small number and $\epsilon_1 \vartheta_1 (\xi)$ is a small deformation. Such a modification to the isodensity condition can be easily propagated throughout the calculation of~\cite{Stein:2013ofa} described in Sec.~\ref{sec:3-hair}, since effectively it reduces to the same integrals but with different harmonic number. The angular integrals, of course, are completely unmodified. The radial integrals can still be rescaled as in Eq.~\eqref{R-rescaling}, except with the replacement 
\ba
\label{replacement}
{\cal{R}}_{n,\ell} &\to& \int^{\xi_1}_0 \left[ \vartheta (\xi) \right]^n \left[ 1 + \epsilon_1 \vartheta_1 (\xi) \right]^n \xi^{\ell+2} d\xi\,, \nn \\
&=&  {\cal{R}}_{n,\ell} + \epsilon_1 n {\cal{R}}_{n,\ell}^{(\vartheta_1)} + {\cal{O}}(\epsilon_1^{2})\,,  
\ea
where we have expanded in $\epsilon_1 \ll 1$ and 
\be
{\cal{R}}_{n,\ell}^{(\vartheta_1)} \equiv \int_0^{\xi_1} [\vartheta (\xi)] ^n \vartheta_1 (\xi) \;  \xi^{\ell +2} d\xi\,.
\ee
We see then that the modification to the density profile has changed the radial integral through the addition of a new term that depends on $\vartheta_1$ and $\epsilon_1$, but also on $n$ and $\ell$. 

Let us now consider how these modifications affect the two universal relations described in Sec.~\ref{sec:3-hair}. First, it is clear that the first relation is not modified at all since Eq.~\eqref{universal-1} does not depend on the radial integral. On the other hand, the second relation, given by Eq.~\eqref{universal-2}, is modified since it depends on the radial integral through $\tilde{R}_{n,\ell}$ in Eq.~\eqref{eq:Rtilde}. Using Eq.~\eqref{replacement}, $\bar{A}_{n,\ell}$ becomes

\ba
\bar{A}_{n,\ell} &=& \frac{(2 \ell +3)^{1/\ell}}{3^{(1+1/\ell)}} \tilde{\mathcal{R}}_{n,\ell}^\LE  \nn \\
& & \left\{ 1  + \epsilon_1 n \left[ \frac{(\ell+1)}{\ell}  \frac{{\cal{R}}_{n,2}^{(\vartheta_1)}}{{\cal{R}}_{n,2}^\LE} - \frac{1}{\ell} \frac{{\cal{R}}_{n,2\ell+2}^{(\vartheta_1)}}{{\cal{R}}_{n,2 \ell+2}^\LE} - \frac{\mathcal{R}_{n,0}^{(\vartheta_1)}}{\mathcal{R}_{n,0}^\LE} \right] \right\} \nn \\
& & + \mathcal{O}(\epsilon_1^2) \,.
\ea
From this result, we can already derive some interesting conclusions. The deviation from the standard universal relations are both \emph{linear} in $\epsilon_1$ and $n$. In fact, how much the relations deviate will simply depend on the magnitude of $\epsilon_1$, since $n = {\cal{O}}(1)$.

\subsection{Relaxing the Self-Similar Isodensity Condition}
\label{sec:relaxing-self-similarity}

Let us relax the self-similar isodensity condition. In particular, one may wish to force the contours to become more spherical as $\tilde{r}$ approaches the NS core, which is in fact what occurs physically inside NSs~\cite{1978trs..book.....T}. Therefore, we promote the eccentricity to be a function of the radial coordinate, i.e.
\be
\label{eq:ecc-replace}
e \to e \left( \tilde{r} \right) = e_0 \;  f\left( \frac{\tilde{r}}{a_1} \right)\,,
\ee
where $e_0$ is the eccentricity \emph{at the surface} and $f$ is an arbitrary function of $\tilde{r}/a_1$ with $f(1)=1$. 
Although the relaxation of the self-similar isodensity condition through Eq.~\eqref{eq:ecc-replace} renders the integrals of the previous subsections non-separable in general, one can still perform the integrals analytically by calculating them as follows.

When the eccentricity becomes a function of the radial coordinate, $\Theta (\mu)$ in Eq.~\eqref{eq:Theta-original} now becomes $\Theta (\tilde{r}, \mu)$. In spite of this, one can still solve the angular integral \emph{exactly}. The mass and spin moments then become
\ba
\label{eq:mass-relax-e}
M_\ell &=& 2 \pi \int_0^{a_1} \rho (\tilde{r}) \tilde{r}^{\ell +2} I_{\ell, 3} (\tilde{r}) d\tilde{r}\,,  \\ 
\label{eq:current-relax-e}
S_\ell &=& \frac{4 \pi \ell}{2 \ell + 1} \Omega \int_0^{a_1} \rho (\tilde{r}) \tilde{r}^{\ell +3} \delta I_{\ell} (\tilde{r}) d\tilde{r}\,, 
\ea
where~\cite{Stein:2013ofa}
\ba
\label{eq:I3}
I_{\ell, 3} (\tilde{r}) &=& (-)^{\frac{\ell}{2}}\frac{2 }{\ell+1}\sqrt{1-e \left( \tilde{r} \right)^2} e \left( \tilde{r} \right)^\ell\,, \\
\label{eq:deltaI}
\delta I_{\ell} (\tilde{r}) &=& (-)^{\frac{\ell-1}{2}}\frac{2 (2 \ell+1) }{\ell (\ell+2)}\sqrt{1-e \left( \tilde{r} \right)^2}  e \left( \tilde{r} \right)^{\ell-1}\,. 
\ea
One can rewrite Eqs.~\eqref{eq:mass-relax-e} and~\eqref{eq:current-relax-e} as
\be
\label{eq:mass-relax-e-2}
M_\ell = 2 \pi I_{\ell,3}^{(0)} R_{\ell}^{(M)}\,, \quad
S_\ell = \frac{4 \pi \ell}{2 \ell + 1} \Omega \delta I_{\ell}^{(0)} R_{\ell+1}^{(S)}\,, 
\ee
where the superscript ${(0)}$ refers to setting eccentricity constant [i.e. $e \to e_0 = {\rm{const}}$ in Eqs.~\eqref{eq:I3} and~\eqref{eq:deltaI}] and 
\ba
R_{\ell}^{(A)} & \equiv & \int_0^{a_1} \rho (\tilde{r}) \tilde{r}^{\ell +2} \left[f \left( \frac{\tilde{r}}{a_1} \right) \right]^{n_A} \sqrt{\frac{1- e_0^2 \left[ f \left( \tilde{r}/a_1 \right)  \right]^2}{1-e_0^2}} d\tilde{r}\,, \nn \\ 
\ea
with $A = (M,S)$, $n_M = \ell$ and $n_S = \ell-2$. Observe the resemblance of Eqs.~\eqref{eq:mass-relax-e-2} and~\eqref{eq:mass-def}. Observe also that $R_{\ell}^{(M)} = R_{\ell} = R_{\ell}^{(S)}$ when $f(\tilde{r}/a_1) = 1$.

\subsubsection{Modification to the Multipole Moments Relation} 
Let us now investigate how the universal relations change by the replacement in Eq.~\eqref{eq:ecc-replace}. First, similar to Eq.~\eqref{R-rescaling}, we define $\mathcal{R}_{n,\ell}^{(A)}$ as
\be
\label{eq:R-scaling2}
R_{\ell}^{(A)} = \rho_{c} \left(\frac{a_{1}}{\xi_{1}}\right)^{\ell+3} {\cal{R}}_{n,\ell}^{(A)}\,,
\ee
with
\ba
\mathcal{R}_{n,\ell}^{(A)} \equiv  \int_0^{\xi_1} \vartheta^n  \xi^{\ell +2} \left[f \left( \frac{\xi}{\xi_1} \right) \right]^{n_A} \sqrt{\frac{1- e_0^2 \left[ f \left( \xi/\xi_1 \right)  \right]^2}{1-e_0^2}} d\xi\,. \nn \\
\ea
We will here assume that $\vartheta$ is a LE function, but omit the index ``LE'' for convenience.
In the slow-rotation limit, the above equation becomes
\be
\mathcal{R}_{n,\ell}^{(A)} =  \left\{ \int_0^{\xi_1} \vartheta^n  \xi^{\ell +2} \left[f \left( \frac{\xi}{\xi_1} \right) \right]^{n_A} d\xi \right\} \left[ 1 + \mathcal{O}\left( e_0^2 \right) \right]\,. \nn \\
\ee

The first universal relation is obtained by substituting Eq.~\eqref{eq:R-scaling2} into Eq.~\eqref{eq:mass-relax-e-2} and taking the ratio of these moments. After nondimensionalizing them through Eqs.~\eqref{eq:Mbar-def} and~\eqref{eq:Sbar-def}, we find
\be
\frac{\bar{M}_{2\ell +2}}{\bar{S}_{2\ell +1}} = \bar{B}^{(f)}_{n,\ell} \bar{M}_2\,,
\ee
where we have defined 
\be
\label{eq:Bbar-noexp}
\bar{B}^{(f)}_{n,\ell} = \frac{\mathcal{R}_{n,2\ell+2}^{(M)} \mathcal{R}_{n,2}^{(S)}}{\mathcal{R}_{n,2\ell+2}^{(S)} \mathcal{R}_{n,2}^{(M)}}\,.
\ee
Notice that when $f(\tilde{r}/a_1)=1$, $\bar{B}^{(f)}_{n,\ell} = 1$ and thus, one recovers the first universal relation of Eq.~\eqref{universal-1-original}. Notice also that $\bar{B}^{(f)}_{n,0} = 1$ irrespective of the functional form of $f(\tilde{r}/a_1)$. 

We follow the same procedure explained in Sec.~\ref{sec:3-hair} to derive the second relation. Namely, we first rewrite $\rho_c$ in terms of $a_1$ and $M$ from Eqs.~\eqref{eq:mass-relax-e-2} and~\eqref{eq:R-scaling2} with $\ell=0$, as in Eq.~\eqref{eq:rhoc}, with the radial integral performed using the LE equation. Next, we express $a_1/\xi_1$ in terms of $S_{2\ell +1}/S_1$ from Eqs.~\eqref{eq:mass-relax-e-2} and~\eqref{eq:R-scaling2}. Then, from these equations, one finds
\be
\bar{M}_{2 \ell + 2} = \bar{A}_{n,\ell}^{(f)} \left(\bar{S}_{2 \ell+1}\right)^{1 + 1/\ell}\,,
\ee
where we have defined
\be
\label{eq:Abar-noexp}
\bar{A}_{n, \ell}^{(f)} = \frac{(3 + 2 \ell)^{1/\ell}}{3^{1+1/\ell}} \frac{{\cal{R}}_{n,2 \ell + 2}^{(M)} \left({\cal{R}}_{n,2}^{(S)}\right)^{1 + 1/\ell}}{{\cal{R}}_{n,0}^{(M)} \left({\cal{R}}_{n,2\ell+2}^{(S)}\right)^{1+1/\ell}}\,.
\ee
Notice again that when $f(\tilde{r}/a_1) = 1$ one recovers the results of~\cite{Stein:2013ofa}. 

\subsubsection{Effect on Universality } 

\begin{figure}[tb]
\centering
\includegraphics[width=\columnwidth,clip=true]{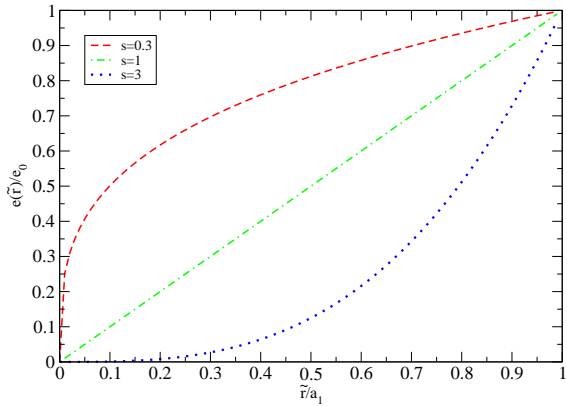}
\caption{
(Color online) 
\label{fig:e} Eccentricity as a function of radius given by Eq.~\eqref{eq:e-assumption} for different choices of $s$.}
\end{figure}

We see clearly that the introduction of a non-trivial radial dependence in the eccentricity has led to different universal relations, but the question is whether such a modification spoils universality. We address this by looking at some example below. Let us consider the simple toy model 
\be
\label{eq:e-assumption}
f \left( \frac{\tilde{r}}{a_1} \right) = \left(\frac{\tilde{r}}{a_{1}}\right)^{s}\,,
\ee
for some power $s>0$ that controls how fast the star becomes spherical as one approaches the core. As shown in Fig.~\ref{fig:e}, when $s = 0.3$, the star is mostly elliptical until $\tilde{r}/a_{1} < 0.1$, at which point $e(\tilde{r}) < e_{0}/2$. On the other hand, when $s = 3$ the star is more spherical, with its eccentricity dropping below half its initial value already at $\tilde{r}/a_{1} = 0.8$.  The actual $e(\tilde{r})$ for Newtonian stars can be estimated by solving the Clairaut-Radau equation~\cite{1978trs..book.....T,brooker-olle,mora-will}, as we will do in Sec.~\ref{sec:3-hair-real-ecc}. Since the purpose of this subsection is to investigate how $e(\tilde{r})$ affects the universality, we do not solve such equation but treat $s$ as arbitrary. 

\begin{figure*}[tb]
\centering
\includegraphics[width=\columnwidth,clip=true]{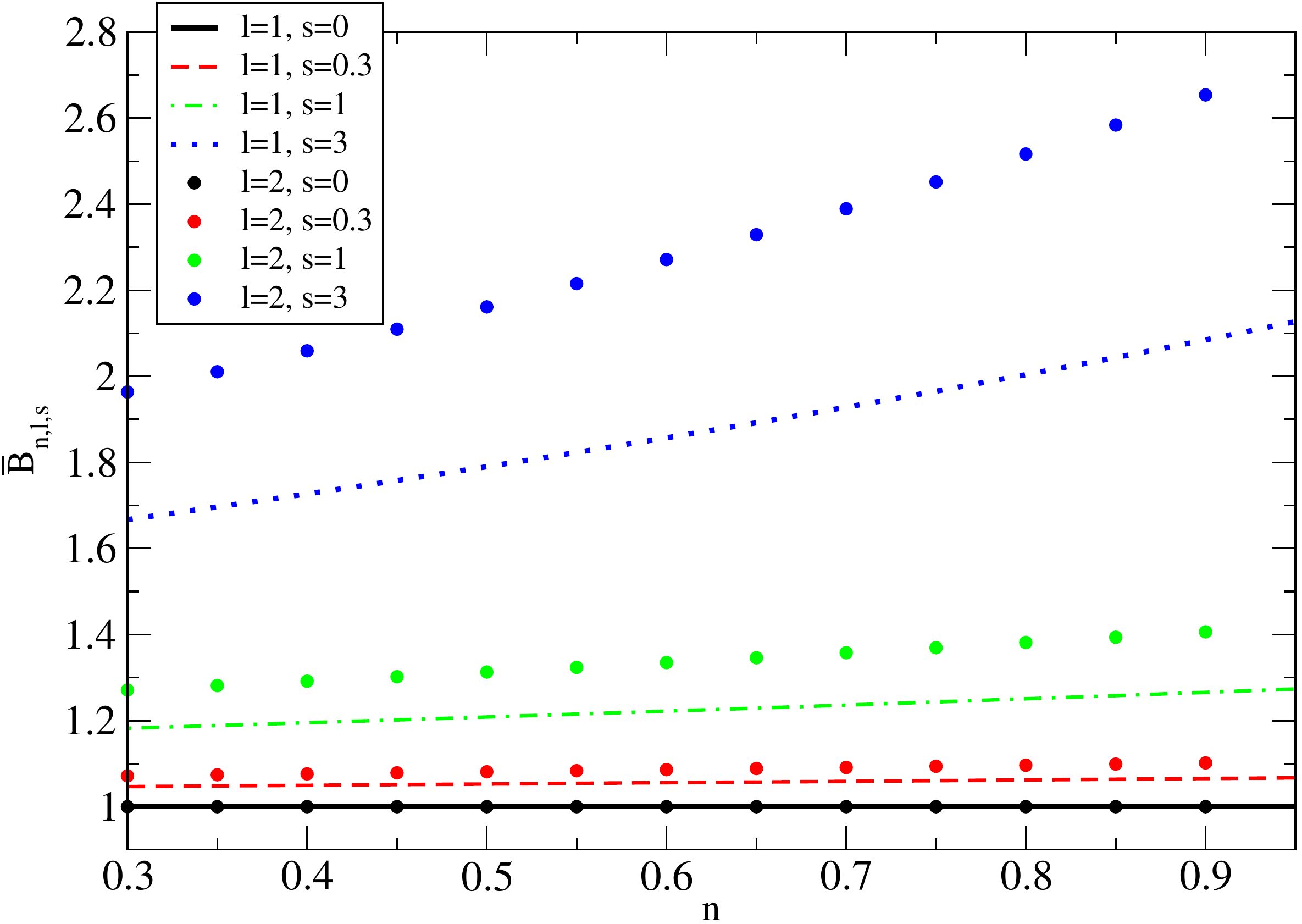}
\includegraphics[width=\columnwidth,clip=true]{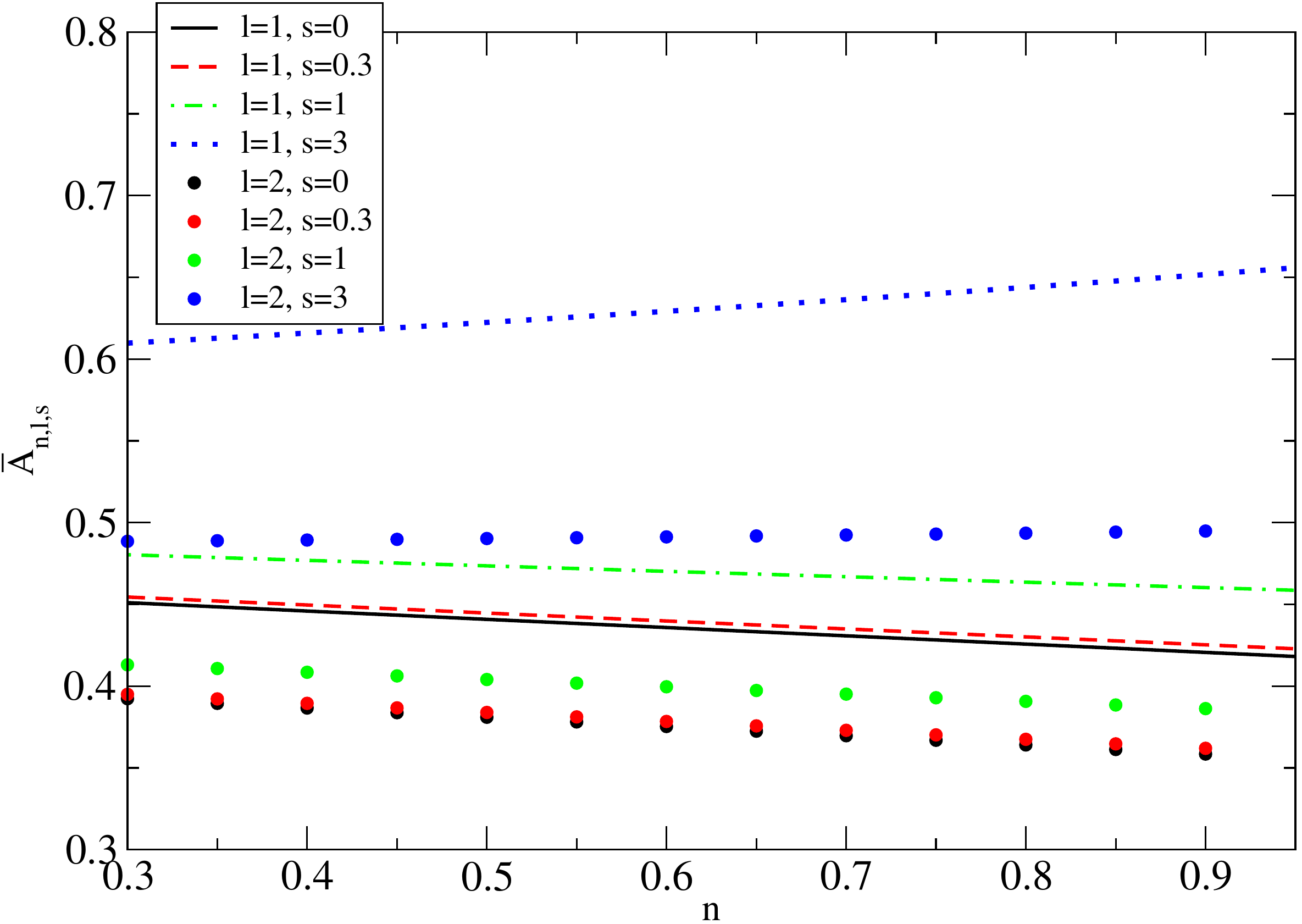}
\caption{
(Color online) 
\label{fig:BbarAbar} $\bar{B}_{n,\ell,s}$ (left) and $\bar{A}_{n,\ell,s}$ (right) given by Eqs.~\eqref{eq:Bbar-nls} and~\eqref{eq:Abar-nls} respectively as a function of $n$ for different choices of $s$ and $\ell$. Observe that as we increase $s$, the universality is lost, especially as $s>1$. Observe also that $\bar{B}_{n,\ell,s}$  and $\bar{A}_{n,\ell,s}$ are barely modified from the $s=0$ case when $s = 0.3$.}
\end{figure*}

For simplicity, let us work in the slow-rotation limit. The coefficients in the universal relations then become
\ba
\label{eq:Bbar-nls}
\bar{B}_{n,\ell}^{(f)} &=& \bar{B}_{n,\ell,s} \equiv  \frac{{\cal{R}}_{n,2}}{{\cal{R}}_{n,2+2s}} \frac{{\cal{R}}_{n,2\ell+2+2s(\ell+1)}}{{\cal{R}}_{n,2\ell+2+2s\ell}}\,, \\
\label{eq:Abar-nls}
\bar{A}_{n, \ell}^{(f)} &=& \bar{A}_{n, \ell, s} \nn \\
& \equiv&  \frac{(3 + 2 \ell)^{1/\ell}}{3^{1+1/\ell}} \frac{1}{\xi_{1}^{2} |\vartheta'_{1}|} \frac{{\cal{R}}_{n,2 \ell + 2 + 2 s(\ell+1)} \left({\cal{R}}_{n,2}\right)^{1 + 1/\ell}}{\left({\cal{R}}_{n,2\ell+2+2s\ell}\right)^{1+1/\ell}}\,. \nn \\
\ea
 Figure~\ref{fig:BbarAbar} shows $\bar{B}_{n,\ell,s}$ (left panel) and $\bar{A}_{n,\ell,s}$ (right panel) as a function of $n$ for different choices of $s$ and $\ell$. Let us first focus on the first universal relation by looking at the left panel of Fig.~\ref{fig:BbarAbar}. Recall that when $\ell=0$, $\bar{B}_{n,0,s} = 1$, which is why we plot only the $\ell \geq 1$ cases. Observe that irrespective of the choice of $\ell$, $\bar{B}_{n,\ell,0.3} \sim \bar{B}_{n,\ell,0}$, while $\bar{B}_{n,\ell,s}$ is very different from $\bar{B}_{n,\ell,0}$ in the more spherical cases when $s>1$. Observe also that when $s=3$, {}the universality is essentially lost, with $\bar{B}_{n,\ell,s}$ changing by as much as $50\%$ over the range of $n$ explored here. This is to be compared to a variability of about $1\%$ when $s=0$ or $0.3$. Let us now look at the second universal relation by focusing on the right panel of Fig.~\ref{fig:BbarAbar}. As in the case of $\bar{B}_{n,\ell,s}$, $\bar{A}_{n,\ell,s}$ is very similar when $s=0$ and when $s=0.3$. Also as before, the universality with $n$ deteriorates as $s$ increases, but the effect is much less pronounced this time.  

In order to understand the behavior of $\bar{B}_{n,\ell}$, let us consider its asymptotic behavior in the $s \rightarrow \infty$ limit for an $n=0$ polytrope, where $\vartheta = 1 - \xi^2/6$ and $\xi_1 = \sqrt{6}$ so that $\mathcal{R}_{0,\ell} = 6^{(\ell + 3)/2}/(\ell + 3)$. First, the mass and current multipole moments in the slow-rotation limit become 
\begin{align}
\label{eq:mass-1}
M_{2 \ell + 2} &= \left(-\right)^{\ell + 1} \frac{4 \pi e_{0}^{2\ell+2} }{2 \ell + 3} \left(\frac{a_{1}}{\xi_{1}}\right)^{2 \ell + 5} \!\!\!\!\! \frac{\rho_{c}}{\xi_{1}^{2 s (\ell+1)}} {\cal{R}}_{0,2 \ell + 2 + 2 s (\ell+1)}\,,
\\
\label{eq:current-1}
S_{2 \ell + 1} &= \left(-\right)^{\ell} \frac{8 \pi \Omega}{2 \ell +3}  e_{0}^{2\ell} \left(\frac{a_{1}}{\xi_{1}}\right)^{2 \ell + 5} \frac{\rho_{c}}{\xi_{1}^{2s \ell}} {\cal{R}}_{0,2 \ell + 2  + 2 s \ell}\,. 
\end{align}
Then, from these equations, one finds
\be
M_{2\ell + 2} \sim \frac{1}{s} \quad (\ell \geq 0)\,, \quad S_{2\ell + 1} \sim \frac{1}{s} \quad (\ell \geq 1)\,, 
\ee
and $M \sim s^0 \sim S_1$. Notice that both $M_{2\ell + 2} (\ell \geq 0)$ and $S_{2\ell + 1} (\ell \geq 1)$ vanish in the $s \to \infty$ limit. This is because in this limit, the density contour is spherical everywhere except at the surface. Since the measure of the integrals is zero, the multipole moments vanish. The above scaling leads to 
\be
\bar{B}_{0,\ell} = \frac{M_{2\ell +1} S_1}{S_{2\ell+1} M_2} \sim s\,
\ee
in the large $s$ limit.
For $n$ close to zero, $\bar{B}_{n,\ell}$ can be expressed as $\bar{B}_{n,\ell} \sim (c_0 + c_1 n)s$ for some constants $c_0$ and $c_1$ and the fractional difference of $\bar{B}_{n,\ell}$ with respect to some mean $n$ is given by
\be
\frac{\bar{B}_{n,\ell}-\langle \bar{B}_{n,\ell} \rangle}{\langle \bar{B}_{n,\ell} \rangle} \sim \frac{c_1 (n - \langle n \rangle)}{c_0 + c_1 \langle n \rangle}\,.
\ee
Therefore, for large $s$, the fractional difference of $\bar{B}_{n,\ell}$ is non-vanishing and can be relatively large.

\section{Eccentricity Profiles for Relativistic Stars}
\label{sec:ecc-profile}

The previous section showed that the elliptical isodensity approximation plays a crucial role in the universality. We now need to investigate the validity of such an approximation for realistic relativistic stars, i.e. study the eccentricity radial profile of such stars, thus addressing the third question in the introduction. 

We will first look at the profile of slowly-rotating NSs and QSs. We will show that the eccentricity only changes by $\sim 10\%$ within the region that matters to the universality. We will then look at the profile of rapidly-rotating NSs and show that the eccentricity variation is always smaller than $(20-30)\%$ even for rapidly-rotating NSs in the region that matters. We will finally re-consider the relation for uniformly-rotating Newtonian polytropes described in Sec.~\ref{sec:relaxing-self-similarity} using a realistic eccentricity profile for NSs. We will show that such modification only affects the 3-hair relation in~\cite{Stein:2013ofa} by less than $10\%$ \emph{relative} to the constant-eccentricity Newtonian results, e.g.~if certain relations are universal to $1\%$  when using the elliptical isodensity approximation, corrections due to a non-constant eccentricity profile induce modifications of $0.1\%$.  

\subsection{Slowly-Rotating Stars}

We first look at eccentricity profiles for slowly-rotating NSs and QSs. We extract the eccentricity of isodensity surfaces from the embedded surface following~\cite{hartlethorne}.
Figure~\ref{fig:ecc-profile} shows the radial profile of the NS eccentricity, which affects the stellar quadrupole moment. Within (50--95)\% of the stellar radius, the eccentricity only changes by $\sim 10\%$. Observe also that the EoS-variation in this radial region is smaller than the one in the region less than 50\% of the radius.

\begin{figure}[tb]
\centering
\includegraphics[width=\columnwidth,clip=true]{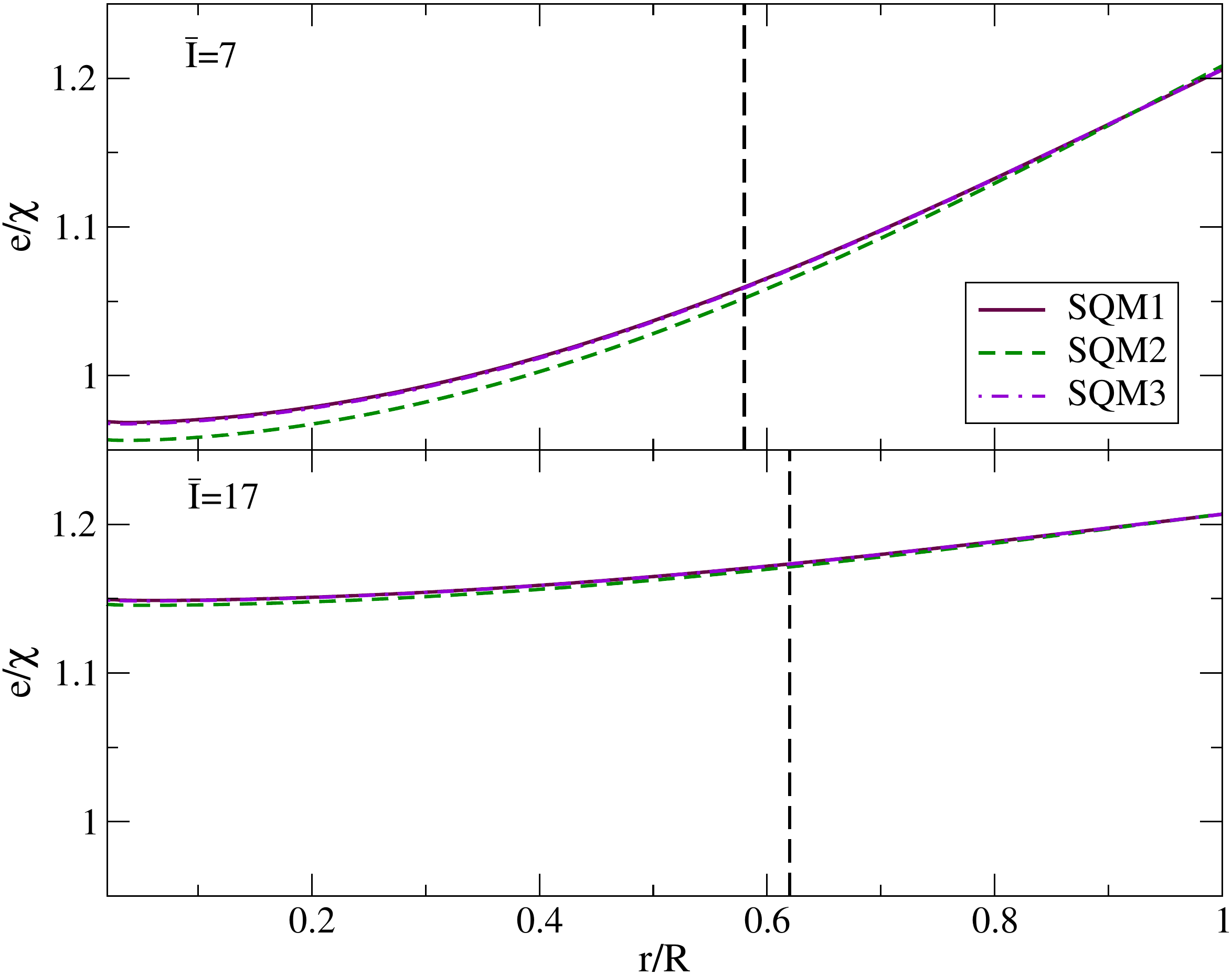}
\caption{
(Color online) 
Eccentricity profile (same as Fig.~\ref{fig:ecc-profile}) but for QSs. The vertical dashed lines are built in the same way as in Fig.~\ref{fig:density-profile-QS}. Observe that the eccentricity does not change much for stars with large $\bar{I}$.
\label{fig:ecc-profile-QS}
}
\end{figure}

Figure~\ref{fig:ecc-profile-QS} shows the eccentricity profile for QSs. Observe that the eccentricity is almost constant for QSs with large $\bar{I}$. For such stars, the elliptical isodensity approximation used in~\cite{Stein:2013ofa} becomes an excellent approximation. Such approximate self-similarity in the stellar eccentricity suppresses the EoS variation in the relations among stellar multipole moments, which in turn realizes the universality.

\subsection{Rapidly-Rotating Stars}

\begin{figure}
\centering
\includegraphics[width=.215\textwidth]{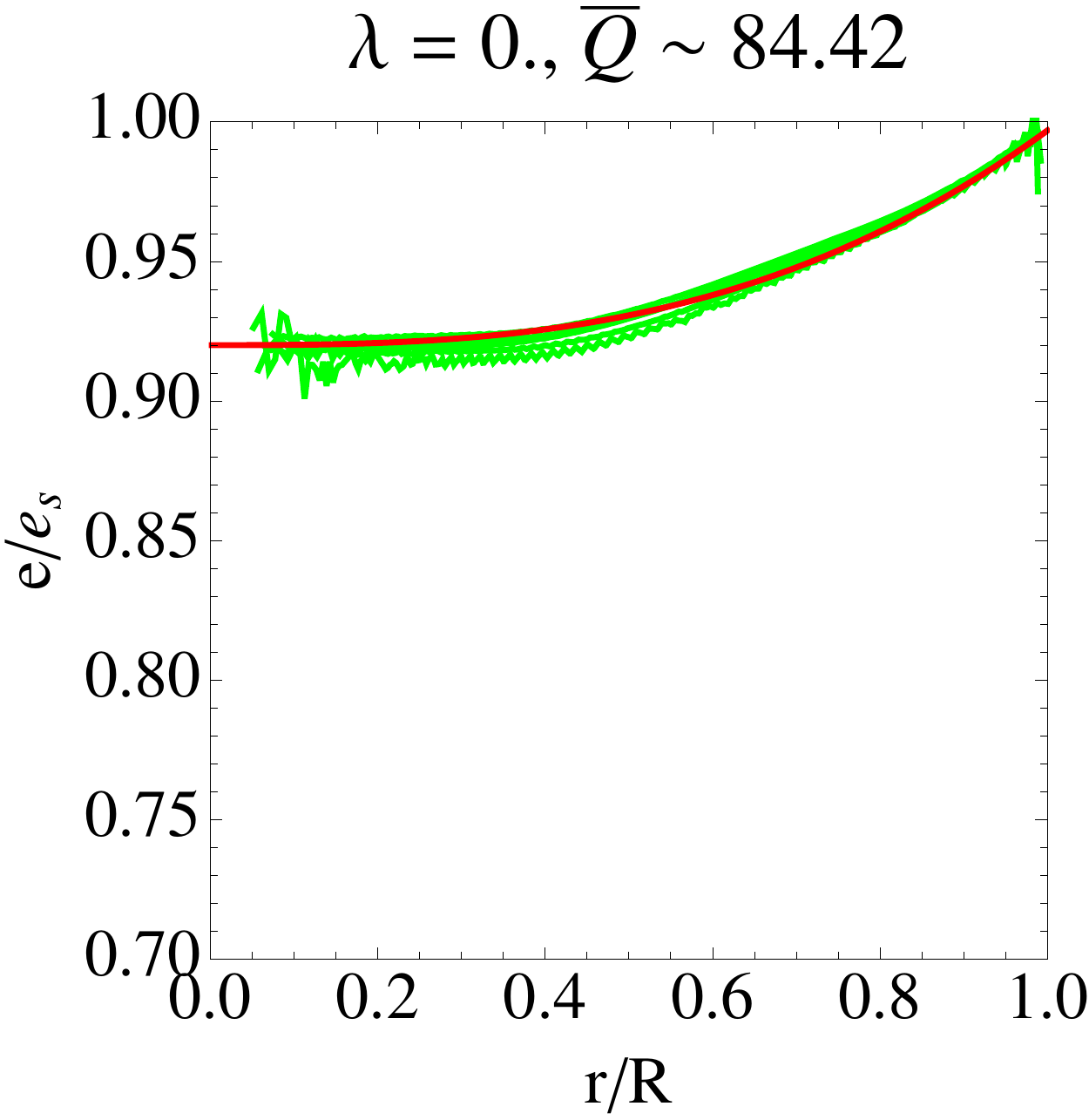}
\includegraphics[width=.215\textwidth]{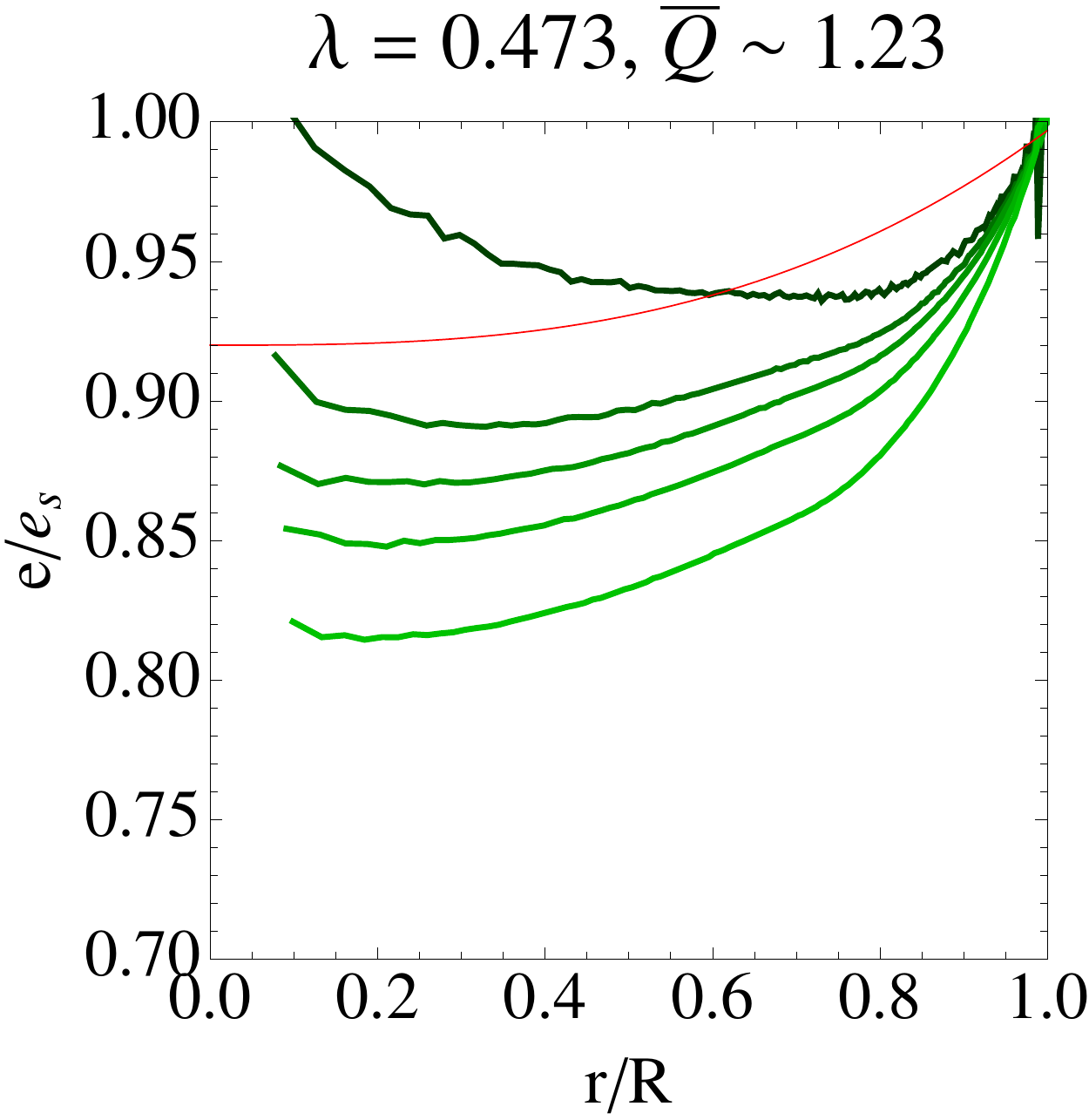}
\\
\includegraphics[width=.215\textwidth]{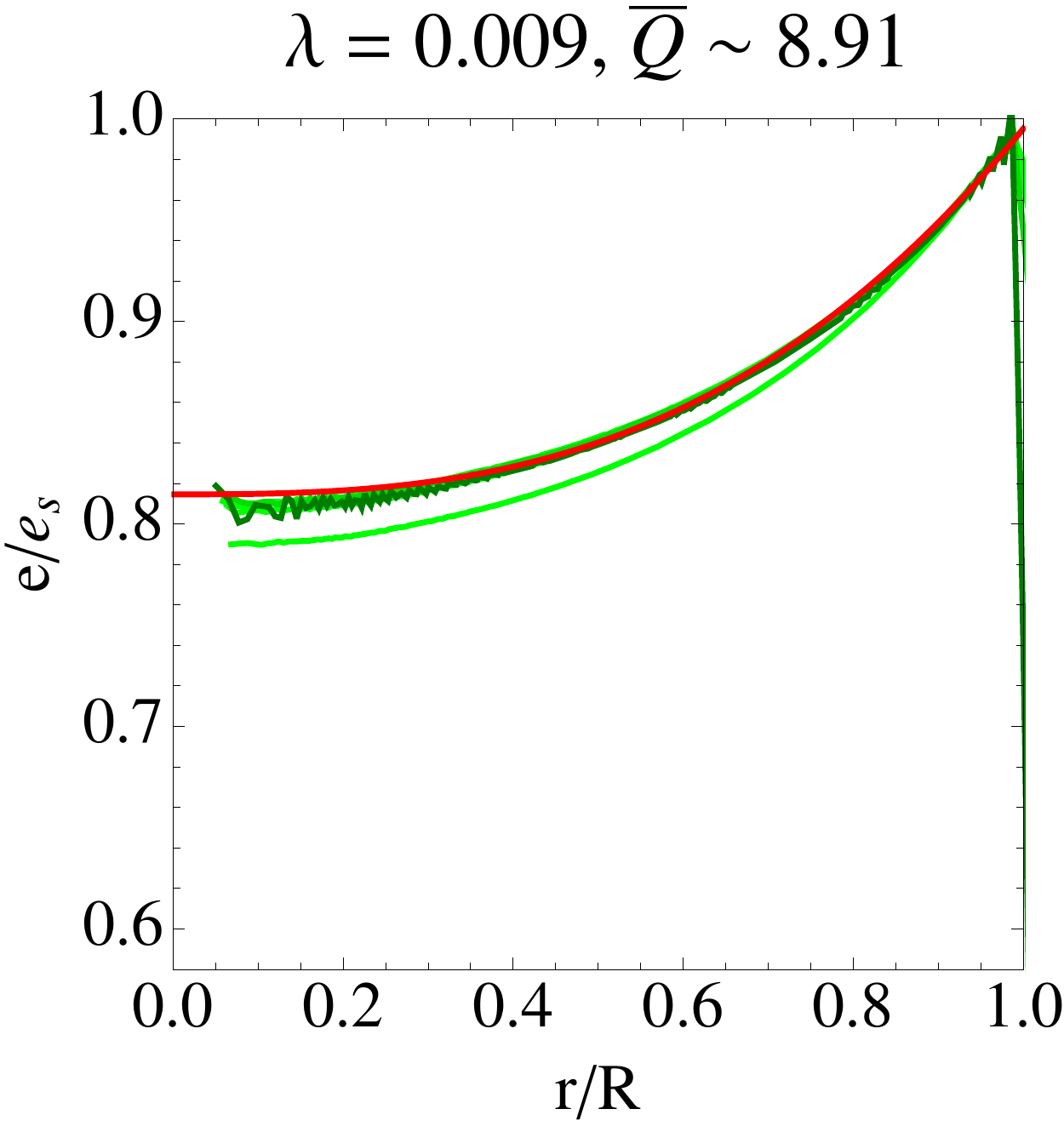}
\includegraphics[width=.215\textwidth]{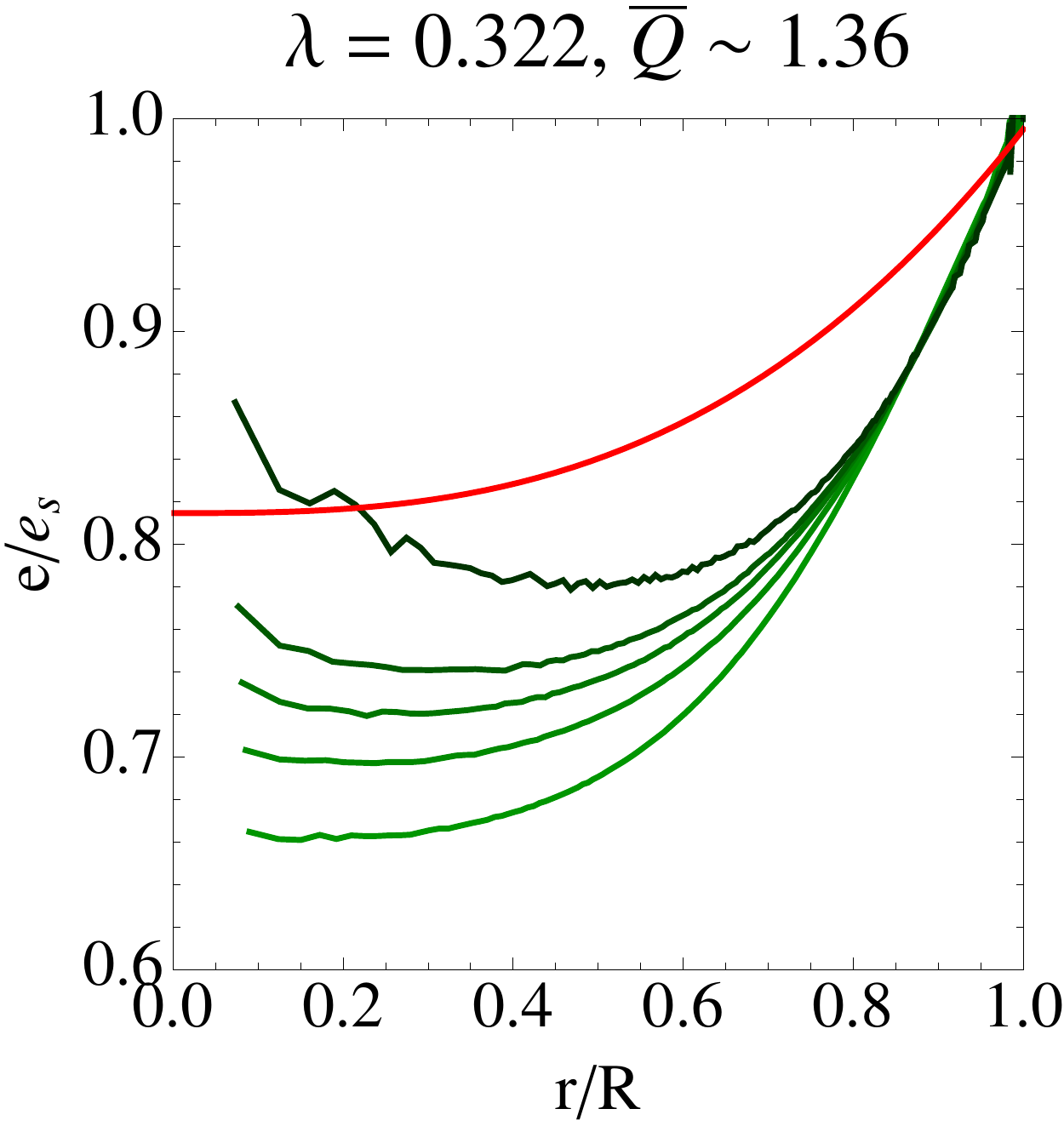}
\caption{
(Color online)
Eccentricity radial profile (normalized by the eccentricity at the stellar surface $e_s$) with a polytropic index $\bar{n}=0.5$ (top) and $\bar{n} = 1$ (bottom). The red curve shows the Newtonian slow-rotation, Clairaut-Legendre model, while the five green curves correspond to the same relativistic rotating models as in Fig.~\ref{RNS-density-N05}. The strange behavior of the eccentricity of
the central regions in the right panels is mainly due to the specific
coordinates used to measure the eccentricity, which have no specific
analogue in Newtonian stars.
\label{RNS-ecc-N05}}
\end{figure}

Let us now study the eccentricity radial profile for rapidly-rotating stars. As in the slow-rotation case, we estimate the eccentricity of constant density surfaces from the embedded surface following~\cite{hartlethorne}. We will present the results in Schwarzschild-like coordinates.
Figure~\ref{RNS-ecc-N05} presents the NS eccentricity profile for $\bar{n}=0.5$ (top) and $\bar{n}=1$ (bottom) polytropic indices, for the various rotation rates of Fig.~\ref{RNS-density-N05}. Observe that in the Newtonian limit (left panels), the Clairaut-Legendre approximation\footnote{The Clairaut-Legendre approximation refers to the expansion calculated by Clairaut using Legendre polynomials~\cite{1978trs..book.....T}, not to be confused with the Clairaut-Radau equation.} (red curves) gives an accurate description of the radial profile of the eccentricity for all the rotating models (green curves). This approximation becomes less accurate as the compactness increases (right panels). Observe that the eccentricity variation due to rotation becomes larger as one increases $\lambda$ and rotation. The relativistic effects with a rapid rotation result in lower central values for the eccentricity and thus higher eccentricity variation throughout the star. The observed behavior of the relativistic eccentricity profile is consistent with the Newtonian picture~\cite{1978trs..book.....T}, where the more centrally condensed polytropes have larger variation in their eccentricity profiles, in agreement with previous results~\cite{1976ApJ...204..561B}.   

Although the relativistic effects lower the central value of the eccentricity compared to the corresponding Newtonian value, the actual variation does not exceed $35\%$, while the variation is $\sim20\%$ throughout a Newtonian $\bar{n}=1$ polytrope. In the region that matters to the universality ($r/R = 0.5$--$0.95$), the eccentricity varies by $(20-30)\%$, validating the elliptical isodensity approximation to $\mathcal{O}(10\%)$ even for rapidly-rotating NSs.

\subsection{Newtonian Multipole Relations with Realistic Eccentricity Profile}
\label{sec:3-hair-real-ecc}

\begin{figure}[tb]
\centering
\includegraphics[width=\columnwidth,clip=true]{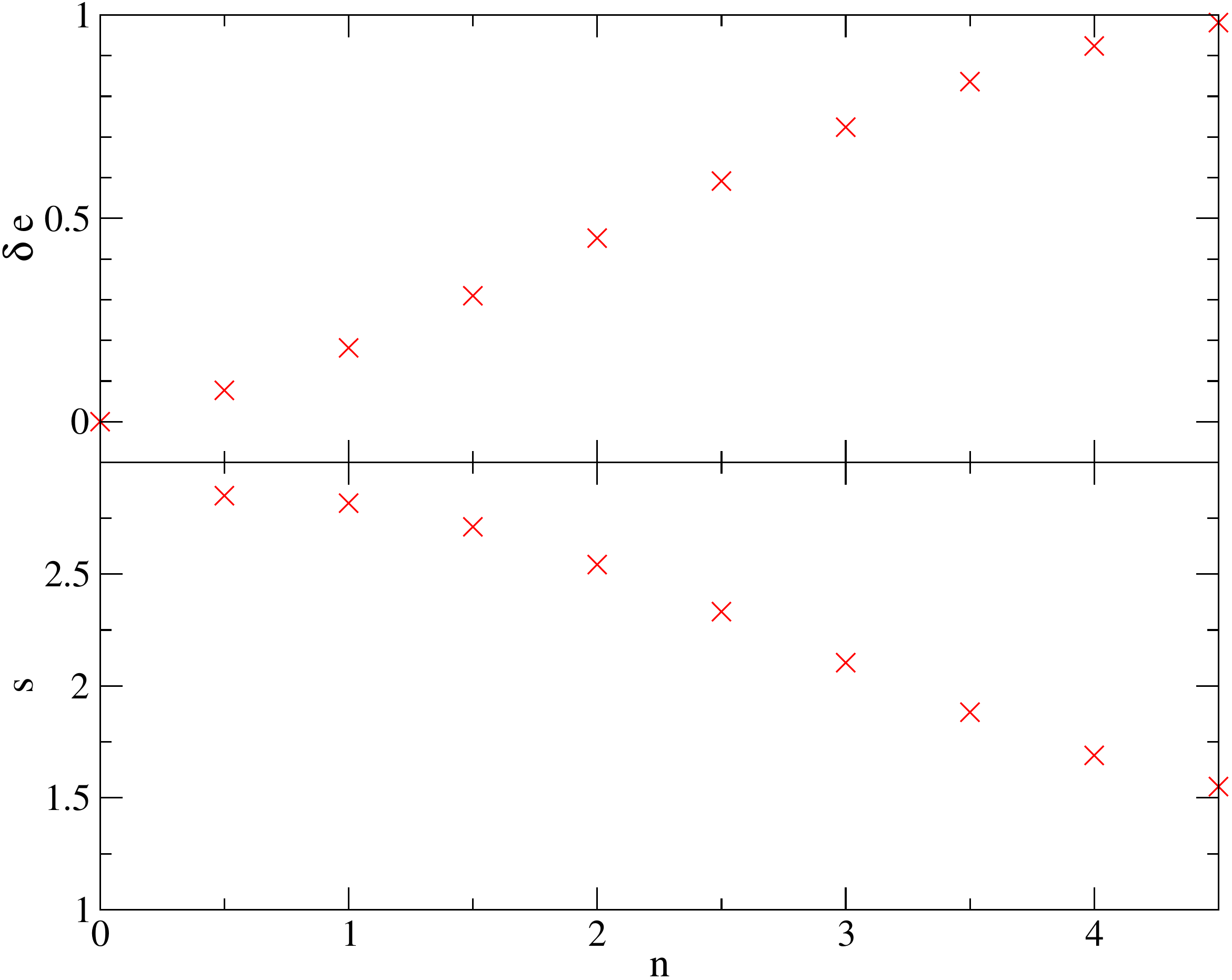}
\caption{
(Color online) 
\label{fig:deltae-s-n} $\delta e$ and $s$ in Eq.~\eqref{eq:e-realistic} as functions of $n$ for Newtonian polytropes.}
\end{figure} 

\begin{figure}[tb]
\centering
\includegraphics[width=\columnwidth,clip=true]{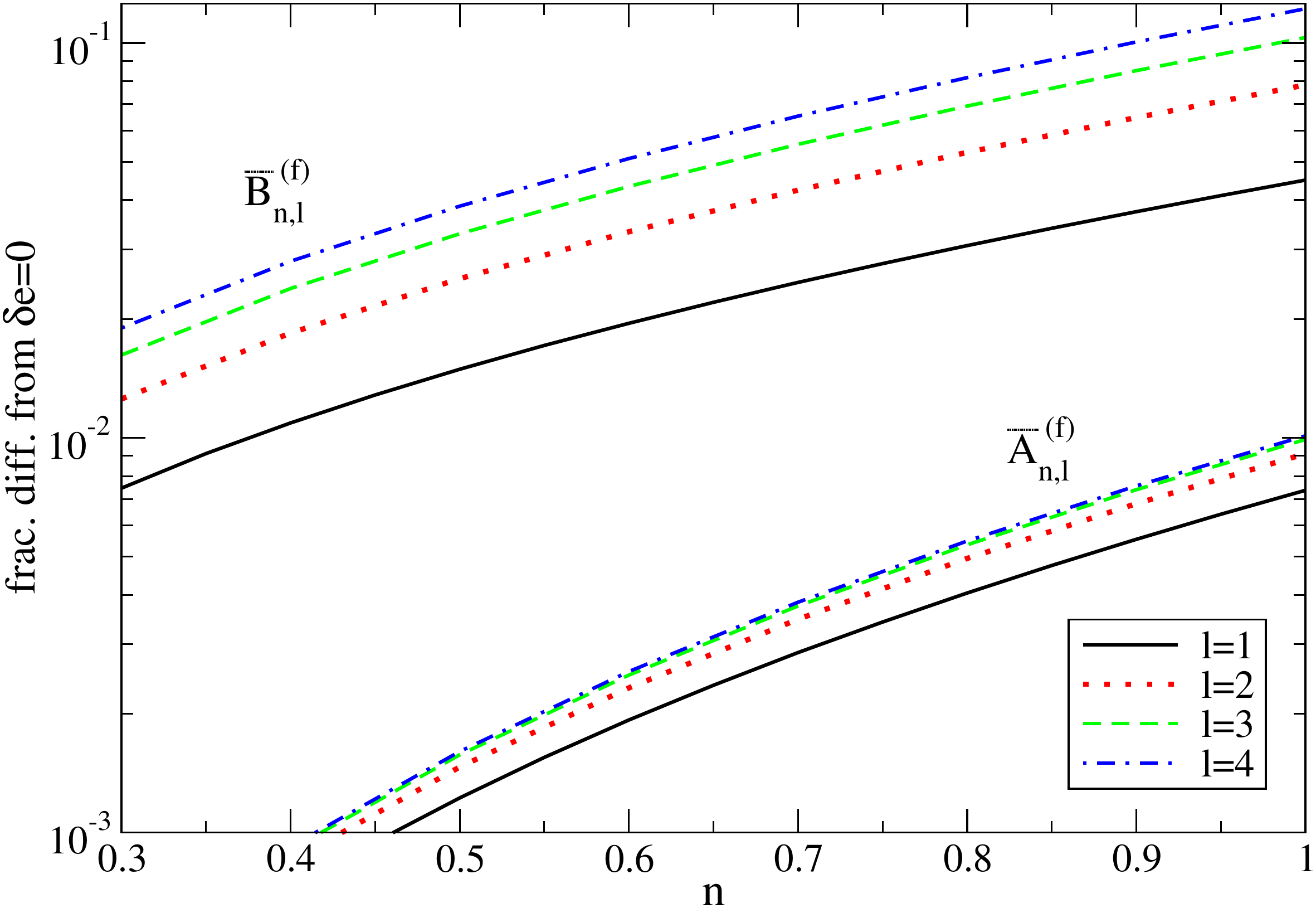}
\caption{
(Color online) 
\label{fig:Abar-Bbar-diff} Fractional difference of $\bar{B}_{n,\ell}^{(f)}$ [Eq.~\eqref{eq:Bbar-noexp}] and $\bar{A}_{n,\ell}^{(f)}$ [Eq.~\eqref{eq:Abar-noexp}] based on Eq.~\eqref{eq:e-realistic} and Fig.~\ref{fig:deltae-s-n} from those with constant eccentricity ($\delta e =0$) as functions of $n$ in the slow-rotation limit.}
\end{figure} 

Let us now consider the Newtonian analysis described in Sec.~\ref{sec:relaxing-self-similarity} but using a realistic eccentricity profile. One sees from Figs.~\ref{fig:ecc-profile-QS} and \ref{RNS-ecc-N05} that the eccentricity for realistic stellar models remains non-vanishing at the stellar center. Therefore, one can promote the eccentricity profile to
\be
\label{eq:e-realistic}
f \left( \frac{\tilde{r}}{a_1} \right) =  1 + \delta e  \left[ \left( \frac{\tilde{r}}{a_1} \right)^s -1 \right]\,,
\ee
where $\delta e$ and $s$ are constants. One can obtain $\delta e$ and $s$ for the slowly-rotating Newtonian polytropes by solving the Clairaut-Radau equation~\cite{1978trs..book.....T,brooker-olle,mora-will}, as shown in Fig.~\ref{fig:deltae-s-n}.

One can use these realistic $\delta e$ and $s$ as functions of $n$ to calculate the coefficients $\bar{B}_{n,\ell}^{(f)}$ and $\bar{A}_{n,\ell}^{(f)}$ in the universal relations. Figure~\ref{fig:Abar-Bbar-diff} shows the fractional difference of $\bar{B}_{n,\ell}^{(f)}$ and $\bar{A}_{n,\ell}^{(f)}$ from those with $f (\tilde{r}/a_1) =1$ (or equivalently, $\delta e=0$) as functions of $n$ for various $\ell$ in the slow-rotation limit. Notice that the fractional difference decreases as one decreases $n$. This is because the elliptical isodensity approximation becomes exact for an $n=0$ polytrope. Notice also that this fractional difference is not a measure of universality, but rather it is a measure of how sensitive the universal relations are to the non-constant eccentricity profile. For example, if in the constant eccentricity case, the relation between the multipoles were accurate to 1\%, then the fractional difference of Fig.~\ref{fig:Abar-Bbar-diff} shows how a non-constant eccentricity profile affects this 1\% accuracy by $\sim0.1\%$ at most.

Figure~\ref{fig:Abar-Bbar-diff} shows that the fractional difference of $\bar{B}_{n,\ell}^{(f)}$ is larger than that of $\bar{A}_{n,\ell}^{(f)}$. One can understand such behavior by expanding the universal relations around $\delta e =0$. To do so, one first needs to expand $\mathcal{R}_{n,\ell}^{(M)}$ and $\mathcal{R}_{n,\ell}^{(S)}$ to yield
\ba
\label{eq:RM-exp}
\mathcal{R}_{n,\ell}^{(M)} &=& \mathcal{R}_{n,\ell} \left[ 1 + \frac{2\ell - (2 \ell +1)e_0^2}{2 (1-e_0^2)} \; \delta e \frac{\mathcal{R}_{n,\ell,s}}{\mathcal{R}_{n,\ell}}  \right] + \mathcal{O} \left( \delta e^2 \right)\,, \nn \\
\\
\label{eq:RS-exp}
\mathcal{R}_{n,\ell}^{(S)} &=& \mathcal{R}_{n,\ell} \left[ 1 + \frac{2 (\ell -2) - (2 \ell -3)e_0^2}{2 (1-e_0^2)} \; \delta e \frac{\mathcal{R}_{n,\ell,s}}{{\mathcal{R}_{n,\ell}}}  \right] \nn \\
& & + \mathcal{O} \left( \delta e^2 \right)\,,
\ea
where
\be
\mathcal{R}_{n,\ell,s} \equiv \int_0^{\xi_1} \vartheta^n \xi^{\ell +2} \left[ \left( \frac{\xi}{\xi_1} \right)^s -1 \right] d\xi\,.
\ee
Substituting Eqs.~\eqref{eq:RM-exp} and~\eqref{eq:RS-exp} into Eqs.~\eqref{eq:Bbar-noexp} and~\eqref{eq:Abar-noexp}, one finds
\ba
\bar{B}_{n,\ell}^{(f)} &=& 1 + 2 \delta e \left( \frac{\mathcal{R}_{n,2\ell +2,s}}{\mathcal{R}_{n,2\ell +2}} - \frac{\mathcal{R}_{n, 2,s}}{\mathcal{R}_{n, 2}} \right) + \mathcal{O} \left( \delta e^2 \right)\,, \\
\bar{A}_{n,\ell}^{(f)} &=& \bar{A}_{n,\ell} \left[ 1 + \frac{e_0^2 \delta e}{2 (1-e_0^2)} \left( \frac{1}{\ell} \frac{\mathcal{R}_{n,2\ell +2,s}}{\mathcal{R}_{n,2\ell +2}} \right. \right. \nn \\
& & \left. \left. - \frac{\ell +1}{\ell} \frac{\mathcal{R}_{n,2,s}}{\mathcal{R}_{n,2}} + \frac{\mathcal{R}_{n,0,s}}{\mathcal{R}_{n,0}} \right) \right] + \mathcal{O} \left( \delta e^2 \right)\,.
\ea
One now sees that the correction to $\bar{B}_{n,\ell}^{(f)}$ is $\mathcal{O} (\delta e)$ while that to $\bar{A}_{n,\ell}^{(f)}$ is $\mathcal{O} (\delta e \; e_0^2)$. 

Figure~\ref{fig:Abar-Bbar-diff} shows that the fractional difference increases as $\ell$ increases. This is somewhat surprising: for higher-$\ell$, the multipole moments are more heavily weighted toward the surface, precisely where the elliptical isodensity approximation should be more accurate, and thus, one expects the fractional difference to decrease as $\ell$ increases. Figure~\ref{fig:Abar-Bbar-diff}, however, shows the opposite behavior. The mass multipole moments are roughly given by $M_\ell \sim \int_0^{a_1} \rho (\tilde{r}) \tilde{r}^{\ell +2} e (\tilde{r})^\ell \; d\tilde{r}$. Because of the $\tilde{r}^{\ell +2}$ factor in the integrand, it is true that the multipole moments have more weight on the outer region inside the star. However, notice that the integrand also contains a factor of $e^\ell$. Since the eccentricity is raised to the $\ell$th power, the multipole moments are more sensitive to the difference between the realistic and constant eccentricity stellar models as $\ell$ increases. Therefore, whether the fractional difference in the universal relations increases as one increases $\ell$ depends on which of these two effects dominate.

In order to quantify the $\ell$-dependence in the fractional difference of the universal relations from the constant eccentricity case, we perform the following scaling estimate in the slow-rotation ($e_0 \ll 1$) and nearly-constant eccentricity ($\delta e \ll 1$) limit. Since all the eccentricity dependence in the universal relations is encoded in $\mathcal{R}_{n,\ell}^{(A)}$, let us focus on this quantity. Using the approximations described above, Eq.~\eqref{eq:RM-exp} leads to  
\be
\mathcal{R}_{n,\ell}^{(M)} = \mathcal{R}_{n,\ell} \left[ 1 +\ell \; \delta e \frac{\mathcal{R}_{n,\ell,s}}{\mathcal{R}_{n,\ell}}  \right] + \mathcal{O} \left( \delta e^2, e_0^2 \right)\,.
\ee
The fractional difference from the constant eccentricity case is then given by
\be
\frac{|\mathcal{R}_{n,\ell}^{(M)} - \mathcal{R}_{n,\ell}|}{\mathcal{R}_{n,\ell}} = \ell \; \delta e \frac{| \mathcal{R}_{n,\ell,s} |}{\mathcal{R}_{n,\ell}}\,.
\ee
As an example, let us use $n \sim 0$, and then
\ba
\mathcal{R}_{n,\ell,s} &\sim & \int_0^{\xi_1} \xi^{\ell +2} \left[ \left( \frac{\xi}{\xi_1} \right)^s -1  \right] d\xi = - \frac{s \; \xi_1^{\ell +3}}{(\ell + 3) (\ell + s + 3)}\,, \nn \\
\mathcal{R}_{n,\ell} &\sim & \int_0^{\xi_1} \xi^{\ell +2} \; d\xi =  \frac{\xi_1^{\ell +3}}{(\ell + 3)}\,,
\ea
and
\be
\frac{|\mathcal{R}_{n,\ell}^{(M)} - \mathcal{R}_{n,\ell}|}{\mathcal{R}_{n,\ell}} \sim \delta e \frac{\ell \; s}{\ell + s + 3}\,,
\ee
which monotonically increases from 0 ($\ell = 0$) to $s$ ($\ell = \infty$). One can easily obtain a similar result for the fractional difference in $\mathcal{R}_{n,\ell}^{(S)}$.  This rough estimate shows that the fractional difference of the universal relations from the constant eccentricity case increases as one increases $\ell$, as shown in Fig.~\ref{fig:Abar-Bbar-diff}. Since the figure shows that the correction from the constant eccentricity case to the universal relations are $\sim 10\%$ at most for $\ell \leq 4$, this justifies the use of the elliptical isodensity approximation to model realistic stars.

\section{Multipole Relations and Eccentricity Profile for Non-compact Stars}
\label{sec:regular-stars}

In order to better understand the relationships satisfied by the
multipole moments of neutron stars, it is instructive to understand
how they are different from regular, non-compact stars. We simulate
regular rotating stars to address two main questions: 
\begin{enumerate}
\item[(i)] Do regular stars satisfy an \ItoQ relation or some other multipole moment
relation? 
\item[(ii)] Is the elliptical isodensity approximation valid for regular stars? 
\end{enumerate}
In this section, we address these questions, which are essentially those posed
in the fourth item of the introduction. The answers to these questions will provide 
further justification of the picture of an emergent symmetry as an explanation for 
the universality, already described in the Introduction and in more detail in the next section.

\begin{figure}[tb]
  \centering
  \includegraphics[width=\columnwidth]{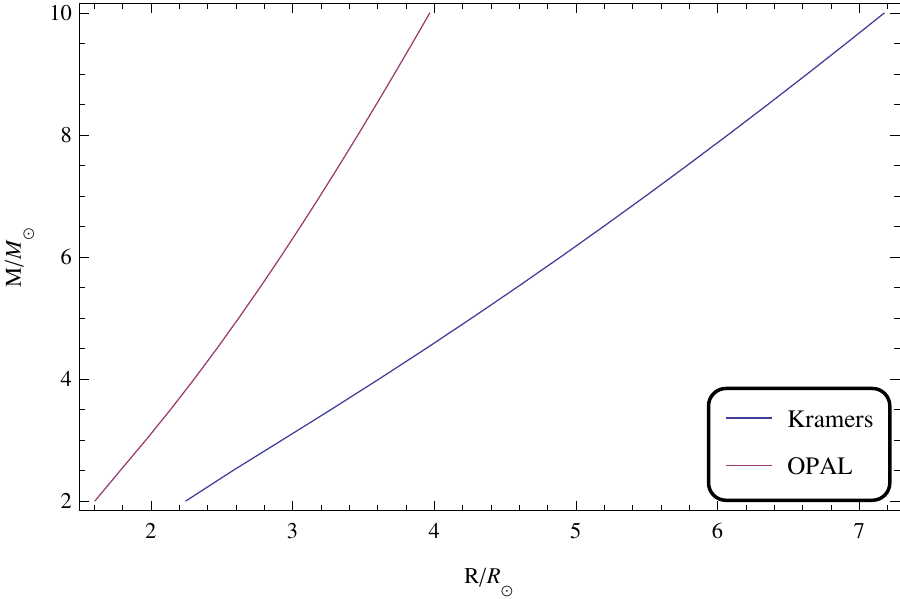}
  \caption{%
    (Color online) The mass-radius relation of non-compact stars with two different opacity laws.
  Observe that different opacity laws corresponds to different effective EoSs.}
  \label{fig:reg-MR}
\end{figure}

We simulate a series of rapidly rotating stellar models, gridded in the mass range $M_{*}\in[2M_{\odot},10
M_{\odot}]$ in increments of $0.5 M_{\odot}$ and in the range of surface equatorial rotation frequencies 
(as a fraction of breakup) $\OmegaEq \in [0.1 \OmegaBk, 0.9 \OmegaBk]$ in increments of $0.1\OmegaBk$. 
Such rotation rates correspond to $\chi = 3-17.5$ for a $3M_\odot$ star.
To perform these simulations we used the publicly-available
\textsc{ESTER} code~\cite{2013LNP...865...49R,2013A&A...552A..35E},
which self-consistently solves the equations of stellar structure for
an axisymmetric, rapidly- and differentially-rotating star with
realistic tabulated EoS.\footnote{ester-project: Evolution STEllaire
  en Rotation, available at
  \url{https://code.google.com/p/ester-project/}. Accessed April
  2014.} We cannot freely vary the equation of state, since it is
well understood for such a star. However, we can vary the opacity law
 as a proxy to understand the EoS
dependence. We adopted two different opacity laws: Kramers' law and
the OPAL tabulated opacity~\cite{1996ApJ...464..943I}.
The latter is more realistic, while we consider the former as a toy model to see the
effective EoS-universality for non-compact stars.
Figure~\ref{fig:reg-MR} shows the mass-radius relation for two different opacity laws.
Observe that changing the opacity law has an effect on the mass-radius
relation similar to changing the EoS.
We address questions (i) and (ii) in
Secs.~\ref{sec:regular-mult-relat} and~\ref{sec:regular-isodensity-contours} respectively.

\subsection{Multipole relationships}
\label{sec:regular-mult-relat}

As the \textsc{ESTER} code is spectral, it is straightforward to
extract the multipole moments from the solutions.\footnote{Program
  available at
  \url{https://github.com/duetosymmetry/ester/tree/master/src/multipoles/}. Accessed
  April 2014.}
So far in this paper, we have used $I = S_{1}/\Omega$
as a definition for the moment of inertia. This definition is
appropriate for rigid rotation, but here we are considering
differentially-rotating stars. Therefore, in this section, we take
advantage of the Newtonian limit and instead use
\begin{equation}
  I^{N} = \frac{2}{3} \int \rho \; r^{2} \; dV\,,
\end{equation}
where the integral is over the entire star; this is
Eq.~\eqref{eq:I-newtonian} without any symmetry assumptions.

{\renewcommand{\arraystretch}{1.2}
\begin{table}[tb]
  \centering
  \begin{tabular}{lr@{$\pm$}lr@{$\pm$}l}
\hline
\hline
\noalign{\smallskip}
\multicolumn{1}{c}{opacity} &
\multicolumn{2}{c}{$b_1$} &
\multicolumn{2}{c}{$b_2$}\\
\hline
\noalign{\smallskip}
Kramers & 0.527 & 0.0051 & --1.39 & 0.11 \\
OPAL & 0.712 & 0.0022 & --2.90 & 0.050 \\
\noalign{\smallskip}
\hline
\hline
  \end{tabular}
  \caption{Fit parameters for the power-law parametric form in
    Eq.~\eqref{eq:logQtologI} for
    the two families with different opacity laws. Each family has an
    approximately (2--3)\% variation from the fit, and the two fits have an
    approximately 40\% variation at most from each other.}
  \label{tab:regular-IQ-fit-params}
\end{table}
}

\begin{figure}[tb]
  \centering
  \includegraphics[width=\columnwidth]{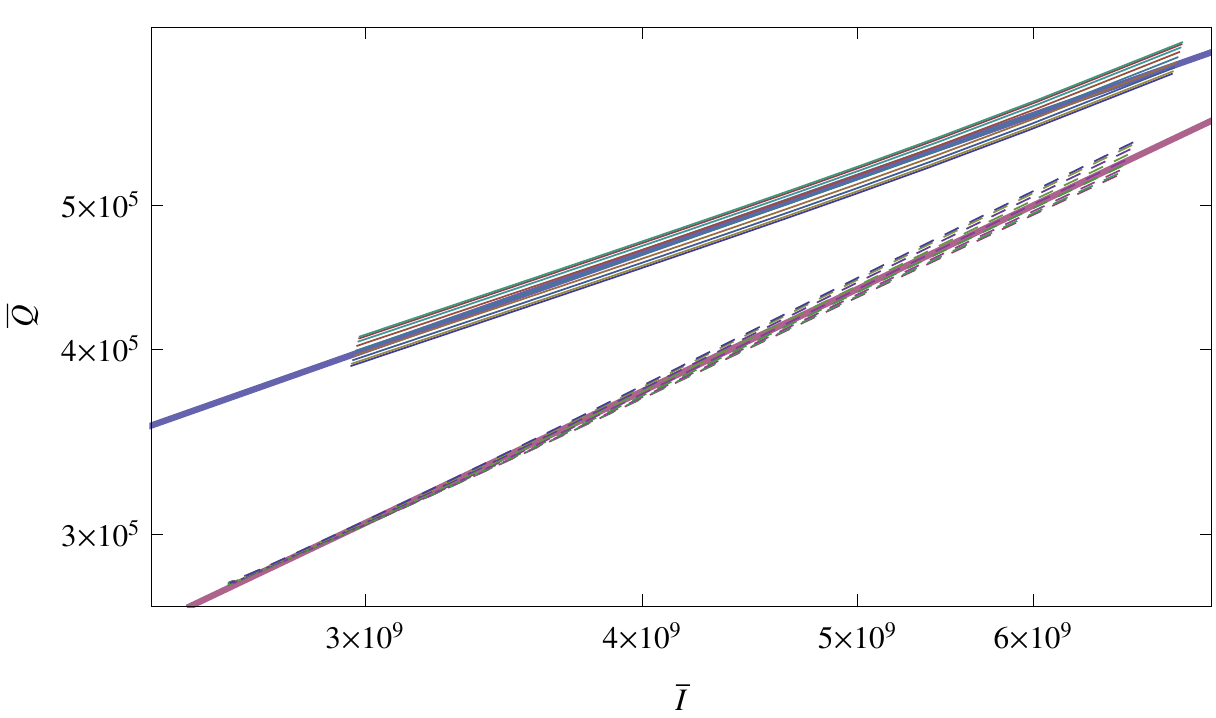}
  \caption{%
    (Color online) \ItoQ relation for regular
    stars of various masses, spins, and opacity laws of Kramers' (solid) and OPAL (dashed).
    Within each opacity law family, each curve has a constant
    $\OmegaEq/\OmegaBk=0.1,0.2,\ldots
    0.9$. Along each curve, total stellar mass increases to the left,
    with $M_{*}=2M_{\odot}$ at the rightmost end and
    $M_{*}=10M_{\odot}$ at the leftmost end. Within each family there is (2--3)\% variation
    from the best fit,
    while the two fits differ from each other by 40\% at most.}
  \label{fig:reg-IQ}
\end{figure}

The \ItoQ relationship for regular stars is shown in 
Fig.~\ref{fig:reg-IQ} with Kramers' (solid) and OPAL (dashed) opacity laws.
Each curve in each opacity-family is at a constant
$\OmegaEq/\OmegaBk$, with $M_{*}$
varying along a curve (and increasing to the left).  We create a fit for each opacity-family
given by a power-law of the
form $\bar{Q}=C \bar{I}^{b_1}$ by fitting a straight line in log space,
i.e.
\begin{equation}
\label{eq:logQtologI}
\log\bar{Q} = b_1 \log\bar{I} + b_2\,.
\end{equation}
The fit parameters are presented in
Table~\ref{tab:regular-IQ-fit-params}. 
The fractional difference between these two fits can be 
as large as 40\%, showing a loss of universality.
Notice that the
dependence on $\Omega$ is weak, about (2--3)\% variation, unlike 
in the relation for compact stars, which has a clear 
spin dependence~\cite{doneva-rapid,Pappas:2013naa,Chakrabarti:2013tca,Yagi:2014bxa}.

Recall that in the Newtonian limit~\cite{Stein:2013ofa,I-Love-Q-PRD},
the \ItoQ relation for compact stars is $\bar{Q}\propto
\bar{I}^{1/2}$. From the $b_1$ parameter in
Table~\ref{tab:regular-IQ-fit-params} we see that non-compact stars lie
on \emph{different} \ItoQ relations than compact stars. We also see
that changing the opacity law, which is our proxy for the EoS, changes
the fit, and so non-compact stars do not possess an \ItoQ relation that is
nearly as universal as that which hold for compact stars. 

\begin{figure}[tb]
  \centering
  \includegraphics[width=\columnwidth]{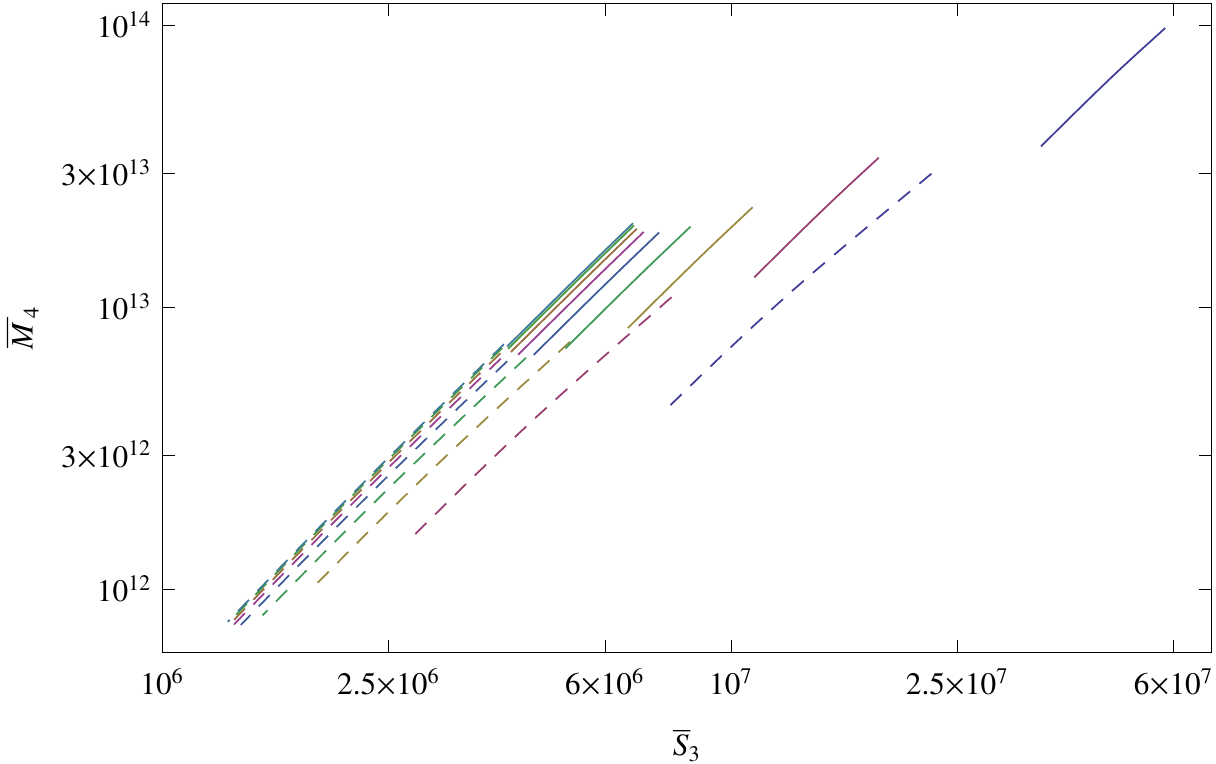}
  \caption{%
    (Color online) The $\bar{S}_{3}$--$\bar{M}_{4}$ relation for
    regular stars with Kramers' (solid) and OPAL (dashed) opacity law, 
    which fails to exhibit universality in both 
    effective EoS (opacity law) and spin. 
    Within each family, each curve is
    for a fixed $\OmegaEq/\OmegaBk=0.1,0.2,\ldots0.9$ which increases
    to the right amongst curves. Along each curve, mass increases to
    the left. Observe how the effective EoS universality is lost especially for 
    a fixed, relatively large $\OmegaEq/\OmegaBk$.}
  \label{fig:reg-S3M4}
\end{figure}

Compact stars also enjoy additional relationships amongst higher
multipole moments that are also approximately universal~\cite{Pappas:2013naa,Stein:2013ofa}; 
is such universality still present for non-compact stars? 
Figure~\ref{fig:reg-S3M4} shows the relation between
$\bar{S}_{3}$ and $\bar{M}_{4}$ for non-compact stars, which should be
the next most-universal combination after \ItoQ~\cite{Stein:2013ofa},
for Kramers' (solid) and OPAL (dashed) opacity laws.
Within each opacity-family, we vary $\OmegaEq/\OmegaBk$.
For slowly-rotating stars, the effective EoS universality
is preserved, while it is lost for rapidly-rotating stars with
a fixed dimensionless spin parameter of $\OmegaEq/\OmegaBk$.
The figure also exhibits a clear spin-dependence in the relations,
which is in contrast to the relation for compact stars~\cite{Pappas:2013naa,Yagi:2014bxa}.
Such loss in the universality could be considered as a consequence of the breakdown of
the elliptical isodensity approximation, as we will see in the next subsection.

\subsection{Isodensity contours}
\label{sec:regular-isodensity-contours}

\begin{figure}[tb]
  \centering
  \includegraphics[width=\columnwidth]{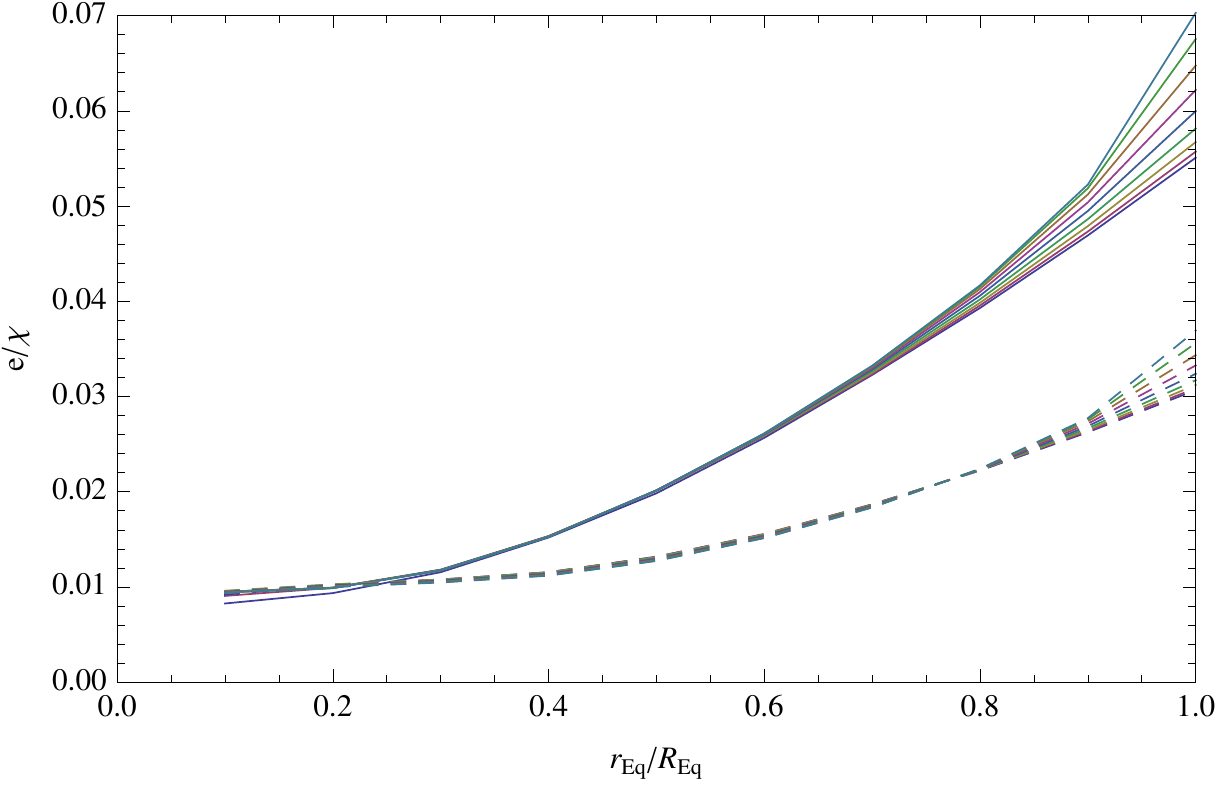}
  \caption{%
    (Color online) Radial dependence of $e/\chi$ values in regular
    stars with Kramers' (solid) and OPAL (dashed) opacity law. 
    $r_\Eq$ and $R_\Eq$ correspond to the radial coordinate in 
  the equatorial plane and the stellar equatorial radius respectively. 
  The values are best-fits to isodensity contours. 
  Within each family, each curve
  corresponds to one value of $\OmegaEq/\OmegaBk=0.1,0.2,\ldots0.9$,
  increasing upwards. We set the mass as $M=5M_{\odot}$, but the result presented in this figure
  is insensitive to the choice of $M$. }
  \label{fig:reg-eOverChi}
\end{figure}

We now turn to the question of the validity of the elliptical
isodensity contour approximation for regular stars. Isodensity contours
were extracted from the \textsc{ESTER} simulations by
root-finding,\footnote{Program available at
  \url{https://github.com/duetosymmetry/ester/tree/master/src/iso_contours}. Accessed
  April 2014. This program makes use of the GNU Scientific
  Library~\cite{galassi2009gnu}.} and then computing the best-fit
ellipse to each contour. Figure~\ref{fig:reg-eOverChi} shows an example of the radial dependence
of $e/\chi$ with Kramers' (solid) and OPAL (dashed) opacity laws with $M=5M_\odot$ for different
spin frequencies. The fits are quite
faithful, with fractional errors between the contour and fit of at worst
(i.e.~at the stellar surface) 7\%, and sub-1\% for spins below
$\OmegaEq < 0.7\OmegaBk$.

From these fits we infer another important difference between compact
and non-compact stars. In Figs.~\ref{fig:ecc-profile} and
\ref{fig:ecc-profile-QS} we saw that the $e/\chi$ does not vary much
in the range of radii that contribute the most to the
moment of inertia and quadrupole moment. By comparison, for regular stars, we can see from
Fig.~\ref{fig:reg-eOverChi} that $e/\chi$ varies by a factor of
$\sim$3--6 throughout the entire star, much more than for compact
stars. Such large eccentricity variation is precisely the reason why 
the effective EoS universality is lost for the \ItoQ and $\bar{S}_3$--$\bar{M}_4$ relations
in general. These results provide further evidence for the picture of an emergent symmetry 
as the origin of universality, which we will explain in more detail in the next section.

\section{Universality as an Emergent Approximate Symmetry}
\label{sec:emergent}

Let us now summarize our work and describe in detail the picture of emergent approximate symmetry that arises as an explanation for the universality.

In this paper, we provided further evidence of why the universal relations hold among stellar multipole moments. We first tackled this problem by looking at the \ItoQ relation using piecewise polytropic EoSs~\cite{Read:2008iy} that can reproduce various realistic EoSs with just five free parameters. Although varying each piecewise parameter by 30\% significantly modifies the mass-radius relation, we found that it only affects the \ItoQ relation by $\mathcal{O}(1\%)$ at most. We also found that the radial profile of the integrand of the moment of inertia and quadrupole moment is hardly affected by the piecewise parameters for a fixed moment of inertia, and most of the contribution in the moment of inertia and quadrupole moment comes from (50--95)\% of the radius for NSs, confirming~\cite{I-Love-Q-Science,I-Love-Q-PRD}. The parameters that control the density in the range $(10^{14}$--$10^{15})$g/cm$^3$ play the most important role in the universality. The variation in the slope of the nuclear EoS within this region can be as large as 17\% and we concluded that this alone cannot explain the universality.

Second, we extended the work in~\cite{Stein:2013ofa} in the Newtonian limit. We relaxed the elliptical isodensity approximation~\cite{Lai:1993ve} to see how this assumption affects the universality. We found that the shape of the isodensity contours does not affect the universality, but relaxing the self-similarity condition does modify the relation significantly. This shows that the fact that isodensity contours of NSs can be approximated by self-similar surfaces plays a crucial role in universality.

Third, we investigated the eccentricity profile for both slowly- and rapidly-rotating relativistic stars. We showed that the eccentricity only varies by $\sim$10\% in the region that matters to the universality for slowly-rotating NSs and QSs, while the eccentricity variation is below (20---30)\% even for rapidly-rotating NSs. This suggests the elliptical isodensity approximation is a very good description for realistic NSs. To verify this, we improved the Newtonian analysis in~\cite{Stein:2013ofa} by using a realistic, radially-dependent eccentricity profile for NSs. We found that, in the slow-rotation limit, the radial dependence of the eccentricity introduces at most a 10\% correction to the \ItoQ relation within the constant eccentricity assumption.

Fourth, we looked at the multipole relations and eccentricity profiles for non-compact, regular stars. 
We found that the eccentricity variation can easily exceed 100\% for such stars, in contrast to relativistic stars that only have $\sim$10\% variation for similar rotation rates. This means that the elliptical isodensity approximation is not good to model non-compact stars. We then studied whether the \ItoQ relations remained universal for non-compact stars when we change its EoS. As a proxy for the latter, we used two different opacity laws, which indeed lead to different slopes in the mass-radius relation. We found that the \ItoQ relations and the relations among higher multipole moments are not EoS-universal for non-compact, regular stars. 

The above results paint an interesting, albeit phenomenological picture of why
universality holds. Consider the multi-dimensional phase space,
spanned by different quantities that characterize stars, such as
their stellar temperature, compactness, rotation rate, strength of the
magnetic field, etc. One corner of this space is inhabited by very hot
and non-compact stars, like supergiants, with stellar compactness of
about $10^{-8}$ and temperatures of $20,000$ K, which may rotate
differentially in some interior region. Since the interior density in
such stars is subnuclear, the EoS can be well-modeled through
simulations, experimental data and helio-seismological
observations. Another corner of this space is inhabited by cold
(relative to their Fermi temperature) and compact stars, like NSs with
compactness of about $10^{-1}$ and temperatures of $10^{-9} T_{F}$,
where $T_{F}$ is the Fermi temperature. These stars rotate rigidly in
the absence of external perturbations, because in the barotropic
limit, vorticity (and thus differential rotation) is unsourced. Since the
interior density in such stars is supra-nuclear, the EoS is not
perfectly well-understood.

In spite of our ignorance of the NS's EoS, the corner of phase space in which they live is effectively dominated by just two dimensions: compactness and EoS effective polytropic index $n$. All other extra dimensions related to microphysics, like temperature, play a small, negligible role in controlling the characteristics of NSs. That is, at sufficiently high compactness, all details of microphysics \emph{efface} away and NSs can be well-described by barotropic EoSs. Effectively, this EoS can be approximately captured by a polytrope, with an effective polytropic index $n \in [0.3,1]$. 

Let us then span this corner of phase space by one axis that measures compactness and one axis that measures the effective polytropic index. In the subregion that corresponds to NSs, i.e.~for $C = {\cal{O}}(0.1)$ and $n \in [0.3,1]$, the isodensity profiles are approximately self-similar in the region inside the star that matters the most for the calculation of multipole moments, i.e.~the eccentricity of isodensity contours remains invariant under a radial remapping. This is not the case for realistic non-compact stars which have large eccentricity variations in their interior. Thus, as compactness increases and one flows to the NS corner of phase space (see Fig.~\ref{fig:schematic-diag-emergence}), an \emph{approximate symmetry} emerges (isodensity contours become approximately self-similar ) and this is responsible for the universality we have observed.  As one approaches the region of phase space inhabited by BHs, the universality becomes exact, as expressed by the no-hair theorems. 

\section{Future Directions}
\label{sec:future}

In this paper, we presented the relations among multipole moments for differentially-rotating stars in the Newtonian limit for the first time. The next task is to see how the universal relations change due to differential rotation in the relativistic case. For this, one could adopt the rotation law given in~\cite{komatsu_eh1989}. It would also be interesting to derive an analytic relation among multipole moments for differentially-rotating stars and compare it with the relativistic results. 

As explained in the introduction, other types of universal relations exist, such as those among higher-$\ell$ tidal Love numbers~\cite{Yagi:2013sva} and NS oscillation modes~\cite{andersson-kokkotas-1998,tsui-leung,lau}. It would be important to carry out a similar analysis as presented in this paper to understand why the universality holds among these observables. One can for example take a piecewise polytropic EoS and see which parameter affects the relations the most. One can also investigate which radial region affects the relations the most, and determine which region in the EoSs controls the universality.  

Another interesting idea would be to study whether the \ItoQ relations remain universal if one breaks the elliptical isodensity approximation. In this paper, we studied how the relation among multipole moments behaves as this approximation is relaxed. To repeat the analysis for the \ItoQ relation, one would have to determine how the angular frequency--eccentricity relation is modified as one promotes the eccentricity to a radial function. Based on the results of this paper, we suspect the \ItoQ relations will be similarly affected if one relaxes the elliptical isodensity approximation.

Finally, one could carry out a similar analysis as that performed in this paper but to study the origin of universality in theories other than GR, such as dynamical Chern-Simons gravity~\cite{I-Love-Q-Science,I-Love-Q-PRD}, Eddington-inspired Born-Infeld gravity~\cite{Sham:2013cya} and scalar-tensor theories~\cite{Pani:2014jra}. It would be interesting to study how the variation in the NS eccentricity in the region that matters changes and see if the explanation found here in GR can also be applied to the multipole relations in such theories.

{\bf Addendum.}---%
After we posted the preprint of this manuscript,
Ref.~\cite{Martinon:2014uua} came out where the authors studied the
I-Love-Q relations for proto-NSs with a non-barotropic EoS. They found
that at the early stage of the NS formation, when the entropy gradient
inside the star is large, the relation differs from the one with
barotropic EoSs by as much as $\sim 30\%$. Moreover, they found that
the eccentricity inside the star varies by $\sim 200\%$ at this early
stage of formation. This supports one of our claims, that the relation
depends strongly on the stellar eccentricity variation. However, we
note that Ref.~\cite{Martinon:2014uua} does not discuss how the
universality itself changes with time for proto-NSs, as the authors
only looked at a single non-barotropic EoS. Namely, the reference does
not provide any evidence on how the EoS-universality itself depends on
the eccentricity variation.

\acknowledgments
We would like to thank Ben Owen, Jan Steinhoff, Terence Delsate and Katerina Chatziioannou for useful comments and suggestions. 
NY acknowledges support from NSF grant PHY-1114374 and the NSF CAREER Award PHY-1250636, as well as support provided by the National Aeronautics and Space Administration from grant NNX11AI49G, under sub-award 00001944. 
LCS acknowledges that support for this work was provided by the National Aeronautics and Space Administration through Einstein Postdoctoral Fellowship Award Number PF2-130101 issued by the Chandra X-ray Observatory Center, which is operated by the Smithsonian Astrophysical Observatory for and on behalf of the National Aeronautics Space Administration under contract NAS8-03060, and further acknowledges support from NSF grant PHY-1068541. 
GP acknowledges financial support from the European Research Council under the European Union's Seventh Framework Programme (FP7/2007-2013) / ERC grant agreement n. 306425 ``Challenging General Relativity''.
Some calculations used the computer algebra-systems \textsc{MAPLE}, in combination with the \textsc{GRTENSORII} package~\cite{grtensor}.

\bibliography{master}
\end{document}